\documentclass[11pt]{article}

\usepackage{amsthm,amsmath}
\RequirePackage[numbers]{natbib}
\RequirePackage[colorlinks,citecolor=blue,urlcolor=blue]{hyperref}
\usepackage[bottom=1.5in,top=0.5in,left=1in, right=1in]{geometry}

\usepackage{algorithm}
\usepackage{enumerate}
\usepackage{ifthen}
\usepackage[utf8]{inputenc}
\usepackage{framed}
\usepackage{chngcntr}

\usepackage[utf8]{inputenc}
\usepackage{makeidx}
\usepackage{amsfonts}
\usepackage{amssymb}
\usepackage{booktabs}

\usepackage{gensymb}
\usepackage{tabularx}
\usepackage{float}
\usepackage{caption}
\usepackage[labelformat=simple]{subcaption}
\usepackage{bbm}
\usepackage{lipsum}
\usepackage{enumitem}
\usepackage{hyperref}
\usepackage[english]{babel} % Document language - required for customizing Section titles
\usepackage{bm} % Required for bold math symbols
\usepackage[T1]{fontenc}	 % For correct hyphenation and T1 encoding
\usepackage{lmodern}
\usepackage{amsthm} % Required for theorem environments
\usepackage{graphicx} % Required for including images in figures
\usepackage{booktabs} % Required for horizontal rules in tables
\usepackage{multicol} % Required for creating multiple columns in slides
\usepackage[toc,page]{appendix}

\newcommand{\footremember}[2]{%
	\footnote{#2}
	\newcounter{#1}
	\setcounter{#1}{\value{footnote}}%
}

\DeclareMathOperator*{\argmax}{arg\,max}

\allowdisplaybreaks
\numberwithin{equation}{section}

\theoremstyle{plain}

\newtheorem{proposition}{Proposition}

\newcommand{\numbereqn}{\addtocounter{equation}{1}\tag{\theequation}} % use \numberthis to add number in align* mode

\allowdisplaybreaks

\begin{document}

\hypersetup{linkcolor=blue}

\date{\today}

\author{Ray Bai\footremember{USCStat}{Department of Statistics, University of South Carolina, Columbia, SC 29208.}\thanks{Email: \href{mailto:RBAI@mailbox.sc.edu}{\tt RBAI@mailbox.sc.edu} }, Mary R. Boland\footremember{StVincent}{Department of Mathematics, Saint Vincent College, Latrobe, PA 15650.}, Yong Chen\footremember{DBEI}{Department of Biostatistics, Epidemiology, and Informatics, University of Pennsylvania, Philadelphia, PA 19104.}}

\title{Scalable high-dimensional Bayesian varying coefficient
	models with unknown within-subject covariance 
}

\maketitle

\begin{abstract}
\noindent Nonparametric varying coefficient (NVC) models are useful for modeling time-varying effects on responses that are measured repeatedly for the same subjects. When the number of covariates is moderate or large, it is desirable to perform variable selection from the varying coefficient functions. However, existing methods for variable selection in NVC models either fail to account for within-subject correlations or require the practitioner to specify a parametric form for the correlation structure. In this paper, we introduce the nonparametric varying coefficient spike-and-slab lasso (NVC-SSL) for Bayesian high-dimensional NVC models. Through the introduction of functional random effects, our method allows for flexible modeling of within-subject correlations without needing to specify a parametric covariance function. We further propose several scalable optimization and Markov chain Monte Carlo (MCMC) algorithms. For variable selection, we propose an Expectation Conditional Maximization (ECM) algorithm to rapidly obtain maximum a posteriori (MAP) estimates. Our ECM algorithm scales linearly in the total number of observations $N$ and the number of covariates $p$. For uncertainty quantification, we introduce an approximate MCMC algorithm that also scales linearly in both $N$ and $p$. We demonstrate the scalability, variable selection performance, and inferential capabilities of our method through simulations and a real data application. These algorithms are implemented in the publicly available \textsf{R} package \texttt{NVCSSL} on the Comprehensive \textsf{R} Archive Network.
\end{abstract}

\section{Introduction} \label{intro}

\subsection{Model set-up} \label{modelsetup}

Consider the nonparametric varying coefficient (NVC) model with $p$ covariates,
\begin{equation} \label{varyingcoefficientmodel}
	y_i(t_{ij}) = \displaystyle \sum_{k=1}^{p} x_{ik} (t_{ij}) \beta_k (t_{ij}) + \varepsilon_{ij}(t_{ij}),~~i = 1, \ldots, n,~j = 1, \ldots, m_i,
\end{equation}
where $y_i(t)$ is the response for the $i$th subject at time point $t \in \mathcal{T}$, $\mathcal{T}$ is the time interval on which the $m_i$ different measurements are taken, $x_{ik}(t)$ is a possibly time-dependent covariate with corresponding smooth coefficient function $\beta_k(t)$, and $\varepsilon_{ij} := \varepsilon_{ij}(t_{ij})$ is random error. Throughout this paper, we denote $N = \sum_{i=1}^{n} m_i$ as the total number of observations.  We also assume that the error terms $\bm{\varepsilon}_i = ( \varepsilon_{i1}, \ldots, \varepsilon_{im_i})^\top$, $i = 1, \ldots, n$, are independent, zero-mean Gaussian processes. That is, $\bm{\varepsilon}_i \sim \mathcal{N} ( \bm{0}, \bm{\Sigma}_i), i=1, \ldots, n$, where $\bm{\Sigma}_i$ is the $m_i \times m_i$ variance-covariance matrix that captures the temporal correlation between the $n_i$ responses, $y_{i}(t_{i1}), \ldots, y_{i}(t_{im_i})$, for the $i$th subject. 

NVC models \eqref{varyingcoefficientmodel} arise in many real applications. A prominent example is in longitudinal data analysis where we aim to model the response for the $i$th experimental subject at $m_i$ different time points \citep{HooverRiceWuYang1998}. NVC models can also be used for functional data analysis where the objective is to model functional responses $y_i(t), i=1, \ldots, n,$ varying over a continuum $t \in \mathcal{T}$ \citep{Rice2004}. \citet{HastieTibshirani1993} and \citet{FanZhang2008} provide some further examples of applications of these models. Under \eqref{varyingcoefficientmodel}, the primary aim is to estimate and conduct inference for the varying coefficients $\beta_k(t), k = 1, \ldots, p$.

There has been extensive frequentist work on fitting NVC models. Typical approaches to fitting \eqref{varyingcoefficientmodel} use local polynomial kernel smoothing \citep{FanZhang2000, WuChiang2000} or basis expansions \citep{HuangWuZhou2004, QuLi2006, XueQu2012} to estimate the varying coefficients. Bayesian approaches to NVC models have also been developed. \citet{LiuBobbHennGenningsSchnaasTellezRojoBellingerAroraWrightCoull2018} and \citet{GuhaniyogiLiSavitskySrivastava2022} endow the varying coefficients with a Gaussian process (GP) prior. \citet{BillerFahrmeir2001} and \citet{HuangLiNottFengNgWong2015} use splines to model the $\beta_k(t)$'s in \eqref{varyingcoefficientmodel} and place multivariate normal priors on the groups of basis coefficients. \citet{LiWangLiWu2015} use a scale-mixture of a multivariate normal distribution as a prior to shrink groups of basis coefficients towards zero. \citet{DeshpandeBaiBalocchiStarlingWeiss2020} use Bayesian additive regression trees (BART) to model the varying coefficients. 

\subsection{Related work} \label{relatedwork}

When the number of covariates $p$ is large, it is often desirable to perform variable selection from the varying coefficient functions. In the frequentist literature, many authors have applied penalty functions such as the group lasso \citep{YuanLin2006} in order to threshold many of the $\beta_k(t)$'s to zero. See, e.g. \citet{WangXia2009}, \citet{WangLiHuang2008}, and \citet{WeiHuangLi2011}. These frequentist penalized NVC models do not account for the within-subject temporal correlations, essentially solving penalized likelihood objective functions with $\bm{\varepsilon} = ( \bm{\varepsilon}_1^\top, \ldots, \bm{\varepsilon}_n^\top)^\top \sim \mathcal{N} (\bm{0}, \bm{I}_N)$ in \eqref{varyingcoefficientmodel}.

In low-dimensional settings and without regularizing the parameter space, \citet{KraftyGimottyHoltzCoukosGuo2008} and \citet{ChenWang2011} incorporated estimation of within-subject correlations into NVC models. However, to the best of our knowledge, no similar extension has been made for high-dimensional, penalized NVC models.  While many researchers \citep{WangXia2009, WangLiHuang2008, WeiHuangLi2011, XueQu2012} have shown that consistent estimation of the $\beta_k$'s and model selection consistency can still be achieved for penalized NVC models, failing to account for the error variances can nevertheless lead to invalid \emph{inferences} in finite samples \citep{LiangZeger1993}. Thus, it seems prudent to explicitly model temporal dependence in NVC models. Furthermore, while point estimates are easily attained, another major limitation of current penalized NVC models is their lack of inferential capabilities.

Unlike frequentist penalized approaches, the Bayesian approaches of \citet{LiuBobbHennGenningsSchnaasTellezRojoBellingerAroraWrightCoull2018}, \citet{LiWangLiWu2015}, \citet{DeshpandeBaiBalocchiStarlingWeiss2020}, and \citet{GuhaniyogiLiSavitskySrivastava2022} explicitly model dependencies by specifying a parametric correlation structure for the residual errors or the cross-covariance correlation function. \citet{LiuBobbHennGenningsSchnaasTellezRojoBellingerAroraWrightCoull2018} employ subject-specific random effects with a random intercept and a random slope,  \citet{DeshpandeBaiBalocchiStarlingWeiss2020} use a compound symmetry covariance structure,   \citet{LiWangLiWu2015} use a first-order autoregressive process, and \citet{GuhaniyogiLiSavitskySrivastava2022} use the exponential or the Gneiting's correlation functions. The choices of covariance structure in \citet{DeshpandeBaiBalocchiStarlingWeiss2020}, \citet{LiWangLiWu2015}, and \citet{GuhaniyogiLiSavitskySrivastava2022} are parameterized by one to three hyperparameters (e.g. an autocorrelation parameter $\rho$). Suitable priors are then placed on these hyperparameters. Apart from being able to properly handle correlations, Bayesian NVC models also allow for natural uncertainty quantification of the varying coefficients through their posterior distributions. 

While the aforementioned Bayesian approaches enable modeling of within-subject correlation, one of their limitations is the need to prespecify a parametric correlation structure. In practice, the final estimates can be sensitive to the choice of kernel function \citep{StephensonGhoshNguyenYurochkinDeshpandeBroderick2022}, and misspecifying the correlation structure may lead to incorrect inferences about the model parameters. For example, if there are long-range dependencies, then correlation functions that decay exponentially with distance (such as the ones used in \citet{GuhaniyogiLiSavitskySrivastava2022}) will not be able to adequately capture the dependence between responses observed at far apart time points.

Another challenge with Bayesian NVC models is computational. In practice, Markov chain Monte Carlo (MCMC) is typically used to fit these Bayesian NVC models. However, when the number of observations $N$ and/or the number of predictors $p$ is large, MCMC can be computationally prohibitive. Recently, there have been efforts to scale up Bayesian NVC models when $N$ is large. \citet{GuhaniyogiLiSavitskySrivastava2022} employ divide-and-conquer MCMC which divides the $N$ data points into subsets of much smaller size, runs MCMC in parallel on each subset, and then combines the posterior samples in a principled manner to approximate the full data posterior. In a separate line of work, \citet{GuhaniyogiBaracaldoBanerjee2023} employed data sketching, which first compresses the dataset to a much smaller size through a random linear transformation and then fits an NVC model to the compressed data using MCMC. 

The approaches in \citet{GuhaniyogiLiSavitskySrivastava2022} and \citet{GuhaniyogiBaracaldoBanerjee2023} do not perform variable selection and are specifically designed to handle the ``large $N$, small $p$'' situation. In contrast, the methodology and algorithms that we introduce in this paper are meant to be applied in the ``small $N$, large $p$'' scenario. This scenario arises often in practice, for example, in genome-wide association studies (GWAS) and other analyses of high-throughput biological data \citep{LiWangLiWu2015}. For example, \citet{LiWangLiWu2015} used NVC modeling to model the changes in body mass index (BMI) for $n=865$ subjects using $p=33{,}239$ single nucleotide polymorphisms (SNPs) from the Framingham Heart Study. In this study, BMI was measured at irregular time points from age 29 to age 61 for each of the $n$ subjects, and a multivariate Laplace prior was used to select the SNP varying coefficients that are significantly associated with BMI.

\subsection{Our contributions} \label{contributions}

The goal of this paper is to introduce a comprehensive methodological and computational framework for high-dimensional Bayesian varying coefficient models. Our framework addresses all of the issues of variable selection, estimation, and uncertainty quantification. Methodologically, we introduce a new approach to Bayesian function selection in NVC models that flexibly accounts for unknown within-subject covariances. Despite the utility of Bayesian methods for inference, there is currently a lack of Bayesian variable selection methods for NVC models that are scalable in the number of \emph{covariates} $p$. This work addresses this gap by proposing scalable optimization and MCMC algorithms when $p$ is large and variable selection is a primary objective for the data analyst. 

In fitting a Bayesian NVC model, we have the following desiderata: 1) our method should perform variable selection, 2) our method should be able to flexibly accommodate a wide variety of unknown within-subject correlation structures, and 3) our method should be scalable for large $p$. To the best of our knowledge, there are no existing Bayesian methods for NVC models that accomplish all three goals. In this paper, we address all of these issues. We focus mainly on methodology and computation. However, theoretical considerations for ``small $N$, large $p$'' Bayesian varying coefficient models are briefly discussed and are reported in much greater detail in a follow-up work by \citet{Baitheory2023}.

Recently, there has been a rapid development in spike-and-slab lasso (SSL) methods to solve various high-dimensional problems, including (generalized) linear models \citep{RockovaGeorge2018, TangShenZhangYi2017GLM, DeshpandeRockovaGeorge2019, BaiMoranAntonelliChenBoland2022, Bai2023}, factor analysis \citep{RockovaGeorge2016, MoranRockovaGeorge2019}, graphical models \citep{GanNarisettyLiang2019, LiMcCormickClark2019, GanYangNarisettyLiang2019}, and nonparametric additive regression \citep{BaiMoranAntonelliChenBoland2022}. SSL methods endow regression coefficients with spike-and-slab priors such that the posterior mode gives exact sparsity. In this work, we extend the SSL methodology to functional and longitudinal data analysis. Our contributions can be summarized as follows:

\begin{itemize}
	\item 
	We introduce \textit{nonparametric varying coefficient spike-and-slab lasso} (NVC-SSL) for Bayesian estimation and variable selection in NVC models. Our method provides several advantages over previously proposed methodology for high-dimensional varying coefficient models. First, unlike existing frequentist penalized NVC models, NVC-SSL incorporates estimation of the within-subject covariance structure and borrows information across functional components through a \textit{non}-separable beta-Bernoulli prior. Second, unlike existing Bayesian approaches, the NVC-SSL model does not assume known within-subject covariance functions. 
	\item
	For scalable variable selection, we propose an ECM algorithm for MAP estimation that scales linearly in both $p$ and $N$. Our approach gives exact sparsity, thereby allowing the MAP estimator to automatically perform selection from the varying coefficient functions.  
	\item
	For scalable uncertainty quantification, we provide both an exact MCMC algorithm and an approximate MCMC algorithm. The exact algorithm scales linearly in $p$ and quadratically in $N$, while the approximate MCMC algorithm scales linearly in \emph{both} $p$ and $N$. The approximate MCMC algorithm is shown to provide massive speed-ups over the exact algorithm as $p$ increases. We quantify the tradeoffs of using the approximate MCMC algorithm in place of the exact algorithm.  
\end{itemize} 
The rest of this paper is structured as follows. In Section \ref{VCSSL}, we introduce the NVC-SSL model. In Section \ref{optimization}, we propose a fast ECM algorithm for rapidly obtaining MAP estimates of the varying coefficients under NVC-SSL. In Section \ref{MCMC}, we provide exact and approximate MCMC algorithms for scalable uncertainty quantification. Section \ref{simulations} presents simulation studies validating the variable selection performance, scalability, and inferential capability of NVC-SSL. Section \ref{dataanalysis} applies NVC-SSL to a real data application of identifying important transcription factors in the yeast cell cycle. Finally, Section \ref{discussion} concludes the paper with a brief discussion. 

We use the following notation in this paper. A Gaussian process with mean function $m := m(x)$ and covariance function $k := k(x,x')$ is denoted as $\mathcal{GP}(m,k)$. For a vector $\bm{v}$, $\lVert \bm{v} \rVert_2$ denotes its $\ell_2$-norm. For two matrices $\bm{A}$ and $\bm{B}$, the Kronecker product is denoted by $\bm{A} \otimes \bm{B}$, and the direct sum of $\bm{A}$ and $\bm{B}$ is denoted by $\bm{A} \oplus \bm{B}$. For a square matrix $\bm{C}$, $\text{det}(\bm{C})$ denotes its determinant and $\text{tr}(\bm{C})$ denotes its trace.  For two square matrices $\bm{C}$ and $\bm{D}$ of the same dimension, $\bm{C} \geq \bm{D}$ means that $\bm{C} - \bm{D}$ is non-negative definite. 

\section{The Nonparametric varying coefficient spike-and-slab lasso} \label{VCSSL}

\subsection{Modeling of unknown within-subject correlations} 

In order to accommodate unknown within-subject correlations, we suppose that we can decompose the error $\varepsilon_i(t_{ij})$ in \eqref{varyingcoefficientmodel} into two terms:  a functional random effect \citep{Guo2002} and a measurement error term. Thus, our model is
\begin{equation} \label{varyingcoefficientmodel2}
	y_i(t_{ij}) = \sum_{k=1}^{p} x_{ik}(t_{ij}) \beta_k(t_{ij}) + \alpha_i(t_{ij}) + r_{ij},~~
	\alpha_i(t) \sim \mathcal{GP}(0, k_i),~~ r_{ij} \sim \mathcal{N}(0, \sigma^2),
\end{equation}
where the $r_{ij}$'s are independent measurement errors at each time point $t_{ij}$, and the $\alpha_i(t)$'s are subject-specific functional random effects that independently follow zero-mean Gaussian processes $\mathcal{GP}(0, k_i)$. In \eqref{varyingcoefficientmodel2}, the covariance function $k_i$ models the $i$th subject's within-subject temporal correlations. In particular, the $m_i$-dimensional random effects vector $\alpha_i(\bm{t}_i) = ( \alpha_i(t_{i1}), \ldots, \alpha_i(t_{im_i}) )^\top$ follows a multivariate Gaussian distribution $\mathcal{N}(\bm{0}, \bm{K}(\bm{t}_i))$, where the $(j,j')$th entry of $\bm{K}(\bm{t}_i)$ is $k_i(t_{ij}, t_{ij'})$, $1 \leq j, j' \leq m_i$. 

The functional random effect $\alpha_i(t)$ in \eqref{varyingcoefficientmodel2} deserves some explanation. Functional random effect models \textit{generalize} mixed effects models with a random intercept $\alpha_i \sim \mathcal{N}(0, \sigma_{\alpha}^2)$ \citep{Guo2002}. The random intercept model is equivalent to the compound symmetry (CS) covariance function, $k(t,t') = \sigma_{e}^2 \mathbb{I}(t=t') + \sigma_{\alpha}^2$. Models that specify a random intercept (and a random slope), e.g. the approach in \citet{LiuBobbHennGenningsSchnaasTellezRojoBellingerAroraWrightCoull2018}, are thus imposing a \emph{specific} parametric covariance function on the model. The more general formulation in \eqref{varyingcoefficientmodel2} where $\alpha_i(\bm{t}_i) \sim \mathcal{N}(\bm{0}, \bm{K}(\bm{t}_i))$ allows for many other covariance functions to characterize the within-subject correlations. For example, if the squared exponential (SE) kernel function is used, then $k_i(t,t') = s^2 \exp \{ -(t-t')^2/2 \ell^2 \}$, where $s$ is the scale factor and $\ell$ is the lengthscale. The SE covariance function assumes that the correlation between time points $t$ and $t'$ decreases exponentially as the distance $|t-t'|$ grows. 

In this work, we treat the within-subject covariance functions $k_i$'s as \emph{completely unknown} and model these $k_i$'s nonparametrically. This makes our model more flexible than previous works which ignore within-subject correlations  \citep{WangXia2009, WangLiHuang2008, WeiHuangLi2011} or which require specific parametric forms for the covariance functions \citep{LiuBobbHennGenningsSchnaasTellezRojoBellingerAroraWrightCoull2018, LiWangLiWu2015, DeshpandeBaiBalocchiStarlingWeiss2020, GuhaniyogiLiSavitskySrivastava2022}.

\subsection{Basis expansion representation of the NVC model}
Following the development in \citet{WangXia2009}, \citet{WangLiHuang2008}, and \citet{WeiHuangLi2011}, we approximate each coefficient function $\beta_k$ in \eqref{varyingcoefficientmodel} by a linear combination of $d$ basis functions, i.e. at a particular time $t$,
\begin{align} \label{basisfunctions}
	\beta_k(t) \approx \displaystyle \sum_{l=1}^{d} \gamma_{kl} B_{kl} (t),
\end{align}
where $B_{kl}(t), l=1, \ldots, d$, are the basis functions with corresponding basis coefficients $\gamma_{kl}$. In addition, we approximate the unknown functional random effect $\alpha_i(t)$ in \eqref{varyingcoefficientmodel2} as  a linear combination of $q$ basis functions with \textit{random} coefficients, i.e. for a particular time $t$,
\begin{align} \label{basisfunctions2}
	\alpha_i(t) \approx \sum_{l=1}^{q} \widetilde{B}_{il}(t) \eta_{il},~~\bm{\eta}_i = (\eta_{i1}, \ldots, \eta_{iq})^\top \sim \mathcal{N}_q(\bm{0}, \bm{\Omega}), 
\end{align}
where $\bm{\Omega}$ is a $q \times q$ positive-definite matrix and $\widetilde{B}_{il}(t), l = 1, \ldots, q$, are the basis functions. Combining \eqref{basisfunctions}-\eqref{basisfunctions2}, the model \eqref{varyingcoefficientmodel2} can be approximated as
\begin{align} \label{approximatemodel}
	y_i (t_{ij}) \approx \displaystyle \sum_{k=1}^{p} \displaystyle \sum_{l=1}^{d} x_{ik} (t_{ij}) \gamma_{kl} B_{kl} (t_{ij}) + \sum_{l=1}^{q} \widetilde{B}_{il}(t_{ij}) \eta_{ik} + r_{ij}.
\end{align}
From \eqref{basisfunctions2}, it is clear that the within-subject covariance function $k_i(t, t')$ is approximated by
\begin{align} \label{approximatecovariance}
	k_i(t, t') \approx \widetilde{\bm{B}}_i^\top(t) \bm{\Omega} \widetilde{\bm{B}}_i(t'),
\end{align} 
where $\widetilde{\bm{B}}_i(t) = ( \widetilde{B}_{i1}(t), \ldots, \widetilde{B}_{iq}(t) )^\top \in \mathbb{R}^{q}$. We see from \eqref{approximatecovariance} that our formulation affords a great deal of flexibility in modeling the unknown within-subject covariances. In particular, the covariance function $k_i(t_{ij}, t_{ij'})$ for the $i$th subject is modeled by a quadratic form of the subject-specific basis function vectors $\widetilde{\bm{B}}_i(t_{ij})$ and $\widetilde{\bm{B}}_i(t_{ij'})$. Therefore, if we choose a flexible family of basis functions for the $\widetilde{\bm{B}}_i$'s, then we can capture a wide variety of within-subject covariance functions. In practice, we use B-splines with equispaced knots as the basis functions for both the $B_{kl}(t)$'s and $\widetilde{B}_{il}(t)$'s, due to their computational simplicity, numerical stability, and excellent local approximation properties \citep{WeiHuangLi2011, YooGhosal2016}. However, other basis functions such as natural splines, trigonometric functions, and wavelets could also be used to model the $B_{kl}(t)$ and $\widetilde{B}_{il}(t)$'s.

Recall that $N = \sum_{i=1}^{n} m_i$ is the total number of observations. Let $\bm{X} = [ \bm{x}_1, \ldots, \bm{x}_p ] \in \mathbb{R}^{N \times p}$, with $\bm{x}_k = (x_{1k}(t_{11}), \ldots, x_{1k}(t_{1 m_1}), \ldots, x_{nk} (t_{n1}), \ldots, x_{nk} ( t_{n m_n}))^\top$.
Further, we define $\bm{B}(t)$ as a $p \times dp$ basis expansion matrix containing the $B_{kl}(t)$'s from \eqref{basisfunctions}, 
\begin{align*} 
	\bm{B}(t) = 
	\begin{pmatrix}
		B_{11}(t) & B_{12}(t) & \ldots & B_{1 d} (t) & 0 & \ldots & 0 & 0 & \ldots & 0 \\
		\vdots & \vdots & & \vdots & \vdots & & \vdots & \vdots & & \vdots \\
		0 & 0 & \ldots & 0 & 0 & \ldots & B_{p1}(t) & B_{p2}(t) & \ldots & B_{pd}(t)
	\end{pmatrix},
\end{align*} 
and we define 
\begin{align} \label{Uimatrix}
	\bm{U}_i = (\bm{u}_{i1}, \ldots, \bm{u}_{im_i})^\top \in \mathbb{R}^{m_i \times dp},
\end{align} where 
\begin{align*} 
	\bm{u}_{ij}^\top = \bm{x}^\top(t_{ij}) \bm{B}(t_{ij})
\end{align*} 
for $i=1, \ldots, n, j = 1, \ldots, m_i$, and $\bm{x}(t_{ij}) \in \mathbb{R}^{p}$ denotes the row of $\bm{X}$ corresponding to the $j$th observation for the $i$th subject. We also define $\bm{Z}_i$ as the matrix with $(j,l)$th entry $\widetilde{B}_{il}(t_{ij})$ from \eqref{basisfunctions2}, i.e. 
\begin{align} \label{Zimatrix}
	\bm{Z}_i = ( \widetilde{\bm{B}}_i(t_{i1}), \ldots, \widetilde{\bm{B}}_i(t_{im_i}) )^\top \in \mathbb{R}^{m_i \times q}. 
\end{align}
Let $\bm{\gamma} = (\bm{\gamma}_1^\top, \ldots, \bm{\gamma}_p^\top)^\top \in \mathbb{R}^{dp}$, where the $k$th subvector $\bm{\gamma}_k = (\gamma_{k1}, \ldots, \gamma_{kd}) \in \mathbb{R}^{d}$ consists of the basis coefficients $\gamma_{kl}$'s in \eqref{basisfunctions} corresponding to the $k$th varying coefficient $\beta_k(t)$. Let $\bm{Y}_i = (y_i(t_{i1}, \ldots, y_i(t_{im_i}))^\top$ and $\bm{r}_i = (r_{i1}, \ldots, r_{im_i})^\top$ denote the $m_i$-dimensional vectors of responses and measurement errors for the $i$th subject. Then \eqref{approximatemodel} can be written in matrix form as
\begin{align} \label{matrixform1}
	\bm{Y}_i = \bm{U}_i \bm{\gamma} + \bm{Z}_i \bm{\eta}_i + \bm{r}_i,~~ \bm{\eta}_i \sim \mathcal{N} (\bm{0}, \bm{\Omega}),~~\bm{r}_i \sim \mathcal{N} (\bm{0}, \sigma^2 \bm{I}_{m_i}),~~i=1, \ldots, n,
\end{align}
where $\bm{U}_i$ and $\bm{Z}_i$ are as in \eqref{Uimatrix} and \eqref{Zimatrix}. Letting $\bm{Y} = (\bm{Y}_1^\top, \ldots, \bm{Y}_n^\top)^\top \in \mathbb{R}^{N}$, $\bm{U} = (\bm{U}_1^\top, \ldots, \bm{U}_n^\top)^\top \in \mathbb{R}^{N \times dp}$, $\bm{Z} = \bm{Z}_1 \oplus \cdots \oplus \bm{Z}_n \in \mathbb{R}^{N \times nq}$, $\bm{\eta} = (\bm{\eta}_1^\top, \ldots, \bm{\eta}_n^\top)^\top \in \mathbb{R}^{nq}$, and $\bm{r} = (\bm{r}_1^\top, \ldots, \bm{r}_n^\top)^\top \in \mathbb{R}^{N}$, we can also express \eqref{matrixform1} for all $N$ observations as
\begin{align} \label{matrixform2}
	\bm{Y} = \bm{U}\bm{\gamma} + \bm{Z} \bm{\eta} + \bm{r},~~\bm{\eta} \sim \mathcal{N} (\bm{0}, \bm{I}_n \otimes \bm{\Omega} ),~~\bm{r} \sim \mathcal{N} (\bm{0}, \sigma^2 \bm{I}_N).
\end{align}
Before introducing the NVC-SSL model, we make a few remarks about our model setup. First, although we focus on continuous responses with Gaussian errors for concreteness, our method can easily be extended to NVC models with discrete or non-Gaussian responses by recasting our model into the generalized linear mixed model (GLMM) framework. In this case, we would employ a monotonically increasing link function $g$ to relate the conditional expectation of $y_i(t_{ij})$ given the $p$ covariates $\bm{x}_{i}(t_{ij}) = (x_{i1}(t_{ij}), \ldots, x_{ip}(t_{ij}))^\top$ to the varying coefficients $\beta_k(t)$'s as  
\begin{align*} \label{GLMM}
	\mathbb{E} [ y_i(t_{ij}) \mid \bm{x}_{i} (t_{ij}) ] & = g^{-1} \left( \sum_{k=1}^{p} x_{ik}(t_{ij}) \beta_k(t_{ij}) + \alpha_i(t_{ij}) \right) \\
	& = g^{-1} \left( \bm{u}_{ij}^\top \bm{\gamma} + \bm{z}_{ij}^\top \bm{\eta}_i \right),~~ \bm{\eta}_i \sim \mathcal{N}(\bm{0}, \bm{\Omega}), \numbereqn
\end{align*}
where $\bm{u}_{ij}$ is the $j$th row of $\bm{U}_i$ in \eqref{Uimatrix} and $\bm{z}_{ij}$ is the $j$th row of $\bm{Z}_i$ in \eqref{Zimatrix}. For example, if the response variables are binary, we can assume that $y_{i}(t_{ij}) \mid \bm{x}_{i}(t_{ij}) \sim \textrm{Bernoulli}(p_{ij})$ and employ the logit link function $g(p_{ij}) = \log(p_{ij}/(1-p_{ij}))$ to obtain a logistic NVC model. By putting the same priors on $(\bm{\gamma}, \bm{\Omega})$ in \eqref{GLMM} as those introduced in Section \ref{ModelFormulation}, we can implement NVC-SSL for logistic NVC models. In particular, the ECM algorithm in Section \ref{optimization} can be extended to logistic NVC models using the approach in \citet{Bai2023}, and the MCMC algorithms in Section \ref{MCMC} can also be extended to GLMMs straightforwardly using P\'{o}lya-gamma data augmentation \citep{polson2013bayesian}. 

Secondly, our method can also be easily extended to NVC models where the varying coefficients are \emph{multivariate} functions, e.g. spatial models where $\beta_k := \beta_k ( \bm{s}), k =1, \ldots, p$, and $\bm{s} \in \mathcal{S}$ where $\mathcal{S}$ is a spatial domain. If the varying coefficients are multivariate functions, we can replace the univariate basis functions $B_{kl}(t)$ and $\widetilde{B}_{il}(t)$ in \eqref{basisfunctions} with tensor products of basis functions (see e.g., \citet{BaiMoranAntonelliChenBoland2022}). For example, if the varying coefficients are functions of two variables, $\beta_k(u,v)$, we can approximate the varying coefficients as
\begin{align*}
	\beta_k(u,v) \approx \sum_{l=1}^{d_u} \sum_{m=1}^{d_v} \gamma_{klm} B_{kl}(u) B_{km}(v),
\end{align*}
and the functional random effects as 
\begin{align*}
	\alpha_i(u,v) \approx \sum_{l=1}^{d_u} \sum_{m=1}^{d_v} \widetilde{B}_{il}(u) \widetilde{B}_{im}(v) \eta_{ilm}.
\end{align*}
We would then proceed to estimate the model parameters, e.g. the basis coefficients, exactly the same way as we would in the case of univariate varying coefficient functions $\beta_k(t)$. 

\subsection{Prior specification for NVC-SSL} \label{ModelFormulation}

Having rewritten our NVC model \eqref{varyingcoefficientmodel2} in matrix form \eqref{matrixform2}, parameter estimation reduces to estimating $(\bm{\gamma}, \bm{\Omega}, \sigma^2)$. We take a Bayesian approach and endow these parameters with suitable priors. In particular, estimating the varying coefficient functions $\beta_k(t)$'s in \eqref{varyingcoefficientmodel2} are straightforward once we have estimates of the basis coefficients $\bm{\gamma}$. By \eqref{basisfunctions}, we can estimate $\widehat{\beta}_k(t) = \sum_{l=1}^{d} \widehat{\gamma}_{kl} B_{kl}(t), k = 1, \ldots, p$, once we have an estimate $\widehat{\bm{\gamma}}$. 

As discussed in Section \ref{relatedwork}, we are interested in not only estimating the varying coefficient functions in \eqref{varyingcoefficientmodel2}, but \emph{also} performing variable selection from them. Under the assumption of sparsity, most of the $\beta_k(t)$'s in \eqref{varyingcoefficientmodel2} should equal zero. To facilitate variable selection, we endow the vector of basis coefficients $\bm{\gamma} = ( \bm{\gamma}_1^\top, \ldots, \bm{\gamma}_p^\top)^\top$ in \eqref{matrixform2} with the \textit{spike-and-slab group lasso} (SSGL) prior of \citet{BaiMoranAntonelliChenBoland2022},
\begin{equation} \label{ssgrouplasso}
	\pi(\bm{\gamma} \mid \theta ) = \prod_{k=1}^{p} \left[ ( 1-\theta) \bm{\Psi} ( \bm{\gamma}_k \mid \lambda_0 ) + \theta  \bm{\Psi} ( \bm{\gamma}_k \mid \lambda_1 ) \right],
\end{equation}
where $\theta \in (0,1)$ is a mixing proportion, or the expected proportion of nonzero $\bm{\gamma}_k$'s, and $\bm{\Psi}( \cdot \mid \lambda)$ denotes a multivariate Laplace density indexed by hyperparameter $\lambda$, 
\begin{equation*} 
	\bm{\Psi} ( \bm{\gamma}_k \mid \lambda ) = \frac{\lambda^{d}  e^{- \lambda \lVert \bm{\gamma}_k \rVert_2 } }{2^{d} \pi^{(d-1)/2} \Gamma ((d+1)/2) }, \hspace{.3cm} k = 1, \ldots, p. 
\end{equation*}
The SSGL prior \eqref{ssgrouplasso}, which we denote as $\mathcal{SSGL}(\lambda_0, \lambda_1, \theta)$ going forward, can be considered a two-group refinement of the group lasso \citep{YuanLin2006}. Under the prior \eqref{ssgrouplasso}, the posterior mode for $\bm{\gamma}$ gives \emph{exact} sparsity (i.e. some of the $\bm{\gamma}_k$ vectors will be exactly $\bm{0}$). This allows $\mathcal{SSGL}(\lambda_0, \lambda_1, \theta)$ to perform joint estimation and variable selection \citep{BaiMoranAntonelliChenBoland2022}. In the present context, if the posterior mode for $\bm{\gamma}_k$ is $\widehat{\bm{\gamma}}_k = \bm{0}$, then the $k$th function will be estimated as $\widehat{\beta}_k(t) = \sum_{l=1}^{d_k} \widehat{\gamma}_{kl} B_{kl}(t) = 0$ and thus thresholded out of the model. 

We typically set $\lambda_0 \gg \lambda_1$ in \eqref{ssgrouplasso}, so that the first mixture component $\bm{\Psi}(\cdot \mid \lambda_0)$ (the spike) is heavily concentrated around the $d$-dimensional zero vector $\bm{0}$ for each $k =1, \ldots, p$. Meanwhile, the slab component $\bm{\Psi}(\cdot \mid \lambda_1)$ stabilizes the posterior estimates of large coefficients, \emph{preventing} them from being downward biased. One of the chief advantages of SSGL over other group penalties such as group lasso, group smoothly clipped absolute deviation (SCAD), or group minimax concave penalty (MCP) \citep{YuanLin2006, BrehenyHuang2015} is the SSGL's ability to perform \emph{adaptive} shrinkage. Group lasso, group SCAD, and group MCP all contain a \emph{single} regularization parameter $\lambda > 0$ controlling the sparsity of the solution. Consequently, if $\lambda$ is large, then all groups may be overshrunk. In contrast, the slab hyperparameter $\lambda_1$ in $\mathcal{SSGL}(\lambda_0, \lambda_1, \theta)$ applies minimal shrinkage to groups with larger coefficients, allowing these groups to escape the pull of the spike.

To model the uncertainty of the mixing proportion $\theta$ in (\ref{ssgrouplasso}), we endow $\theta$ with a beta prior, 
\begin{equation} \label{thetaprior}
	\theta \sim \mathcal{B}(a,b),
\end{equation}
where $a>0$ and $b>0$ are fixed positive constants. Unlike the group lasso, group SCAD, and group MCP, this prior \eqref{thetaprior} on $\theta$ ultimately renders our Bayesian penalty \textit{non}-separable in the sense that the groups $\bm{\gamma}_k, k=1, \ldots, p$ are a priori \emph{dependent}. This non-separability provides several benefits. First, the prior on $\theta$ allows the NVC-SSL model to \textit{share} information across functional components and self-adapt to ensemble information about sparsity. Second, with appropriate choices for the hyperparameters in $\theta \sim \mathcal{B}(a, b)$, namely $a=1, b=p$, our prior performs an automatic multiplicity adjustment \citep{ScottBerger2010} and favors parsimonious models in high dimensions. This helps NVC-SSL to avoid the curse of dimensionality for large $p$.

To complete the NVC-SSL prior specification, we place independent conditionally conjugate priors on the parameters $(\bm{\Omega}, \sigma^2)$ in \eqref{matrixform2}. Namely, we endow $\bm{\Omega}$ with the prior,
\begin{equation} \label{Omegaprior}
	\bm{\Omega} \sim \text{Inverse-Wishart}(\nu, \bm{\Phi}),
\end{equation}
where the degrees of freedom $\nu > q-1$ and the scale matrix $\bm{\Phi}$ is positive-definite. Finally, we endow the measurement error variance $\sigma^2$ with the prior, 
\begin{equation} \label{sigma2prior}
	\sigma^2 \sim \text{Inverse-Gamma} (c_0/2, d_0/2),
\end{equation}
where $c_0 > 0, d_0 > 2$. 

\subsection{Theoretical considerations} 

In the literature on Bayesian asymptotics, a common theme is to study the posterior contraction rate, or the speed at which the posterior distribution converges to a point mass at the true parameter as sample size $N$ grows to infinity. Recently, in the ``fixed $p$'' regime, posterior contraction rates have been derived for Bayesian NVC models by \citet{DeshpandeBaiBalocchiStarlingWeiss2020}, \citet{GuhaniyogiLiSavitskySrivastava2022}, and \citet{GuhaniyogiBaracaldoBanerjee2023}. However, these papers do \emph{not} consider the case where $p$ is allowed to diverge with $n$. 

In a follow-up paper to this article, \citet{Baitheory2023} derives sufficient conditions for posterior contraction in high-dimensional Bayesian NVC models when $p \gg n$ and $p$ grows subexponentially with $n$. To summarize briefly, the prior distribution is required to be heavily concentrated near zero (to capture sparsity) and to have a sufficiently heavy tail (to capture the true nonzero varying coefficients). With appropriately chosen hyperparameters, \citet{Baitheory2023} shows that the NVC-SSL prior can achieve adaptive posterior contraction to the true varying coefficients. The NVC-SSL prior is adaptive in the sense that it adapts to the unknown sparsity level \textit{and} the unknown smoothness of the varying coefficients. These sufficient conditions are not specific to the NVC-SSL prior; other multivariate priors that satisfy the conditions in \citet{Baitheory2023} would also theoretically achieve adaptive posterior contraction. This general theory for Bayesian NVC models when $p > n$ is described in detail in \citet{Baitheory2023}.

\section{Scalable MAP estimation for variable selection} \label{optimization}

\subsection{ECM algorithm} \label{ECM}

We now detail how to implement NVC-SSL, i.e. the model \eqref{approximatemodel} with priors \eqref{ssgrouplasso}-\eqref{sigma2prior}, for variable selection. We first present a very fast ECM algorithm which targets the posterior mode. Once we have obtained the MAP estimator $\widehat{\bm{\gamma}}$, the varying coefficients can then be estimated as $\widehat{\beta}_k(t) = \sum_{l=1}^{d} \widehat{\gamma}_{kl} B_{kl}(t), k = 1, \ldots, p$. As discussed in Section \ref{ModelFormulation}, the MAP estimator under NVC-SSL is \emph{exactly} sparse, with many $\bm{\gamma}_k$'s thresholded to zero. This enables the MAP estimator to perform \emph{automatic} variable selection, since $\widehat{\beta}_k(t) = 0$ if $\widehat{\bm{\gamma}}_k = \bm{0}$. 

Let $\bm{\Xi}$ denote the collection $\bm{\Xi} = \{ \bm{\gamma}, \theta, \bm{\eta}, \bm{\Omega}, \sigma^2 \}$. Based on \eqref{matrixform2} and the prior densities \eqref{ssgrouplasso}-\eqref{sigma2prior}, the log-posterior density for $\bm{\Xi}$ (up to an additive constant) is given by
\begin{align*}  \label{logposterior}
	\log \pi ( \bm{\Xi} \mid \bm{Y}) = &  - \frac{N}{2} \log \sigma^2 - \frac{\lVert \bm{Y} - \bm{U} \bm{\gamma} - \bm{Z} \bm{\eta} \rVert^2}{2 \sigma^2}  + \frac{n}{2} \text{log} (\text{det}( \bm{\Omega}^{-1})) - \frac{1}{2} \sum_{i=1}^{n} \bm{\eta}_i^\top \bm{\Omega}^{-1} \bm{\eta}_i \\
	& + \sum_{k=1}^{p} \log \left( (1-\theta) \lambda_0^{d} e^{-\lambda_0 \lVert \bm{\gamma}_k \rVert_2} + \theta \lambda_1^d e^{-\lambda_1 \lVert \bm{\gamma}_k \rVert_2} \right) \\
	& + (a-1) \log \theta + (b-1) \log(1-\theta) \\
	& + \frac{\nu+q+1}{2} \log \left( \text{det}(\bm{\Omega}^{-1}) \right) - \frac{1}{2} \text{tr}(\bm{\Phi} \bm{\Omega}^{-1}) - \left( \frac{c_0+2}{2} \right) \log \sigma^2 - \frac{d_0}{2 \sigma^2}.  \numbereqn
\end{align*}
Our objective is to maximize the log-posterior \eqref{logposterior} with respect to $\bm{\Xi}$. We first introduce latent 0-1 indicators, $\bm{\tau} = (\tau_1, \ldots, \tau_p)^\top$, i.e. $\tau_k \in \{ 0, 1\}$ for $k = 1, \ldots, p$. Then the $\mathcal{SSGL}(\lambda_0, \lambda_1, \theta)$ prior \eqref{ssgrouplasso} can be expressed as the marginal prior under a hierarchical beta-Bernoulli prior,
\begin{equation} \label{SSGLrepar}
	\begin{array}{rl} 
		\pi ( \bm{\gamma} \mid \bm{\tau} ) = & \displaystyle \prod_{k=1}^{p} \left[ (1-\tau_k) \bm{\Psi}(\bm{\gamma}_k \mid \lambda_0) + \tau_k \bm{\Psi} ( \bm{\gamma}_k | \lambda_1) \right], \\
		\pi ( \bm{\tau} \mid \theta ) = & \displaystyle \prod_{k=1}^{p} \theta^{\tau_k} (1-\theta)^{1-\tau_k}.
	\end{array}
\end{equation}
With the augmented log-posterior $\log \pi(\bm{\Xi}, \bm{\tau} \mid \bm{Y})$, we can now implement an ECM algorithm to find the MAP estimator $\widehat{\bm{\Xi}} = \{ \widehat{\bm{\gamma}},  \widehat{\theta}, \widehat{\bm{\eta}}, \widehat{\bm{\Omega}}, \widehat{\sigma}^2 \}$. We first initialize the parameters $\bm{\Xi}^{(0)}$, and then in each $t$th iteration, we iterate between the E-step and the CM-steps until convergence. In the E-step, we treat the latent indicator variables $\bm{\tau}$ in \eqref{SSGLrepar} as missing data and compute $F^{(t)} ( \bm{\gamma}, \theta, \bm{\eta}, \bm{\Omega}, \sigma^2 ) = \mathbb{E}_{\bm{\tau}} \left[ \log ( \bm{\Xi}, \bm{\tau} \mid \bm{Y} ) \mid \bm{\Xi}^{(t-1)} \right]$. In the CM-step, we then optimize $F^{(t)} ( \bm{\gamma}, \theta, \bm{\eta}, \bm{\Omega}, \sigma^2 )$ with respect to $\bm{\Xi}$ by performing two iterative updates:
\begin{enumerate}
	\item  Update $(\theta, \bm{\eta})$, holding $(\bm{\gamma}, \bm{\Omega}, \sigma^2)$ fixed at their previous values, i.e. solve
	\begin{align*}
		(\theta^{(t)}, \bm{\eta}^{(t)} ) = \argmax_{\theta, \bm{\eta}} F^{(t)} ( \bm{\gamma}^{(t-1)}, \theta, \bm{\eta}, \bm{\Omega}^{(t-1)}, \sigma^{2(t-1)} ).
	\end{align*}
	\item Update  $(\bm{\gamma}, \bm{\Omega}, \sigma^2)$, holding $(\theta, \bm{\eta})$ fixed at their current values, i.e. solve
	\begin{align*}
		( \bm{\gamma}^{(t)}, \bm{\Omega}^{(t)}, \sigma^{2(t)} ) = \argmax_{\bm{\gamma}, \bm{\Omega}, \sigma^2} F^{(t)} ( \bm{\gamma}, \theta^{(t)}, \bm{\eta}^{(t)}, \bm{\Omega}, \sigma^2 ).
	\end{align*}
\end{enumerate}
To be more specific, in the E-step, we compute $\mathbb{E}_{\bm{\tau}} [ \tau_k \mid \bm{Y}, \bm{\Xi}^{(t-1)} ] = p_k^{\star} (\bm{\gamma}_k^{(t-1)}, \theta^{(t-1)}), k = 1, \ldots, p$, where 
\begin{equation} \label{pstark}
	p_k^{\star} ( \bm{\gamma}_k, \theta) = \frac{ \theta \bm{\Psi} ( \bm{\gamma}_k | \lambda_1)}{ \theta \bm{\Psi} (\bm{\gamma}_k | \lambda_1) + (1-\theta) \bm{\Psi} ( \bm{\gamma}_k | \lambda_0 )}, 
\end{equation}
is the conditional posterior probability that $\bm{\gamma}_k$ is drawn from the slab distribution rather than from the spike. We then compute $\lambda_k^{\star}, k=1, \ldots, p$, where
\begin{align} \label{lambdastark}
	\lambda_k^{\star} = \mathbb{E}_{\bm{\tau}} \left[ \log \left( (1-\theta) \lambda_0^{d} e^{-\lambda_0 \lVert \bm{\gamma}_k \rVert_2} + \theta \lambda_1^d e^{-\lambda_1 \lVert \bm{\gamma}_k \rVert_2} \right) \mid \bm{Y}, \bm{\Xi} \right] = \lambda_1 p_k^{\star} + \lambda_0 (1-p_k^{\star}).
\end{align}
Based on \eqref{pstark}-\eqref{lambdastark},
\begin{align*} \label{Mobj}
	& F^{(t)} ( \bm{\gamma}, \theta, \bm{\eta}, \bm{\Omega}, \sigma^2 )  = \mathbb{E}_{\bm{\tau}} \left[ \log \pi ( \bm{\Xi} \mid \bm{Y} ) \mid \bm{\Xi}^{(t-1)} \right]  \\
	& \qquad = - \frac{N}{2} \log \sigma^2  - \frac{\lVert \bm{Y} - \bm{U} \bm{\gamma} - \bm{Z} \bm{\eta} \rVert_2^2}{2 \sigma^2} + \frac{n}{2} \log \left( \text{det}(\bm{\Omega}^{-1}) \right) - \frac{1}{2} \sum_{i=1}^{n} \bm{\eta}_i^\top \bm{\Omega}^{-1} \bm{\eta}_i \\
	& \qquad \qquad + \sum_{k=1}^{p} \lambda_k^{\star} \lVert \bm{\gamma}_k \rVert_2 + \left( a - 1 + \sum_{k=1}^{p} p_k^{\star} \right) \log \theta + \left( b - 1 + p - \sum_{k=1}^{p} p_k^{\star} \right) \log (1-\theta) \\
	& \qquad \qquad + \frac{\nu + q + 1}{2} \log \left( \text{det}(\bm{\Omega}^{-1}) \right) - \frac{1}{2} \text{tr} \left( \bm{\Phi} \bm{\Omega}^{-1} \right) - \left( \frac{c_0+2}{2} \right) \log \sigma^2 - \frac{d_0}{2 \sigma^2}. \numbereqn
\end{align*}
The CM-step maximizes the objective \eqref{Mobj}. First, holding $(\bm{\gamma}^{(t-1)}, \bm{\Omega}^{(t-1)}, \sigma^{2(t-1)})$ fixed, $\theta$ has the following closed form update,
\begin{equation} \label{thetaupdate}
	\theta^{(t)} = \frac{a - 1 + \sum_{k=1}^{p} p_k^{\star}}{a+b+p-2}. 
\end{equation}
Meanwhile, each $\bm{\eta}_i, i = 1, \ldots, n$ in $\bm{\eta}$ can be updated individually in closed form as
\begin{equation} \label{etaupdate}
	\bm{\eta}_i^{(t)} = \bm{B}_{\bm{Z}_i} \left( \bm{Y}_i - \bm{U}_i \bm{\gamma}^{(t-1)} \right),
\end{equation} 
where
\begin{align*}
	\bm{B}_{\bm{Z}_i} = \left( \bm{Z}_i^\top \bm{Z}_i + \sigma^{2(t-1)} \bm{\Omega}^{(t-1)} \right)^{-1} \bm{Z}_i^\top,
\end{align*} 
and $\bm{Y}_i$, $\bm{U}_i$, and $\bm{Z}_i$ are as in \eqref{matrixform1}.

Next, we update $(\bm{\gamma}, \bm{\Omega}, \sigma^2)$ holding $(\theta^{(t)}, \bm{\eta}^{(t)})$ fixed. First, in order to update $\bm{\gamma}$, we solve the following optimization:
\begin{equation} \label{objgamma}
	\bm{\gamma}^{(t)} = \argmax_{\bm{\gamma}} -\frac{1}{2} \lVert \widetilde{\bm{Y}} - \bm{U} \bm{\gamma} \rVert_2^2 - \sum_{k=1}^{p} \sigma^{2(t-1)} \lambda_k^{\star} \lVert \bm{\gamma}_k \rVert_2,
\end{equation}
where $\widetilde{\bm{Y}} = \bm{Y} - \bm{Z} \bm{\eta}^{(t)}$ and the $\lambda_k^{\star}$'s are as in \eqref{lambdastark}. It can be seen that \eqref{objgamma} is an adaptive group lasso problem with group-specific weights $\sigma^{2(t-2)} \lambda_k^{\star}$. This optimization can be solved with any standard group lasso algorithm \citep{YuanLin2006,  BrehenyHuang2015}. We opt to use the coordinate ascent algorithm of \citet{BrehenyHuang2015} due to its speed and numerical stability. 

Although each CM step requires solving the optimization \eqref{objgamma}, it should be reiterated that $\lambda_k^{\star} = \lambda_1 p_k^{\star} + \lambda_0 (1-p_k^{\star})$ in \eqref{objgamma}. Under sparsity, most of the $p_k^{\star}$'s in \eqref{pstark} are very close to zero (i.e. most of the $\bm{\gamma}_k$'s come from the spike density in \eqref{SSGLrepar}), and thus, most of the $\lambda_k^{\star}$'s satisfy $\lambda_k^{\star} \approx \lambda_0$. Therefore, as long as the spike hyperparameter $\lambda_0$ is large, \emph{most} of the group-specific weights in \eqref{objgamma} will \emph{also} be large. Since a larger penalty is applied these respective groups, most of them will be thresholded to zero very early on and then remain at zero in the group optimization algorithm. This ensures that the coordinate ascent algorithm for solving \eqref{objgamma} converges very rapidly. At the same time, for the few nonzero varying coefficients, we have $p_k^{\star} \approx 1$ and $\lambda_k^{\star} \approx \lambda_1$ (where $\lambda_1 \ll \lambda_0$). We are therefore able to apply a \emph{weaker} penalty to basis coefficients $\bm{\gamma}_k$ with \emph{larger} entries. 

Finally, the updates for $\bm{\Omega}$ and $\sigma^2$ have the closed forms,
\begin{equation} \label{Omegaupdate}
	\bm{\Omega}^{(t)} = \frac{1}{n+\nu+q+1} \left( \bm{\Phi} + \sum_{i=1}^{n} \bm{\eta}_i^{(t)} (\bm{\eta}_i^{(t)} )^\top \right),
\end{equation}
and
\begin{align} \label{sigma2update}
	\sigma^{2(t)} = \frac{ \lVert \bm{Y} - \bm{U} \bm{\gamma}^{(t)} - \bm{Z} \bm{\eta}^{(t)} \rVert_2^2 + d_0}{N + c_0 + 2}.
\end{align}
In particular, as long as the scale matrix $\bm{\Phi}$ in \eqref{Omegaupdate} is chosen to be positive-definite, the update for $\bm{\Omega}$ is also guaranteed to be positive-definite. 

The complete algorithm for the NVC-SSL model is given in Algorithm \ref{algorithm1}. Convergence can be assessed using a criterion $\lVert \bm{\gamma}^{(t)} - \bm{\gamma}^{(t-1)} \rVert_2^2 / \lVert \bm{\gamma}^{(t-1)} \rVert_2^2 < 10^{-6}$. Let $\bm{t} = (t_{11}, \ldots, t_{1 m_1}, \ldots, t_{n1}, \ldots, t_{nm_n})^\top$ be the vector of all observation times for all subjects. Once we have obtained the final MAP estimate $\widehat{\bm{\gamma}}$, we can estimate the varying coefficients as $\widehat{\beta}_k (\bm{t}) = \sum_{l=1}^{d} \widehat{\gamma}_{kl} B_{kl} (\bm{t}), k=1, \ldots, p$, where $\widehat{\beta}_k(\bm{t})$ is the $N \times 1$ vector of $\widehat{\beta}_k$ evaluated at all $N$ time points in $\bm{t}$. 

Since the ECM algorithm has the ascent property, our algorithm is guaranteed to converge to a local mode. However, the NVC-SSL log-posterior \eqref{logposterior} is a nonconvex function of its parameters, and hence, Algorithm \ref{algorithm1} is not guaranteed to converge to the global mode. Nevertheless, we have not found local convergence to be a practical problem, and the (local) MAP estimate under the NVC-SSL model performs very well in practice, as we demonstrate in Sections \ref{simulations} and \ref{dataanalysis}. 

\begin{algorithm}[t!]
	\begin{flushleft}
		\textbf{Input:} Initial values $\bm{\gamma}^{(0)}$, $\theta^{(0)}$, $\bm{\Omega}^{(0)}$, $\sigma^{2(0)}$, $t=0$ \\
		\textbf{Output:} Estimated varying coefficients $\widehat{\beta}_k(\bm{t}),~k = 1, \ldots, p$
		\vspace{.2cm}
		
		\textbf{while} $\text{diff} > \epsilon$ \textbf{do}
		\begin{enumerate}
			\item Increment $t$
			\item
			\textbf{E-step:} 
			
			\textbf{for} $k = 1, \ldots, p$ \textbf{do}
			\begin{enumerate}
				\item Compute $p_k = p^{\star} ( \bm{\gamma}_k^{(t-1)}, \theta^{(t-1)})$ as in \eqref{pstark}
				\item Set $\lambda_k^{\star} = \lambda_1 p_k + \lambda_0 (1-p_k)$
			\end{enumerate}
			\item
			\textbf{M-step:}
			\begin{enumerate}
				\item Update $\theta^{(t)}$ according to \eqref{thetaupdate}
				\item For $i = 1, \ldots, n$, update $\bm{\eta}_i^{(t)}$ according to \eqref{etaupdate}
				\item Update $\bm{\gamma}^{(t)}$ by solving \eqref{objgamma}
				\item Update $\bm{\Omega}^{(t)}$ according to \eqref{Omegaupdate}
				\item Update $\sigma^{2(t)}$ according to \eqref{sigma2update}
			\end{enumerate}
			\item Set $\textsf{diff} = \lVert \bm{\gamma}^{(t)} - \bm{\gamma}^{(t-1)} \rVert_2^2 / \lVert \bm{\gamma}^{(t-1)} \rVert_2^2 $
		\end{enumerate}
		\item 
		\textbf{return} $\widehat{\beta}_k(\bm{t}) = \sum_{l=1}^{d} \widehat{\gamma}_{kl} B_{kl}(\bm{t})$, $k=1, \ldots, p$
	\end{flushleft}
	\caption{ECM algorithm for MAP estimation under NVC-SSL} \label{algorithm1}
\end{algorithm}

An alternative way to perform variable selection is to fit NVC-SSL with MCMC and to use the MCMC samples to estimate the posterior inclusion probabilities $P(\tau_k = 1 \mid \bm{Y}), k = 1, \ldots, p$. Selection can then be performed by thresholding these probabilities. For example, we can use the median probability model (MPM) \citep{BarbieriBerger2004, BarbieriBA2021} and select $\beta_k(t)$ if $P(\tau_k = 1 \mid \bm{Y}) \geq 0.5$. In Appendix \ref{App:A}, we explore the use of MPM for selection. We found that the MPM approach was inferior to using the MAP estimator for variable selection, and the MAP estimator was better able to detect weak signals. 

In addition to its superior selection performance, another advantage of MAP estimation is that it tends to be much faster than MCMC. As we show in Section \ref{simulationsIII}, the ECM algorithm took on average less than one minute to complete for $p=5000$ varying coefficients (and 40,000 total unknown basis coefficients in $\bm{\gamma}$). On the other hand, it took on average 18.83 minutes to run the approximate MCMC algorithm for 1000 iterations when $p=5000$ and $\bm{\gamma} \in \mathbb{R}^{40{,}000}$. Thus, if \emph{variable selection} is a primary objective of the data analyst, we recommend using the ECM algorithm presented in this section to select the varying coefficient functions. If inference is also desirable, then the analyst can additionally use the MCMC algorithms in Section \ref{MCMC} to obtain uncertainty intervals.

The ECM algorithm that we introduced in this section is based specifically on a beta-Bernoulli hierarchical construction \eqref{SSGLrepar} for the $\mathcal{SSGL}(\lambda_0, \lambda_1, \theta)$ prior \eqref{ssgrouplasso}, and therefore, it is not applicable to other NVC models. On the other hand, if one is able to obtain good estimates for $(\bm{\Omega}, \sigma^2)$ in \eqref{matrixform2}, then it may be possible to use the profile likelihood approach of \citet{FanLi2012} to perform (non-Bayesian) variable selection under a penalized regression framework for \eqref{matrixform2}. When there is \textit{no} sparsity in the $\beta_k(t)$'s, restricted maximum likelihood (REML) can be used to estimate $(\bm{\Omega}, \sigma^2)$ \citep{Guo2002}. However, under sparsity and/or high dimensions, it is not as straightforward to estimate variance components in the penalized regression framework \citep{ReidTibshiraniFriedman2016}. In contrast, estimating $(\bm{\Omega}, \sigma^2)$ is fairly simple under a fully Bayesian framework, where we can endow these parameters with appropriate priors \eqref{Omegaprior}-\eqref{sigma2prior} and use our ECM algorithm to estimate them.

\subsection{Computational complexity} \label{ECMcomplexity}

Recall that $m_i$ is the number of repeated measurements for subject $i$, and for $n$ subjects, there are a total of $N = \sum_{i=1}^{n} m_i$ observations. With $p$ varying coefficients, each represented by a basis expansion \eqref{basisfunctions} with $d$ basis functions, the number of unknown parameters in $\bm{\gamma}$ is $dp$. Meanwhile, we have $nq$ unknown parameters in $\bm{\eta}$ and $q^2$ unknown parameters in $\bm{\Omega}$, where $q$ is the number of basis functions in \eqref{basisfunctions2}. The computational cost of performing the E-step is $\mathcal{O}(d^2p)$ operations, which arises from computing $p$ $\ell_2$-norms $\lVert \bm{\gamma}_k \rVert_2, k = 1, \ldots, p$, in $\bm{\Psi} ( \bm{\gamma}_k \mid \lambda_1 ) \propto \exp( - \lambda_1 \lVert \bm{\gamma}_k \rVert_2 )$ and $\bm{\Psi} ( \bm{\gamma}_k \mid \lambda_0 ) \propto \exp( - \lambda_0 \lVert \bm{\gamma}_k \rVert_2 )$ to obtain the $p_k^{\star}$'s in \eqref{pstark}. In the M-step, the costs of updating $\theta$, $\bm{\eta}$, $\bm{\gamma}$, $\bm{\Omega}$, and $\sigma^2$ are respectively $\mathcal{O}(p)$, $\mathcal{O}(Ndp+Nq^2+q^3)$, $\mathcal{O}(Ndpr)$, $\mathcal{O}(nq^2)$, and $\mathcal{O}(Ndp + Nnq + N^2)$, where $r$ is the number of iterations it takes for the group coordinate ascent algorithm of \citet{BrehenyHuang2015} to converge.

Assuming that $\max \{d, q\} \ll n$ and $dp > N$, the most expensive step in the ECM algorithm is therefore solving for $\bm{\gamma}$ in \eqref{objgamma}. This gives Algorithm \ref{algorithm1} an overall computational complexity of $\mathcal{O}(Ndpr)$. However, as discussed in Section \ref{ECM}, $r$ (i.e. the number of iterations it takes for the group coordinate ascent algorithm to solve \eqref{objgamma}) tends to be small provided that $\lambda_0$ is sufficiently large. Thus, our ECM algorithm not only scales linearly in both $N$ and $p$, but it is also quite efficient in practice. We verify the speed and efficiency of our ECM algorithm in Section \ref{simulationsIII}.

\subsection{Choice of hyperparameters} \label{hyperparameters}

We now provide our recommendations for setting the hyperparameters in the NVC-SSL prior \eqref{ssgrouplasso}-\eqref{sigma2prior}.  We fix the slab hyperparameter $\lambda_1$ in the SSGL prior \eqref{ssgrouplasso} to be $\lambda_1 = 1$. This allows the slab density $\bm{\Psi}(\cdot \mid \lambda_1)$ to be fairly diffuse so that it is able to prevent overshrinkage of important covariates. We also set the shape parameters in the prior \eqref{thetaprior} on the mixing proportion $\theta$ to be $a=1$ and $b=p$. This ensures that $\theta$ is small with high probability, and therefore, most of the $\bm{\gamma}_k$'s will belong to the spike density $\bm{\Psi}(\cdot \mid \lambda_0)$ in the SSGL prior \eqref{ssgrouplasso}. Finally, to ensure that the priors on the variance parameters $\bm{\Omega}$ and $\sigma^2$ are weakly informative, we set $\nu = q+2$ and $\bm{\Phi} = \bm{I}_q$ in \eqref{Omegaprior} and $c_0=1$ and $d_0 =1$ in \eqref{sigma2prior}. 

The spike hyperparameter $\lambda_0$ in the SSGL prior \eqref{ssgrouplasso} plays the role of a regularization parameter on $\bm{\gamma}$, with larger values of $\lambda_0$ leading to more basis coefficients being thresholded to zero. Hence, the practical performance of NVC-SSL is governed heavily by the choice of $\lambda_0$. While one could fix $\lambda_0$ \textit{a priori} to a positive value where $\lambda_0 \gg \lambda_1$, the speed of the ECM algorithm that we introduced in Section \ref{ECM} makes it computationally feasible to determine a more optimal choice of $\lambda_0$ from a set of candidate values.

To this end, we fit NVC-SSL using several choices of $\lambda_0$ from a grid of $L$ decreasing $\lambda_0$ values $\lambda_0^{1} > \lambda_0^{2} > \ldots > \lambda_0^{L}$. Following \citet{WeiHuangLi2011}, we choose $\lambda_0^{l}, 1 \leq l \leq L$, using the Bayesian information criterion (BIC) of \citet{SchwarzBIC1978}. We have from \eqref{matrixform2} that the marginal likelihood of $\bm{Y}$ is $\bm{Y} \sim \mathcal{N}(\bm{U}\bm{\gamma}, \bm{Z} ( \bm{I}_n \otimes \bm{\Omega} ) \bm{Z}^\top + \sigma^2 \bm{I}_N )$. Let $\ell(\bm{\gamma}, \bm{\Omega}, \sigma^2)$ denote the log-marginal likelihood of $\bm{Y}$. For a given $\lambda_0$, the BIC in our present context is
\begin{equation} \label{BIC}
	\text{BIC}(\lambda_0) = -2~\ell(\widehat{\bm{\gamma}}_{\lambda_0}, \widehat{\bm{\Omega}}_{\lambda_0}, \widehat{\sigma}_{\lambda_0}^2) + \log N \times \# \text{ of nonzero elements in } \widehat{\bm{\gamma}}_{\lambda_0},
\end{equation}
where $(\widehat{\bm{\gamma}}_{\lambda_0}, \widehat{\bm{\Omega}}_{\lambda_0}, \widehat{\sigma}_{\lambda_0}^2)$ are the MAP estimates for $(\bm{\gamma}, \bm{\Omega}, \sigma^2)$ with $\lambda_0$ as the spike hyperparameter in \eqref{ssgrouplasso}. We select the $\lambda_0 \in \{ \lambda_0^{1}, \ldots, \lambda_0^{L} \}$ which minimizes the BIC \eqref{BIC}. An alternative to minimizing BIC is to choose $\lambda_0$ using cross-validation (CV). However, CV requires refitting the model $KL$ times, where $K$ is the number of folds. In contrast, BIC only requires solving $L$ optimization problems, one for each $\lambda_0$ in the grid. Thus, using CV is roughly $K$ times slower than minimizing BIC. 

In order to accelerate the computation of the $L$ optimizations for $\lambda_0 \in \{ \lambda_0^1, \ldots, \lambda_0^L \}$, we employ a warm starting strategy, where for each $\lambda_0^{l}, 2 \leq \l \leq L$, we initialize the ECM algorithm with $\bm{\gamma}^{(0)} = \widehat{\bm{\gamma}}(\lambda_0^{l-1})$, where $\widehat{\bm{\gamma}}(\lambda_0^{l-1})$ denotes the MAP estimator obtained from fitting NVC-SSL with the previous $\lambda_0$ in the grid. Since $\widehat{\bm{\gamma}}(\lambda_0^{l-1})$ serves as a reasonable initialization for $\bm{\gamma}^{(0)}$, the ECM algorithm for each $\lambda_0^{l}$ converges very quickly to a local mode. As shown in Section \ref{simulationsIII}, the ECM algorithm typically converges in 10 or fewer iterations. In all of our numerical experiments and real data applications, we found that tuning $\lambda_0$ from the equispaced grid $\{ 300, 290, \ldots, 20, 10 \}$ worked well in practice. 

Finally, using B-splines as the basis functions in \eqref{basisfunctions}-\eqref{basisfunctions2}, we determined that it is sufficient to fix the basis dimensions to be $d = q = 8$. While it is possible to further tune these values (for instance, we could use the BIC \eqref{BIC} to select $(\lambda_0, d, q)$ from a grid of triplets), we found that increasing $d$ and $q$ to be greater than eight offered little to no benefits in terms of improved estimation or variable selection. Thus, we recommend only tuning the spike hyperparameter $\lambda_0$, while keeping all other hyperparameters in the priors \eqref{ssgrouplasso}-\eqref{sigma2prior} and the basis dimensions in the model \eqref{approximatemodel} fixed at the default values suggested in this section.

\section{MCMC for scalable uncertainty quantification} \label{MCMC}

Apart from the ease with which one can incorporate unknown within-subject covariances through appropriate prior distributions, another advantage of Bayesian NVC models over penalized frequentist NVC models is their ability to provide natural uncertainty quantification through their posterior distributions. However, posterior sampling can be very challenging if $p$ is large. In this section, we demonstrate how NVC-SSL is amenable to fast posterior sampling for uncertainty quantification. 

We emphasize that the algorithms in this section are intended to be used in cases where $p \gg N$. If $p$ is small and $N$ is very large, then it is more advisable to use the scalable approaches introduced in \citet{GuhaniyogiLiSavitskySrivastava2022} and \citet{GuhaniyogiBaracaldoBanerjee2023} instead. In Section \ref{discussion}, we discuss some avenues for future work where \emph{both} $N$ and $p$ could be very large.

\subsection{Exact Gibbs sampling algorithm} \label{exactMCMC}

We first introduce an exact Gibbs sampling algorithm for fitting the NVC-SSL model. In order to obtain closed form updates in the Gibbs sampler, it will be convenient to reparameterize the $\mathcal{SSGL}(\lambda_0, \lambda_1, \theta)$ prior \eqref{SSGLrepar} as a Gaussian scale mixture density. First, note that for $\bm{\gamma}_k \sim \bm{\Psi}(\bm{\gamma}_k \mid \lambda) \propto \lambda^d \exp(-\lambda \lVert \bm{\gamma}_k \rVert_2)$, $\bm{\gamma}_k$ is the marginal density of the scale mixture,
\begin{equation*}
	\bm{\gamma}_k \mid \xi_k \sim \mathcal{N}(\bm{0}, \xi_k \bm{I}_d),~~\xi_k \sim \text{Gamma} \left( \frac{d+1}{2}, \frac{\lambda^2}{2} \right).
\end{equation*} 
Consequently, for $\bm{\xi} = (\xi_1, \ldots, \xi_p)^\top$, we can rewrite the prior \eqref{SSGLrepar} as the hierarchical model,
\begin{equation} \label{SSGLrepar2}
	\begin{array}{ll}
		\bm{\gamma} \mid \bm{\xi} \sim \mathcal{N}(\bm{0}, \bm{D}_{\bm{\xi}}),& \text{where } \bm{D}_{\bm{\xi}} = \text{Bdiag} (\xi_1 \bm{I}_d, \ldots, \xi_p \bm{I}_d), \\
		\xi_k \mid \tau_k \sim \text{Gamma} \left( \frac{d+1}{2}, \frac{(\lambda_k^{\star})^2}{2} \right), & \text{where } \lambda_k^{\star} = \tau_k \lambda_1 + (1-\tau_k) \lambda_0, \\
		\tau_k \mid \theta \sim \text{Bernoulli}(\theta), & k = 1, \ldots, p,
	\end{array}
\end{equation}
and $\text{Bdiag}$ denotes a block-diagonal matrix. With the likelihood function for \eqref{matrixform2}, the hierarchical priors in \eqref{SSGLrepar2} for $\bm{\gamma}$, and the priors \eqref{Omegaprior}-\eqref{sigma2prior} on $(\bm{\Omega}, \sigma^2)$, we obtain as the joint posterior for all parameters,

\begin{align*} \label{fullposterior}
	\pi( \bm{\gamma}, \bm{\xi}, \bm{\tau}, \theta, \bm{\eta}, \bm{\Omega}, \sigma^2 \mid \bm{Y}) & \propto (\sigma^2)^{-N/2} \exp \left( - \frac{\lVert \bm{Y} - \bm{U} \bm{\gamma} - \bm{Z} \bm{\eta} \rVert_2^2}{2 \sigma^2} \right) \\
	& \qquad \times \left( \text{det}(\bm{\Omega}) \right)^{-1/2} \exp \left( - \frac{\bm{\eta}^\top \bm{\Omega}^{-1} \bm{\eta}}{2} \right) \\
	& \qquad \times \pi(\bm{\gamma} \mid \bm{\xi}) \times \pi(\bm{\xi} \mid \bm{\tau}) \times \pi(\bm{\tau} \mid \theta) \times \pi(\bm{\Omega}) \times \pi (\sigma^2), \numbereqn
\end{align*}
where the terms in the last line of the display denote the prior densities for the respective parameters in \eqref{SSGLrepar2}, \eqref{Omegaprior}, and \eqref{sigma2prior}. Based on \eqref{fullposterior}, we can derive an exact Gibbs sampling algorithm where all conditional distributions are available in closed form. This MCMC algorithm is given in Algorithm \ref{algorithm2}. 

\begin{algorithm}[t!]
	\begin{flushleft}
		\textbf{Input:} Initial values $\bm{\gamma}^{(0)}$, $\theta^{(0)}$, $\bm{\Omega}^{(0)}$, $\bm{\xi}^{(0)}$, $\sigma^{2(0)}$, $T$ (number of MCMC samples to run),  \\
		\hspace{1.2cm} $B$ (number of samples to discard as burn-in) \\
		\textbf{Output:} MCMC samples for varying coefficients $\beta_k(\bm{t}),~k = 1, \ldots, p$
		\vspace{.5cm}
		
		\textbf{for} $t = 1, \ldots, T$ \textbf{do}
		\begin{enumerate}
			\item \textbf{for} $i = 1, \ldots, n$ \textbf{do} \\
			\hspace{.5cm} Sample $\bm{\eta}_i^{(t)} \sim \mathcal{N}(\bm{\zeta}_i, ~\bm{\Xi}_i)$, where $\bm{\Xi}_i = \left\{ \bm{Z}_i^\top \bm{Z}_i / \sigma^{2(t-1)} + (\bm{\Omega}^{(t-1)})^{-1} \right\}^{-1}$ and \\
			\hspace{.5cm} $\bm{\zeta}_i = \bm{\Xi}_i \bm{Z}_i^\top (\bm{Y}_i - \bm{U}_i \bm{\gamma}^{(t-1)}) / \sigma^{2(t-1)}$
			\item Sample $\bm{\Omega}^{(t)} \sim \textrm{Inverse-Wishart} \left(\nu + n,  ~\bm{\Phi} + \sum_{i=1}^{n} \bm{\eta}_i^{(t)} (\bm{\eta}_i^{(t)})^\top \right)$
			\item Sample $\sigma^{2(t)} \sim \textrm{Inverse-Gamma} \left( \frac{N+c_0}{2}, ~\frac{\lVert \bm{Y} - \bm{U} \bm{\gamma}^{(t-1)} - \bm{Z} \bm{\eta}^{(t)} \rVert_2^2 + d_0}{2} \right)$
			\item \textbf{for} $k = 1, \ldots, p$ \textbf{do} \\
			\begin{enumerate}
				\item Sample $\tau_k^{(t)} \sim \text{Bernoulli} \left( \frac{\pi_1}{\pi_1 + \pi_0} \right)$, where $\pi_1 = \theta^{(t-1)} \lambda_1^{d+1} \exp \{ - \lambda_1^2 \xi_k^{(t-1)} / 2 \}$ and $\pi_0 = (1-\theta^{(t-1)}) \lambda_0^{d+1} \exp \{ - \lambda_0^2 \xi_k^{(t-1)} / 2 \}$
				\item Sample $\xi_k^{(t)} \sim \text{Generalized-Inverse-Gaussian} \left( \frac{1}{2}, \lVert \bm{\gamma}_k^{(t-1)} \rVert_2^2, (\lambda_k^{\star})^2 \right)$, where $\lambda_k^{\star} = \tau_k^{(t)} \lambda_1 + (1-\tau_k^{(t)}) \lambda_0$
			\end{enumerate}
			\item Sample $\theta^{(t)} \sim \text{Beta} \left( a+ \sum_{k=1}^{p} \tau_k^{(t)},~b + p - \sum_{k=1}^{p} \tau_k^{(t)} \right)$
			\item Sample $\bm{\gamma}^{(t)} \sim \mathcal{N} \left( \bm{\mu}_{\bm{\gamma}}, \bm{\Sigma}_{\bm{\gamma}} \right)$, where $\bm{\mu}_{\bm{\gamma}} = \bm{\Sigma}_{\bm{\gamma}} \bm{U}^\top \left(\bm{Y} - \bm{Z} \bm{\eta}^{(t)} \right) / \sigma^{2(t)}$, $ \bm{\Sigma}_{\bm{\gamma}} = \{ \bm{U}^\top \bm{U} / \sigma^{2(t)} + \bm{D}_{\bm{\xi}}^{-1} \}^{-1}$, and $\bm{D}_{\bm{\xi}} = \text{Bdiag}~( \xi_1^{(t)} \bm{I}_d, \ldots, \xi_p^{(t)} \bm{I}_d )$
		\end{enumerate}
		\item 
		\textbf{return} $\beta_k^{(B+1)}(\bm{t}) = \sum_{l=1}^{d} \gamma_{kl}^{(B+1)} B_{kl}(\bm{t}), \ldots, \widehat{\beta}_k^{(T)}(\bm{t}) = \sum_{l=1}^{d} \gamma_{kl}^{(T)} B_{kl}(\bm{t})$ for $k=1, \ldots, p$
	\end{flushleft}
	\caption{MCMC algorithm for NVC-SSL} \label{algorithm2}
\end{algorithm}

As shown in Algorithm \ref{algorithm2}, the main computational bottleneck in the exact Gibbs sampler is sampling the basis coefficients $\bm{\gamma} \in \mathbb{R}^{dp}$ in Step 6, which overwhelms the cost of sampling from any of the other parameters when $dp > N$. Note that sampling from the $\bm{\eta}_i$ vectors and $\bm{\Omega}$ is not expensive, since we only need to sample $q$-dimensional vectors and a $q \times q$ matrix respectively in this case. Typically, $q$ (the number of basis functions) is not overwhelming large. On the other hand, if the number of covariates $p$ is large, then it can be computationally demanding to sample a $dp$-dimensional random Gaussian vector. Typical methods based on Cholesky decomposition require $\mathcal{O}(d^3 p^3)$ operations to compute the Cholesky decomposition for the covariance matrix $\bm{\Sigma}_{\gamma}$ \citep{bhattacharya2016fast}.  

In order to alleviate the cost of directly sampling from $\mathcal{N}(\bm{\mu}_{\bm{\gamma}}, \bm{\Sigma}_{\bm{\gamma}})$ in Step 6 of Algorithm \ref{algorithm2} when $dp > N$, we can employ the fast sampling method of \citet{bhattacharya2016fast}. The algorithm of \citet{bhattacharya2016fast} indirectly samples from $\mathcal{N}(\bm{\mu}_{\bm{\gamma}}, \bm{\Sigma}_{\bm{\gamma}})$ by solving a system of linear equations with $N$ equations and is much more efficient than methods based on Cholesky decomposition. This algorithm for sampling $\bm{\gamma}$ is given in Algorithm 3. 

The computational complexity of Algorithm \ref{algorithm3} is $\mathcal{O}(N^2 dp)$ when $dp > N$. Here, the main bottleneck is computing the matrix product $\bm{U} \bm{D}_{\bm{\xi}} \bm{U}^\top/\sigma^2$ in Step 3 which requires $\mathcal{O}(N^2 dp)$ operations and is more expensive than even inverting the matrix $\bm{K} = \bm{U} \bm{D}_{\bm{\xi}} \bm{U}^\top/\sigma^2 + \bm{I}_N$ in Step 4, which requires $\mathcal{O}(N^3)$ operations. 

\begin{algorithm}[t!]
	\begin{flushleft}
		\textbf{Input:} Most recent MCMC samples of $\bm{\xi}$, $\bm{\eta}$, and $\sigma$ \\
		\textbf{Output:} An exact sample of $\bm{\gamma}$ from Step 6 of Algorithm \ref{algorithm2}
		
		\begin{enumerate}
			\item Sample $\bm{m} \sim \mathcal{N}(\bm{0}, \bm{D}_{\bm{\xi}})$ and $\bm{\delta} \sim \mathcal{N}(\bm{0}, \bm{I}_N)$ independently
			\item Set $\bm{v} = (\bm{U}/\sigma) \bm{m} + \bm{\delta}$ and $\bm{v}^{\star} = (\bm{Y} - \bm{Z}\bm{\eta}) / \sigma - \bm{v}$ 
			\item Set $\bm{K} = \bm{U} \bm{D}_{\bm{\xi}} \bm{U}^\top / \sigma^2 + \bm{I}_N$ 
			\item Set $\bm{w} = \bm{K}^{-1} \bm{v}^{\star}$
			\item Set $\bm{\gamma} = \bm{m} + \bm{D}_{\bm{\xi}} \bm{U}^\top \bm{w} / \sigma$
		\end{enumerate}
		
		\textbf{return} $\bm{\gamma}$
	\end{flushleft}
	\caption{Exact algorithm for sampling $\bm{\gamma}$ in Step 6 of Algorithm \ref{algorithm2} when $dp > n$} \label{algorithm3}
\end{algorithm}

Using Algorithm \ref{algorithm3} to sample from $\bm{\gamma}$ enables NVC-SSL to scale linearly in $p$ per MCMC iteration rather than cubically. However, the fact that this exact algorithm scales quadratically in terms of total sample size $N$ may still be problematic. In particular, the multiplicative factor of $N^2$ in $\mathcal{O}(N^2 dp)$ might still render it costly to run the exact NVC-SSL Gibbs sampler if $p$ is large. For many large-scale problems, it is desirable to sample from the marginal posteriors in \emph{linear} time with respect to sample size $N$. This motivates us to develop an approximate MCMC algorithm in the next section that has a computational complexity of $\mathcal{O}(Ndp)$ per iteration.

\subsection{Approximate Gibbs sampling algorithm and its computational complexity} \label{approxMCMC}

Before introducing our approximate MCMC algorithm, we first review several other recently proposed approaches for sampling from spike-and-slab models that also scale linearly in both the number of covariates and the sample size. \citet{BiswasMackeyMeng2022} proposed an exact sampling method that has order $\max \{ N^2 p_t, Np \}$, where $p_t$ is the number of covariates that switch between the spike and the slab states between iterations $t-1$ and $t$. Typically, $p_t$ is much smaller than $p$, potentially offering substantial speed-ups over Algorithm \ref{algorithm3}. However, the algorithm of \citet{BiswasMackeyMeng2022} \emph{requires} both the spike and slab densities to be Gaussian and \emph{cannot} be used for Laplace densities (or other scale-mixture densities) like the ones employed by $\mathcal{SSGL}(\lambda_0, \lambda_1, \theta)$. For NVC-SSL, we use multivariate Laplace densities in the prior \eqref{ssgrouplasso} so that the MAP estimator is \emph{exactly} sparse (a feature that is not the case for Gaussian spike-and-slab priors). This rules out the approach of \citet{BiswasMackeyMeng2022}. 

In another line of work, \citet{NarisettyShenHe2019} proposed the skinny Gibbs algorithm. In each iteration of skinny Gibbs, the components of the regression coefficients vector $\bm{\beta}$ are partitioned as $\bm{\beta} = (\bm{\beta}_S^\top, \bm{\beta}_{S^c}^\top)^\top$, where $S$ denotes the set of variables belonging to the slab density and $S^c$ denotes the set of those belonging to the spike density. Suppose that the set $S$ is of size $s$. Skinny Gibbs decreases the cost of sampling $\bm{\beta}$ to $\mathcal{O}(Np)$ by ``sparsifying'' the covariance matrix for the conditional distribution of $\bm{\beta}$. That is, the cross-covariances in the conditional distribution of $\bm{\beta}$ are set to be zero, i.e. $\text{cov}(\bm{\beta}_S, \bm{\beta}_{S^c}) = \bm{0}$, and the off-diagonal entries of $\textrm{cov}(\bm{\beta}_{S^c})$ are also set to be zero. Thus, skinny Gibbs independently samples $\bm{\beta}_S$ from an $s$-dimensional multivariate Gaussian distribution and then independently samples the $p-s$ the entries in $\bm{\beta}_{S^c}$ from univariate Gaussian distributions.

While the skinny Gibbs algorithm has computational complexity that is linear in both $N$ and $p$, ignoring the correlations between $\bm{\beta}_S$ and $\bm{\beta}_{S^c}$ actually changes the prior on $\bm{\beta}$. As shown in Theorem 1 of \citet{NarisettyShenHe2019}, the posterior distribution under the skinny Gibbs algorithm becomes different from the posterior for the original, ``non-sparsified'' model as a result. In our case, we do \emph{not} want to completely change the NVC-SSL prior \eqref{SSGLrepar} for $\bm{\gamma}$, which would be the effect of employing the skinny Gibbs approach. Instead, we aim to \emph{approximate the transition kernel} of the NVC-SSL Markov chain.

To achieve an order $N$ speed-up, we adopt a similar idea as \citet{JohndrowOrensteinBhattacharya2020} who devised an approximate MCMC algorithm for the horseshoe prior \citep{CarvahoPolsonScott2010} in univariate Gaussian regression. However, unlike spike-and-slab priors, the horseshoe prior does \emph{not} naturally partition the covariates into ``significant'' and ``insignificant'' groups. Consequently, \citet{JohndrowOrensteinBhattacharya2020} require artificially segregating the covariates into two groups by using a user-specified threshold $\delta > 0$ to determine significant groups (i.e. regression coefficients with magnitude larger than $\delta$ are deemed to be ``significant''). In practice, it can be difficult to specify an appropriate threshold for $\delta$. In contrast, the binary indicators $\bm{\tau}$ in our spike-and-slab prior \eqref{SSGLrepar} naturally partition the groups of basis coefficients $\bm{\gamma}_k$'s in the \eqref{matrixform2} as either belonging to the spike or the slab. The automatic partitioning scheme allows us to construct a suitable approximate MCMC algorithm.

First, for the indicator variables $\bm{\tau} = (\tau_1, \ldots, \tau_p)^\top$ in \eqref{SSGLrepar2}, define the set $S = \{ k : \tau_k = 1 \}$, i.e. $S$ is the set of indices of the varying coefficients belonging to the slab density in \eqref{SSGLrepar}. Suppose that the cardinality of $S$ is $|S| = s$. Then $S^c = \{ 1, \ldots, p\} \setminus S$ denotes the indices of the varying coefficients belonging to the spike density in \eqref{SSGLrepar}, and $|S^c| = p-s$. Let $\bm{U}_{S}$ denote the $N \times ds$ submatrix of $\bm{U}$ in \eqref{Uimatrix} whose columns correspond to $S$, i.e. for each $k \in S$, $\bm{U}_S$ contains the $d$ columns of $\bm{U}$ that correspond to the $k$th varying coefficient. Similarly, let $\bm{D}_{S} = \text{Bdiag} \{ \xi_k \bm{I}_d  \}_{k \in S}$ denote the $ds \times ds$ block-diagonal submatrix of $\bm{D}_{\bm{\xi}}$ in \eqref{SSGLrepar2} whose diagonal blocks correspond to the $s$ indices of $S$.  

\begin{algorithm}[t!]
	\begin{flushleft}
		\textbf{Input:} Most recent MCMC samples of $\bm{\xi}$, $\bm{\eta}$, $\bm{\tau}$ and $\sigma$ \\
		\textbf{Output:} An approximate sample of $\bm{\gamma}$ from Step 6 of Algorithm \ref{algorithm2}
		
		\begin{enumerate}
			\item Sample $\bm{m} \sim \mathcal{N}(\bm{0}, \bm{D}_{\bm{\xi}})$ and $\bm{\delta} \sim \mathcal{N}(\bm{0}, \bm{I}_N)$ independently
			\item Set $\bm{v} = (\bm{U}/\sigma) \bm{m} + \bm{\delta}$ and $\bm{v}^{\star} = (\bm{Y} - \bm{Z}\bm{\eta}) / \sigma - \bm{v}$ 
			\item Set $\widetilde{\bm{K}} = \bm{I}_N - \bm{U}_S (\bm{U}_S^\top \bm{U}_S / \sigma^2 + \bm{D}_S^{-1})^{-1} \bm{U}_S^\top / \sigma^2$ 
			\item Set $\bm{w} = \widetilde{\bm{K}}^{-1} \bm{v}^{\star}$, where $\widetilde{\bm{K}}^{-1}$ is computed as in \eqref{Ktildeinverse}
			\item Set $\bm{\gamma} = \bm{m} + \widetilde{\bm{D}}_{\bm{\xi}} \bm{U}^\top \bm{w} / \sigma$, where $\widetilde{\bm{D}}_{\bm{\xi}} = \text{Bdiag}\{ \xi_k \bm{I}_d \mathbbm{I} (\tau_k = 1) \}$
		\end{enumerate}
		
		\textbf{return} $\bm{\gamma}$
	\end{flushleft}
	\caption{Approximate algorithm for sampling $\bm{\gamma}$ in Step 6 of Algorithm \ref{algorithm2} when $dp > n$} \label{algorithm4}
\end{algorithm}

When $\tau_k = 0$ in \eqref{SSGLrepar2}, $\lambda_k^{\star} \approx \lambda_0$ and $\lVert \bm{\gamma}_k \rVert_2 \approx 0$ in step 4(b) of Algorithm \ref{algorithm2}. Hence, $\xi_k \approx 0$ for $k \in S^c$. As a result, the $N \times N$ matrix product $\bm{U} \bm{D}_{\bm{\xi}} \bm{U}^\top$ can be well-approximated by $\bm{U}_S \bm{D}_{S} \bm{U}_S^\top$ \citep{JohndrowOrensteinBhattacharya2020}. Likewise, if we let $\widetilde{\bm{D}}_{\bm{\xi}} = \text{Bdiag}\{ \xi_k \bm{I}_d \mathbbm{I} (\tau_k = 1) \}$ denote the block-diagonal matrix where we replace the $\xi_k, k \in S$, in $\bm{D}_{\bm{\xi}}$ with zero, then the matrix product $\bm{D}_{\bm{\xi}} \bm{U}^\top$ is well-approximated by $\widetilde{\bm{D}}_{\bm{\xi}} \bm{U}^\top$. This suggests that we can make the following replacements in Algorithm \ref{algorithm3} to obtain an \emph{approximate} MCMC algorithm:
\begin{itemize}
	\item We can approximate $\bm{K}$ in Step 3 with $\widetilde{\bm{K}}$, where $\widetilde{\bm{K}} = \bm{U}_S \bm{D}_{S} \bm{U}_S^\top / \sigma^2 + \bm{I}_N$.
	\item We can approximate $\bm{K}^{-1}$ in Step 4 with $\widetilde{\bm{K}}^{-1}$.
	\item We can approximate $\bm{D}_{\bm{\xi}}$ in Step 5 with $\widetilde{\bm{D}}_{\bm{\xi}}$.   
\end{itemize} 
In particular, using the Woodbury matrix identity, we have that
\begin{equation} \label{Ktildeinverse}
	\widetilde{\bm{K}}^{-1} = \bm{I}_N - \bm{U}_S \left( \bm{U}_S^\top \bm{U}_S / \sigma^2 + \bm{D}_S^{-1} \right)^{-1} \bm{U}_S^\top / \sigma^2,
\end{equation}
i.e. computing the inverse of $\widetilde{\bm{K}}$ requires inverting only an $ds \times ds$ matrix now instead of an $N \times N$ matrix. Under sparsity and a relatively small basis dimension $d$, we typically have that $ds \ll N$. In addition, the matrix multiplication $\widetilde{\bm{D}}_{\bm{\xi}} \bm{U}^\top \mathbf{w} / \sigma$ costs $\mathcal{O}(Nds)$ operations, instead of the $\mathcal{O}(Ndp)$ operations that are required to compute the product $\bm{D}_{\bm{\xi}} \bm{U}^\top \bm{w} / \sigma$ in Algorithm \ref{algorithm3}. 

The complete approximate MCMC algorithm for approximately sampling from $\bm{\gamma}$ is given in Algorithm \ref{algorithm4}. By examining Algorithm \ref{algorithm4}, we see that the most expensive operation is now computing the matrix-vector product $(\bm{U}/\sigma) \bm{m}$ in Step 2, which requires $\mathcal{O}(Ndp)$ operations. If $dp$ is very large, then $\mathcal{O}(Ndp)$ represents a \emph{substantial} cost reduction from the $\mathcal{O}(N^2 dp)$ cost of the exact sampling scheme in Algorithm \ref{algorithm3}. 

In short, Algorithm \ref{algorithm4} allows us to \emph{approximately} sample from the basis coefficients $\bm{\gamma}$ in the NVC-SSL model with a runtime per MCMC iteration that is linear in \textit{both} the number of covariates $p$ \textit{and} the total sample size $N$.  The approximate MCMC algorithm has an even faster per iteration runtime than the ECM algorithm introduced in Section \ref{ECM}, which has time complexity of $\mathcal{O}(Ndpr)$, $r > 1$, per iteration. However, the ECM algorithm typically converges after a few iterations, whereas we may need to run MCMC for a much larger number of iterations to obtain enough posterior samples for good inference.

\subsection{Trade-offs between exact and approximate Gibbs sampling algorithms} \label{tradeoffs}

Algorithm \ref{algorithm4} reduces the per iteration cost of posterior sampling by a factor of $N$. In Section \ref{simulationsIII}, we demonstrate that as $dp$ increases, this offers a very significant reduction in MCMC runtime. However, the trade-off for faster computation is that the pointwise uncertainty intervals for the varying coefficients are slightly more conservative. This finding stands in contrast to that of \citet{JohndrowOrensteinBhattacharya2020} who claimed that inference from their approximate MCMC algorithm for the horseshoe prior was ``virtually indistinguishable'' from inference under the exact MCMC algorithm. In this section, we precisely quantify this trade-off.

Specifically, we investigate the implications of Algorithm \ref{algorithm4} on the posterior mean and variance of the conditional distribution for $\bm{\gamma}$. Without loss of generality, assume that $S = \{ 1, \ldots, s \}$ and $S^c = \{ s+1, \ldots, p \}$. Then we can partition the basis coefficients $\bm{\gamma}$ as  $\bm{\gamma} = ( \bm{\gamma}_S^\top , \bm{\gamma}_{S^c}^\top )^\top$, where $\bm{\gamma}_S$ consists of the first $s$ groups ($ds$ entries) in $\bm{\gamma}$, while $\bm{\gamma}_{S^c}$ consists of the last  $p-s$ groups ($d(p-s)$ entries) in $\bm{\gamma}$. Instead of exactly sampling from $\mathcal{N}(\bm{\mu}_{\gamma}, \bm{\Sigma}_{\bm{\gamma}})$ in Step 6 of Algorithm \ref{algorithm2}, the steps in Algorithm \ref{algorithm4} amount to sampling $\bm{\gamma}$ from the conditional distribution $\mathcal{N}(\widetilde{\bm{\mu}}_{\bm{\gamma}}, \widetilde{\bm{\Sigma}}_{\bm{\gamma}})$ \citep{JohndrowOrensteinBhattacharya2020}, where
\begin{equation} \label{mutilde}
	\widetilde{\bm{\mu}}_{\bm{\gamma}} = \begin{pmatrix} \widetilde{\bm{\mu}}_{\bm{\gamma}_S} \\ \widetilde{\bm{\mu}}_{\bm{\gamma}_{S^c}} \end{pmatrix} = \begin{pmatrix} \left( \bm{U}_S^\top \bm{U}_S / \sigma^2 + \bm{D}_{S}^{-1} \right)^{-1} \bm{U}_S^\top (\bm{Y} - \bm{Z} \bm{\eta} ) / \sigma^2 \\ \bm{0}_{(dp-ds) \times 1} \end{pmatrix},
\end{equation} 
and 
\begin{equation} \label{Sigmatilde}
	\widetilde{\bm{\Sigma}}_{\bm{\gamma}} = \begin{pmatrix} \widetilde{\bm{\Sigma}}_{\bm{\gamma}_S} & \widetilde{\bm{\Sigma}}_{\bm{\gamma}_{S,S^c}} \\ \widetilde{\bm{\Sigma}}_{\bm{\gamma}_{S, S^c}}^\top & \widetilde{\bm{\Sigma}}_{\bm{\gamma}_{S^c}} \end{pmatrix} =  \begin{pmatrix} \left( \bm{U}_S^\top \bm{U}_S / \sigma^2 + \bm{D}_S^{-1} \right)^{-1} & - \bm{D}_S \bm{U}_S^\top \widetilde{\bm{K}}^{-1} \bm{U}_{S^c} \bm{D}_{S^c} / \sigma^2 \\ - \bm{D}_{S^c} \bm{U}_{S^c}^\top \widetilde{\bm{K}}^{-1} \bm{U}_S \bm{D}_s / \sigma^2 & \bm{D}_{S^c} \end{pmatrix}.
\end{equation}
We see from \eqref{mutilde} that the marginal conditional distribution of $\bm{\gamma}_S$ has the same mean as the mean of the conditional distribution for $\bm{\gamma}$ if we were to fit NVC-SSL to \textit{only} the first $s$ varying coefficients. Meanwhile, the conditional mean for $\bm{\gamma}_{S^c}$ is a zero vector of length $d(p-s)$. Under sparsity, the exact MCMC algorithm will also result in MCMC samples for $\bm{\gamma}_{S^c}$ that are centered around a mean that is very close to zero. Based on these observations, we expect fairly negligible differences between the estimated posterior mean of $\bm{\gamma}$ under the exact and approximate MCMC algorithms. This is confirmed in our numerical experiments in Section \ref{simulationsII}.

Unlike the skinny Gibbs algorithm \citep{NarisettyShenHe2019}, we also observe from \eqref{Sigmatilde} that Algorithm \ref{algorithm4} preserves the cross-correlations between $\bm{\gamma}_{S}$ and $\bm{\gamma}_{S^c}$, i.e. $\text{cov}(\bm{\gamma}_S, \bm{\gamma}_{S^c}) = \widetilde{\bm{\Sigma}}_{\bm{\gamma}_{S,S^c}} \neq \bm{0}_{ds \times (dp-ds)}$. However, while there is negligible bias for the posterior \emph{means}, the marginal posterior \emph{variances} for $\bm{\gamma}$ under the approximate MCMC algorithm are slightly inflated.  This is formalized in the next proposition.

\begin{proposition} \label{prop:1}
	Let $\bm{\Sigma}_{\bm{\gamma}_S}$ and $\bm{\Sigma}_{\bm{\gamma}_{S^c}}$ denote the marginal covariance matrices for the conditional distributions of $\bm{\gamma}_S$ and $\bm{\gamma}_{S^c}$ respectively under the exact MCMC algorithm (Algorithm \ref{algorithm3}). Meanwhile, let $\widetilde{\bm{\Sigma}}_{\bm{\gamma}_S}$ and $\widetilde{\bm{\Sigma}}_{\bm{\gamma}_{S^c}}$ denote the marginal covariance matrices for the conditional distributions of $\bm{\gamma}_S$ and $\bm{\gamma}_{S^c}$ under the approximate MCMC algorithm (Algorithm \ref{algorithm4}), as defined in \eqref{Sigmatilde}. Then $\widetilde{\bm{\Sigma}}_{\bm{\gamma}_S} \geq \bm{\Sigma}_{\bm{\gamma}_S}$ and $\widetilde{\bm{\Sigma}}_{\bm{\gamma}_{S^c}} \geq \bm{\Sigma}_{\bm{\gamma}_{S^c}}$. 
\end{proposition}	

The proof of Proposition \ref{prop:1} is given in Appendix \ref{App:B}, which also gives precise expressions for $\widetilde{\bm{\Sigma}}_{\bm{\gamma}_S} - \bm{\Sigma}_{\bm{\gamma}_S}$ and $\widetilde{\bm{\Sigma}}_{\bm{\gamma}_{S^c}} - \bm{\Sigma}_{\bm{\gamma}_{S^c}}$. The implication of Proposition \ref{prop:1} is that the variances in the marginal posteriors for  the entries in $\bm{\gamma}$ will tend to be \emph{larger} under the approximate MCMC algorithm than under the exact MCMC algorithm. As a consequence, the pointwise posterior credible intervals for the varying coefficients $\beta_k(t)$ will also tend to be wider. To see this, suppose that, based on the posterior samples for $\bm{\gamma}$, we form the credible intervals $[ \gamma_{kl}^L, \gamma_{kl}^U], k = 1, \ldots, p, l = 1, \ldots, d$ with a prescribed level of probability $1 - \alpha, \alpha \in (0, 1)$. Then the $(1-\alpha) \times 100 \%$ posterior credible intervals for $\beta_k(t)$ will be $[\sum_{l=1}^{d} \gamma_{kl}^{L} B_{kl}(t), \sum_{l=1}^{d} \gamma_{kl}^{U} B_{kl}(t)]$. As a result of Proposition \ref{prop:1}, the endpoints $[ \gamma_{kl}^L, \gamma_{kl}^U]$ will tend to be further apart under Algorithm \ref{algorithm4}, leading to wider pointwise credible intervals for the varying coefficients $\beta_k(t)$'s than those under Algorithm \eqref{algorithm3}.

Nevertheless, our approximate MCMC algorithm for NVC-SSL is still a practical choice if computational time is a primary concern. In Section \ref{simulationsIII}, we show that when $dp = 40{,}000$, the approximate MCMC algorithm reduces the average runtime for 1000 MCMC iterations from 5.6 hours (for the exact algorithm) to 18.82 minutes.
This is a substantial computational gain and suggests that Algorithm \ref{algorithm4} is much more scalable for large $p$ than Algorithm \ref{algorithm3}. Sparse and/or low-rank approximations like Algorithm \ref{algorithm4} are also routinely employed in practice to improve the scalability and computational feasibility of fully Bayesian inference, at the expense of not performing exact inference. See, for example, the Gaussian process and spatial statistics literature \citep{Rasmussen2006, BanerjeeDunsonTokdar2013, DattaBanerjeeFinleyGelfand2016, HughesHaran2013}.

Moreover, if one is interested in \emph{simultaneous} coverage of the varying coefficient functions rather than pointwise coverage, then wider credible intervals are actually preferred. We show in Section \ref{simulationsII} that the approximate MCMC algorithm has higher simultaneous coverage of the true varying coefficients than the exact MCMC algorithm. The approximate MCMC algorithm also manages to produce uncertainty intervals that capture the true shape of the varying coefficients, as shown in the right three panels of Figure \ref{fig:exact_vs_approx_MCMC}.  

\section{Simulation studies} \label{simulations}

Here, we conduct simulation studies for NVC-SSL to validate its variable selection and estimation performance, inferential capabilities, and scalability. All of the methods in this section were implemented in the publicly available \textsf{R} package \texttt{NVCSSL}, which can be found on the Comprehensive \textsf{R} Archive Network.

\subsection{Variable selection and estimation performance} \label{simulationsI}

We first assessed the performance of the NVC-SSL MAP estimator obtained from the ECM algorithm in Section \ref{optimization}. In particular, we investigated the MAP estimator's ability to:
\begin{enumerate}
	\item capture different shapes for the varying coefficients, including nonzero but constant (i.e. non-time varying) functions;
	\item detect weak signals, i.e. varying coefficients with small magnitudes for $\lVert \beta_k(t) \rVert_{\infty}$;
	\item perform well under a variety of unknown within-subject covariance functions, including those that allow for long-range correlation and zero covariance functions (i.e. zero within-subject correlations).
\end{enumerate}
We generated data for $n=50$ subjects from model \eqref{varyingcoefficientmodel2} as follows. To simulate the observation times $t_{ij}$'s, we first sampled from $\{1, 2, \ldots, 20 \}$, where each time point has a 60 percent chance of being skipped. This way, we had very irregularly spaced data, with $m_i$ being different for different subjects. We then added random perturbation from $\mathcal{U}(-0.5,0.5)$ to the non-skipped time points. 

To model the high-dimensional scenario, we set $p=500$, with the first six variables $x_{i1}, \ldots, x_{i6}$ being the relevant ones. The first covariate $x_{i1}$ was simulated from $\mathcal{U}(t/10, 2+t/10)$ for any given time point $t$. The covariates $x_{ik}, k=2, \ldots, 5$, conditioned on $x_{i1}$, were i.i.d. drawn from a normal distribution with mean zero and variance $(1+x_{i1})/(2+x_{i1})$. The covariate $x_{i6}$, independent of $x_{ik}, k=1, \ldots, 5$, was normal with mean $1.5 \exp(t/40)$ and variance $1$. Finally, for $k=7,\ldots, 500$, each $x_{ik}$, independent of the others, was drawn from a multivariate normal distribution with covariance structure $\textrm{cov}(x_{ik}(t), x_{ik}(s)) = \rho^{-|t-s|}$, with $\rho=0.5$. The true coefficient functions were
\begin{align*}
	& \beta_1(t) = 10 \sin \left( \frac{\pi t}{15} \right),~~\beta_2(t) = -0.6t + 6,~~\beta_3(t) = -1+2 \sin \left( \frac{\pi(t-25)}{8} \right), \\
	& \beta_4(t) = 1+2 \cos \left( \frac{\pi (t-25)}{15} \right),~~\beta_5(t) = 2 +\frac{10}{1+e^{10-t}},~~\beta_6(t) = -5, \\
	& \beta_7(t) = \cdots = \beta_p(t) = 0.
\end{align*}
In particular, $\beta_3$ and $\beta_4$ are weak signals with small magnitudes for $\lVert \beta_k(t) \rVert_{\infty}$. Meanwhile, $\beta_2$ is a linear function, $\beta_5$ is a sigmoid curve with flat regions, and $\beta_6$ is a nonzero constant (non-time varying) function. For the measurement error term in \eqref{varyingcoefficientmodel2}, we fixed the noise variance $\sigma^2 = 1$. For the unknown within-subject covariances, we considered the following five experimental settings for the covariance function $k(t, t')$ in \eqref{varyingcoefficientmodel2}:
\begin{itemize}
	\item Experiment 1: first-order autoregressive (AR(1)). We set $k_i(t_{ij}, t_{ij'}) = s_i^2 \rho_i^{| t_{ij} - t_{ij'}|}$, $1 \leq j, j' \leq m_i$.
	\item Experiment 2: compound symmetry (CS). We set $k_i (t_{ij}, t_{ij'}) = s_i^2 \{ \mathbbm{1}(t_{ij}=t_{ij'}) + \rho_i \mathbbm{1} (t_{ij}=t_{ij'}) \}, 1 \leq j, j' \leq m_i$.
	\item Experiment 3: squared exponential (SE). We set $k_i(t_{ij}, t_{ij'}) = s_i^2 \exp ( - (t_{ij}-t_{ij'})^2/\ell_i^2 )$, $1 \leq j, j' \leq m_i$.
	\item Experiment 4: periodic. We set $k_i(t_{ij}, t_{ij'}) = s_i^2 \exp ( -2 [ \sin^2 (\pi |t_{ij} - t_{ij'}| / p_i ) ] / \ell_i^2 )$, $1 \leq j, j' \leq m_i$.
	\item Experiment 5: zero (i.id. errors). We set $k_i (t_{ij}, t_{ij'}) = 0, 1 \leq j, j' \leq m_i$.
\end{itemize}
We randomly sampled the within-subject variance hyperparameters $s_i^2 \in \{ 0.5, 0.75, 1, 1.25, 1.5 \}$, the autocorrelation hyperparameters $\rho_i \in \{ 0.2, 0.4, 0.6, 0.8 \}$, the lengthscale hyperparameters  $\ell_i \in \{ 0.3, 0.6, 0.9, 1.2, 1.5, 1.8 \}$, and the period hyperparameters $p_i \in \{ 0.5, 1, 1.5, 2 \}$.

We briefly discuss our choices of covariance functions. The AR(1) and SE covariance functions in Experiments 1 and 3 embody the belief that correlations between two time points $t$ and $t'$ decreases exponentially as $|t - t'|$ increases. The behavior implied by these kernel functions is not appropriate when there could be strong correlations between far apart time points. In contrast, the CS covariance function in Experiment 2 implies that the correlation between $t$ and $t'$ is the \emph{same} for all $t \neq t'$, making it suitable for capturing long-range correlations. The periodic covariance function in Experiment 4 exhibits periodic oscillation, meaning that as $|t - t'|$ increases, the correlation could be \emph{either} weak \textit{or} strong between $t$ and $t'$. Finally, the zero covariance function in Experiment 5 simply means that there are no within-subject correlations, and we are operating under the assumption of i.i.d. errors.

To implement the NVC-SSL method, we tuned the spike hyperparameter $\lambda_0$ from the grid $\{ 300, 290, \ldots, 10 \}$ and selected $\lambda_0$ using the BIC criterion \eqref{BIC}. Meanwhile, we fixed all the other hyperparameters to the ones recommended in Section \ref{hyperparameters} and fixed the basis dimensions as $d=q=8$ in \eqref{approximatemodel}.
We compared our approach to the group lasso (gLASSO), group smoothly clipped absolute deviation (gSCAD), and group minimax concave penalty (gMCP) \citep{YuanLin2006, BrehenyHuang2015}. For high-dimensional NVC models, these methods solve the following optimization problem:
\begin{equation*}
	\bm{\widehat{\gamma}} = \argmax_{\bm{\gamma}} \frac{1}{2} \lVert \bm{Y} - \bm{U} \bm{\gamma} \rVert_2^2 + \sum_{k=1}^{p} pen_{\lambda}(  \bm{\gamma}_k ),
\end{equation*}
where $\bm{U}$ is defined as in \eqref{Uimatrix} and $pen_{\lambda} (\cdot)$ is a penalty function that depends on a tuning parameter $\lambda$. These penalized approaches, which we refer to NVC-gLASSO, NVC-gSCAD, and NVC-gMCP respectively, have been considered by numerous authors in the varying coefficient literature \citep{WangXia2009, WangLiHuang2008, WeiHuangLi2011}. For these methods, we also fixed the basis dimension $d=8$ and tuned $\lambda$ using the BIC criterion.

\begin{table}[t!]
	\centering
	\caption{Simulation results for NVC-SSL, NVC-gLASSO, NVC-gSCAD, and NVC-gMCP, averaged across 200 replicates. To better highlight the differences in estimation performance, we rescale the MSE by 100, i.e. we report $\text{MSE} \times 100$ in the first column. The empirical standard error is reported in parentheses following the average.}
	\label{table:point_estimator_results}
	\medskip
	\resizebox{\textwidth}{!}{
		\begin{tabularx}{1.1\linewidth}{l *{11}X}
			\multicolumn{6}{c}{\textbf{Experiment 1: AR(1) covariance function}} \\ \toprule
			& MSE $\times$ 100 & MSPE & Sens & Spec & MCC \\ 
			\hline \hline
			NVC-SSL & \textbf{0.114} (0.049) & \textbf{4.057} (1.976) & \textbf{0.988} (0.052) & 0.999 (0.002) & \textbf{0.948} (0.069) \\
			NVC-gLASSO & 0.989 (0.128) & 8.559 (2.948) & 0.688 (0.057) & \textbf{1} (0) & 0.827 (0.033) \\
			NVC-gSCAD & 0.291 (0.030) & 4.752 (2.043) & 0.667 (0) & \textbf{1} (0) & 0.815 (0) \\ 
			NVC-gMCP & 0.284 (0.025) & 4.729 (2.050) & 0.667 (0) & \textbf{1} (0) & 0.815 (0) \\
			\bottomrule
	\end{tabularx}}
	
	\medskip
	
	\resizebox{\textwidth}{!}{
		\begin{tabularx}{1.1\linewidth}{l *{11}X}
			\multicolumn{6}{c}{\textbf{Experiment 2: CS covariance function}} \\ \toprule
			& MSE $\times$ 100 & MSPE & Sens & Spec & MCC \\ 
			\hline \hline
			NVC-SSL & \textbf{0.113} (0.046) & \textbf{4.255} (2.254) & \textbf{0.989} (0.041) & 0.999 (0.002) & \textbf{0.947} (0.067) \\
			NVC-gLASSO & 0.990 (0.116) & 8.884 (3.175) & 0.681 (0.050) & 0.999 (0.001) & 0.823 (0.029) \\
			NVC-gSCAD & 0.291 (0.029) & 4.956 (2.266) & 0.667 (0) & \textbf{1} (0) & 0.815 (0) \\ 
			NVC-gMCP & 0.284 (0.024) & 4.930 (2.274) & 0.667 (0) & \textbf{1} (0) & 0.815 (0) \\
			\bottomrule
	\end{tabularx}}
	
	\medskip
	
	\resizebox{\textwidth}{!}{
		\begin{tabularx}{1.1\linewidth}{l *{11}X}
			\multicolumn{6}{c}{\textbf{Experiment 3: SE covariance function}} \\ \toprule
			& MSE $\times$ 100 & MSPE & Sens & Spec & MCC \\ 
			\hline \hline
			NVC-SSL & \textbf{0.112} (0.056) & \textbf{4.448} (2.536) & \textbf{0.989} (0.053) & 0.999 (0.002) & \textbf{0.946} (0.071) \\
			NVC-gLASSO & 0.980 (0.129) & 9.347 (3.784) & 0.681 (0.050) & 0.999 (0.001) & 0.823 (0.029) \\
			NVC-gSCAD & 0.287 (0.031) & 5.056 (2.486) & 0.667 (0) & \textbf{1} (0) & 0.815 (0) \\ 
			NVC-gMCP & 0.280 (0.024) & 5.030 (2.498) & 0.667 (0) & \textbf{1} (0) & 0.815 (0) \\
			\bottomrule
	\end{tabularx}}
	
	\medskip
	
	\resizebox{\textwidth}{!}{
		\begin{tabularx}{1.1\linewidth}{l *{11}X}
			\multicolumn{6}{c}{\textbf{Experiment 4: Periodic covariance function}} \\ \toprule
			& MSE $\times$ 100 & MSPE & Sens & Spec & MCC \\ 
			\hline \hline
			NVC-SSL & \textbf{0.106} (0.042) & \textbf{4.062} (2.134) & \textbf{0.981} (0.045) & 0.999 (0.002) & \textbf{0.965} (0.054) \\
			NVC-gLASSO & 0.996 (0.128) & 8.706 (3.213) & 0.688 (0.055) & 0.999 (0.001) & 0.827 (0.033) \\
			NVC-gSCAD & 0.291 (0.029) & 4.737 (2.106) & 0.667 (0) & \textbf{1} (0) & 0.815 (0) \\ 
			NVC-gMCP & 0.286 (0.026) & 4.722 (2.096) & 0.667 (0) & \textbf{1} (0) & 0.815 (0) \\
			\bottomrule
	\end{tabularx}}
	
	\medskip
	
	\resizebox{\textwidth}{!}{
		\begin{tabularx}{1.1\linewidth}{l *{11}X}
			\multicolumn{6}{c}{\textbf{Experiment 5: Zero covariance function (i.i.d. errors)}} \\ \toprule
			& MSE $\times$ 100 & MSPE & Sens & Spec & MCC \\ 
			\hline \hline
			NVC-SSL & \textbf{0.062} (0.024) & \textbf{3.027} (2.269) & \textbf{0.999} (0.012) & 0.999 (0.002) & \textbf{0.962} (0.055) \\
			NVC-gLASSO & 0.966 (0.111) & 7.784 (3.123) & 0.680 (0.045) & \textbf{1} (0) & 0.823 (0.026) \\
			NVC-gSCAD & 0.268 (0.023) & 3.798 (2.284) & 0.667 (0) & \textbf{1} (0) & 0.815 (0) \\ 
			NVC-gMCP & 0.261 (0.017) & 3.775 (2.278) & 0.667 (0) & \textbf{1} (0) & 0.815 (0) \\
			\bottomrule
	\end{tabularx}}
\end{table}

While \citet{WangLiHuang2008}, \citet{WangLiHuang2008}, and \citet{WeiHuangLi2011} demonstrated competitive performance of their methods even when failing to account for within-subject correlations, we wanted to see whether explicitly incorporating estimation of unknown within-subject correlations (as with the NVC-SSL model) could improve estimation and variable selection. Moreover, Experiment 5 (i.e. zero covariance function) was conducted in order to fairly compare NVC-SSL to NVC-gLASSO, NVC-gSCAD, and NVC-gMCP, since the setting of Experiment 5 is the exact NVC model implied by the latter three methods. 

To compare these methods, we evaluated the estimation error, out-of-sample prediction error, and variable selection performance. For estimation error, we computed the mean squared error (MSE),
\begin{align*}
	\textrm{MSE} = \frac{1}{Np} \sum_{k=1}^{p} \sum_{i=1}^{n} \sum_{j=1}^{m_i} \left[ \widehat{\beta}_k(t_{ij}) - \beta_{0k}(t_{ij}) \right]^2.
\end{align*} 
For out-of-sample prediction, we generated 30 new observations $(\bm{Y}_{new}, \bm{t}_{new}, \bm{X}_{new})$, calculated a new $\bm{U}$ matrix \eqref{Uimatrix}, and computed the mean squared prediction error (MSPE),
\begin{align*}
	\textrm{MSPE} = \frac{1}{N} \lVert \bm{Y}_{new} - \bm{U}_{new} \widehat{\bm{\gamma}} \rVert_2^2.
\end{align*} 
Finally, to evaluate variable selection performance, we compared the sensitivity (Sens), specificity (Spec), and Matthews correlation coefficient (MCC), given by
\begin{align*}
	& \textrm{Sens} = \frac{\textrm{TP}}{\textrm{TP}+\textrm{FN}},~~~\textrm{Spec} = \frac{\textrm{TN}}{\textrm{TN}+\textrm{FP}},\\
	& \textrm{MCC} = \frac{\textrm{TP} \times \textrm{TN} - \textrm{FP} \times \textrm{FN}}{\sqrt{ (\textrm{TP}+\textrm{FP})(\textrm{TP}+\textrm{FN})(\textrm{TN}+\textrm{FP})(\textrm{TN}+\textrm{FN})}},
\end{align*}
where TP, TN, FP, and FN are the number of true positives, true negatives, false positives, and false negatives respectively. MCC takes values between -1 and 1, with higher values indicating better overall variable selection performance.

\begin{table}[t!]
	\centering 
	\caption{Simulation results for estimation and variable selection of the nonzero varying coefficients $\beta_k(t), k = 1, \ldots, 6$, for NVC-SSL, NVC-gLASSO, NVC-gSCAD, and NVC-gMCP, averaged across 200 replicates. ``Proportion'' gives the proportion of replicates that selected the varying coefficient. } \label{table:ind_beta_point_estimate_results}
	\medskip
	\resizebox{.94\textwidth}{!}{  
		\begin{tabular}{lccccccccccccc} 
			\multicolumn{14}{c}{\textbf{Experiment 1: AR(1) covariance function}} \\
			\midrule
			\phantom{abc} & \multicolumn{6}{c}{MSE} & \phantom{abc}& \multicolumn{6}{c}{Proportion} \\
			\cmidrule{2-7} \cmidrule{9-14} 
			& $\beta_1$ & $\beta_2$ & $\beta_3$ & $\beta_4$ & $\beta_5$ & $\beta_6$ & &  $\beta_1$ & $\beta_2$ & $\beta_3$ & $\beta_4$ & $\beta_5$ & $\beta_6$ \\ \midrule 
			NVC-SSL & \textbf{0.060} & \textbf{0.081} & \textbf{0.093} & \textbf{0.103} & \textbf{0.113} & \textbf{0.042} & & 1 & 1 & \textbf{0.97} & \textbf{0.96} & 1 & 1 \\
			NVC-gLASSO & 0.870 & 1.082 & 0.468 & 0.565 & 1.068 & 0.894 & & 1 & 1 & 0.045 & 0.08 & 1 & 1 \\ 
			NVC-gSCAD & 0.074 & 0.136 & 0.476 & 0.577 & 0.136 & 0.058 & & 1 & 1 & 0 & 0 & 1 & 1 \\
			NVC-gMCP & 0.073 & 0.105 & 0.476 & 0.577 & 0.134 & 0.057 & & 1 & 1 & 0 & 0 & 1 & 1 \\
			\bottomrule
		\end{tabular}
	}
	
	\medskip
	\resizebox{.94\textwidth}{!}{  
		\begin{tabular}{lccccccccccccc} 
			\multicolumn{14}{c}{\textbf{Experiment 2: CS covariance function}} \\
			\midrule
			\phantom{abc} & \multicolumn{6}{c}{MSE} & \phantom{abc}& \multicolumn{6}{c}{Proportion} \\
			\cmidrule{2-7} \cmidrule{9-14} 
			& $\beta_1$ & $\beta_2$ & $\beta_3$ & $\beta_4$ & $\beta_5$ & $\beta_6$ & &  $\beta_1$ & $\beta_2$ & $\beta_3$ & $\beta_4$ & $\beta_5$ & $\beta_6$ \\ \midrule 
			NVC-SSL & \textbf{0.066} & \textbf{0.080} & \textbf{0.094} & \textbf{0.094} & \textbf{0.114} & \textbf{0.045} & & 1 & 1 & \textbf{0.965} & \textbf{0.97} & 1 & 1 \\
			NVC-gLASSO & 0.890 & 1.068 & 0.472 & 0.566 & 1.036 & 0.916 & & 1 & 1 & 0.025 & 0.06 & 1 & 1 \\
			NVC-gSCAD & 0.085 & 0.129 & 0.475 & 0.577 & 0.132 & 0.056 & & 1 & 1 & 0 & 0 & 1 & 1 \\
			NVC-gMCP & 0.084 & 0.096 & 0.475 & 0.577 & 0.131 & 0.056 & & 1 & 1 & 0 & 0 & 1 & 1 \\
			\bottomrule
		\end{tabular}
	}
	
	\medskip
	\resizebox{.94\textwidth}{!}{  
		\begin{tabular}{lccccccccccccc} 
			\multicolumn{14}{c}{\textbf{Experiment 3: SE covariance function}} \\
			\midrule
			\phantom{abc} & \multicolumn{6}{c}{MSE} & \phantom{abc}& \multicolumn{6}{c}{Proportion} \\
			\cmidrule{2-7} \cmidrule{9-14} 
			& $\beta_1$ & $\beta_2$ & $\beta_3$ & $\beta_4$ & $\beta_5$ & $\beta_6$ & &  $\beta_1$ & $\beta_2$ & $\beta_3$ & $\beta_4$ & $\beta_5$ & $\beta_6$ \\ \midrule 
			NVC-SSL & \textbf{0.063} & \textbf{0.076} & \textbf{0.087} & \textbf{0.098} & \textbf{0.106} & \textbf{0.043} & & 1 & 1 & \textbf{0.975} & \textbf{0.96} & 1 & 1 \\
			NVC-gLASSO & 0.848 & 1.059 & 0.473 & 0.562 & 1.051 & 0.908 & & 1 & 1 & 0.015 & 0.07 & 1 & 1 \\
			NVC-gSCAD & 0.078 & 0.128 & 0.474 & 0.574 & 0.122 & 0.058 & & 1 & 1 & 0 & 0 & 1 & 1 \\
			NVC-gMCP & 0.076 & 0.100 & 0.474 & 0.574 & 0.121 & 0.056 & & 1 & 1 & 0 & 0 & 1 & 1 \\
			\bottomrule
		\end{tabular}
	}
	
	\medskip
	\resizebox{.94\textwidth}{!}{  
		\begin{tabular}{lccccccccccccc} 
			\multicolumn{14}{c}{\textbf{Experiment 4: Periodic covariance function}} \\
			\midrule
			\phantom{abc} & \multicolumn{6}{c}{MSE} & \phantom{abc}& \multicolumn{6}{c}{Proportion} \\
			\cmidrule{2-7} \cmidrule{9-14} 
			& $\beta_1$ & $\beta_2$ & $\beta_3$ & $\beta_4$ & $\beta_5$ & $\beta_6$ & &  $\beta_1$ & $\beta_2$ & $\beta_3$ & $\beta_4$ & $\beta_5$ & $\beta_6$ \\ \midrule 
			NVC-SSL & \textbf{0.065} & \textbf{0.079} & \textbf{0.092} & \textbf{0.098} & \textbf{0.110} & \textbf{0.042} & & 1 & 1 & \textbf{0.97} & \textbf{0.975} & 1 & 1 \\
			NVC-gLASSO & 0.899 & 1.047 & 0.471 & 0.562 & 1.122 & 0.878 & & 1 & 1 & 0.035 & 0.09 & 1 & 1 \\
			NVC-gSCAD & 0.089 & 0.125 & 0.476 & 0.578 & 0.133 & 0.056 & & 1 & 1 & 0 & 0 & 1 & 1 \\.
			NVC-gMCP & 0.089 & 0.101 & 0.476 & 0.578 & 0.132 & 0.056 & & 1 & 1 & 0 & 0 & 1 & 1 \\
			\bottomrule
		\end{tabular}
	}
	
	\medskip
	\resizebox{.94\textwidth}{!}{  
		\begin{tabular}{lccccccccccccc} 
			\multicolumn{14}{c}{\textbf{Experiment 5: Zero covariance function (i.i.d. errors)}} \\
			\midrule
			\phantom{abc} & \multicolumn{6}{c}{MSE} & \phantom{abc}& \multicolumn{6}{c}{Proportion} \\
			\cmidrule{2-7} \cmidrule{9-14} 
			& $\beta_1$ & $\beta_2$ & $\beta_3$ & $\beta_4$ & $\beta_5$ & $\beta_6$ & &  $\beta_1$ & $\beta_2$ & $\beta_3$ & $\beta_4$ & $\beta_5$ & $\beta_6$ \\ \midrule 
			NVC-SSL & \textbf{0.039} & \textbf{0.041} & \textbf{0.045} & \textbf{0.048} & \textbf{0.082} & \textbf{0.026} & & 1 & 1 & \textbf{1} & \textbf{0.995} & 1 & 1 \\
			NVC-gLASSO & 0.839 & 1.032 & 0.470 & 0.570 & 1.010 & 0.907 & & 1 & 1 & 0.045 & 0.035 & 1 & 1 \\ 
			NVC-gSCAD & 0.054 & 0.093 & 0.475 & 0.576 & 0.099 & 0.040 & & 1 & 1 & 0 & 0 & 1 & 1 \\
			NVC-gMCP & 0.053 & 0.064 & 0.475 & 0.576 & 0.097 & 0.039 & & 1 & 1 & 0 & 0 & 1 & 1 \\
			\bottomrule
		\end{tabular}
	}
\end{table}

We repeated Experiments 1-5 for 200 replications each. In Table \ref{table:point_estimator_results}, we report our results averaged across the 200 replicates. We see that under all of the different unknown within-subject covariance structures, NVC-SSL had the lowest MSE and the lowest MSPE, indicating the best estimation and predictive performance. For variable selection, NVC-SSL had the highest average sensitivity, indicating the best ability to detect the true nonzero varying coefficient functions. NVC-gSCAD and NVC-gMCP had higher specificity, but NVC-SSL's average specificity was still quite good, at 0.999. Moreover, NVC-SSL had the highest average MCC, indicating the best overall ability to correctly select the true nonzero functions while excluding the spurious ones. Our results suggest that accounting for within-subject correlations can greatly improve estimation, prediction, and variable selection.

\begin{figure}[t!]
	\centering
	\includegraphics[scale=0.65]{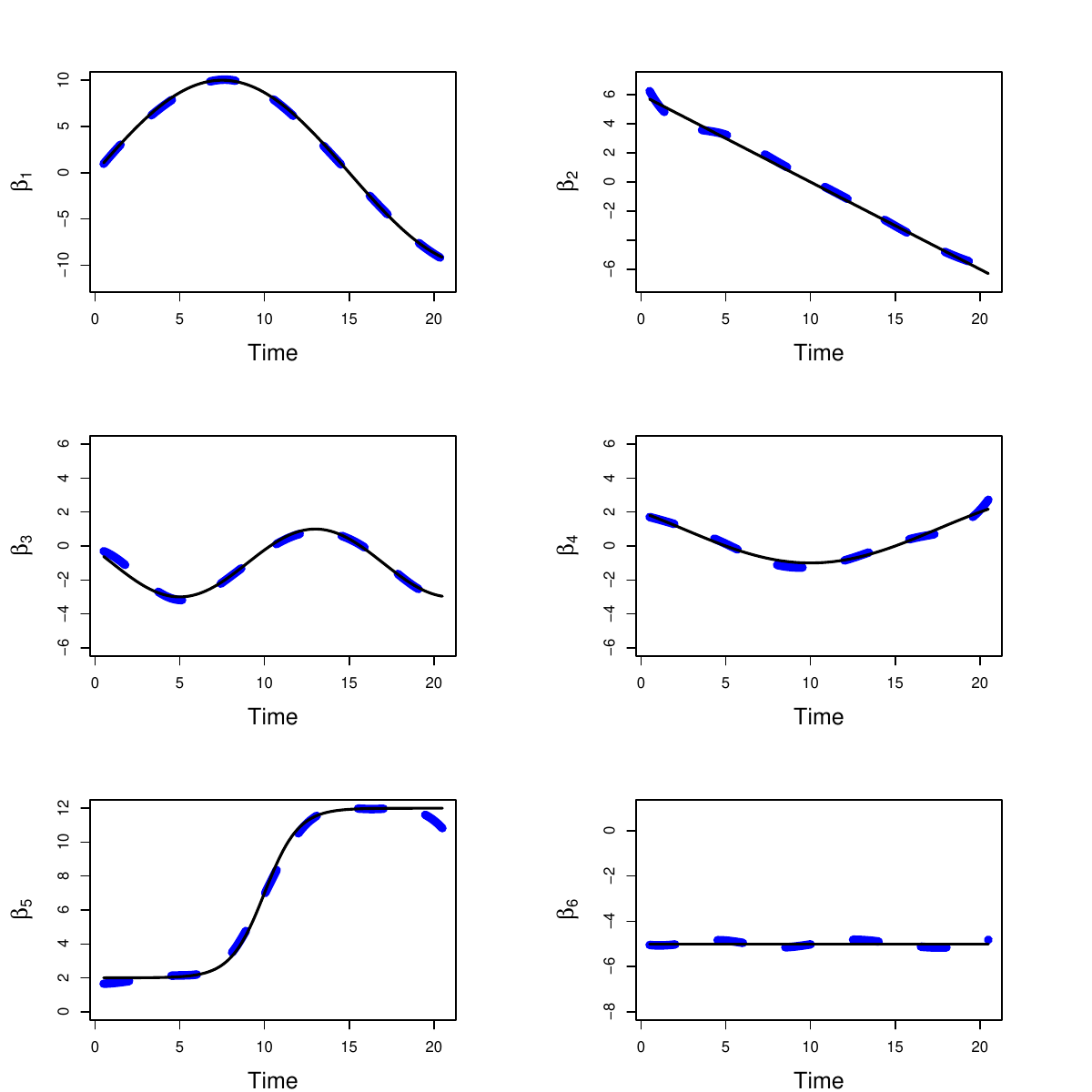} 
	\caption{Plots of the MAP estimates (dashed blue lines) for $\beta_k(t), k=1, \dots, 6$, under the NVC-SSL model from one replication of Experiment 1 (i.e. AR(1) within-subject covariance function). The true functions are the solid black lines.}
	\label{fig:MAP_estimates}
\end{figure}

An interesting to point out is that in Experiment 5 (i.e. zero within-subject correlations), NVC-SSL \textit{still} managed to outperform NVC-gLASSO, NVC-gSCAD, and NVC-gMCP. One might assume that the latter three methods would be competitive in this setting, since the implied model under these methods is i.i.d. errors with noise variance $\sigma^2 = 1$. However, NVC-SSL continued to have lower MSE and MSPE and higher MCC in this scenario. This suggests the following two things. First, the NVC-SSL method is flexible enough to \emph{also} estimate covariance functions equal to zero, and thus, NVC-SSL can work well in the case where the residual errors truly are i.i.d. Secondly, our results also point to the practical benefit having a slab density $\bm{\Psi}(\cdot \mid \lambda_1)$, in addition to a spike density $\bm{\Psi}(\cdot \mid \lambda_0)$ in the $\mathcal{SSGL}(\lambda_0, \lambda_1, \theta)$ prior \eqref{ssgrouplasso}. Meanwhile, gLASSO, gMCP, and gSCAD only have one regularization parameter $\lambda > 0$ controlling the level of sparsity. As a result, these other methods may overshrink many of the functions to zero when $\lambda$ is large.

\begin{figure}[t!]
	\centering
	\includegraphics[scale=0.7]{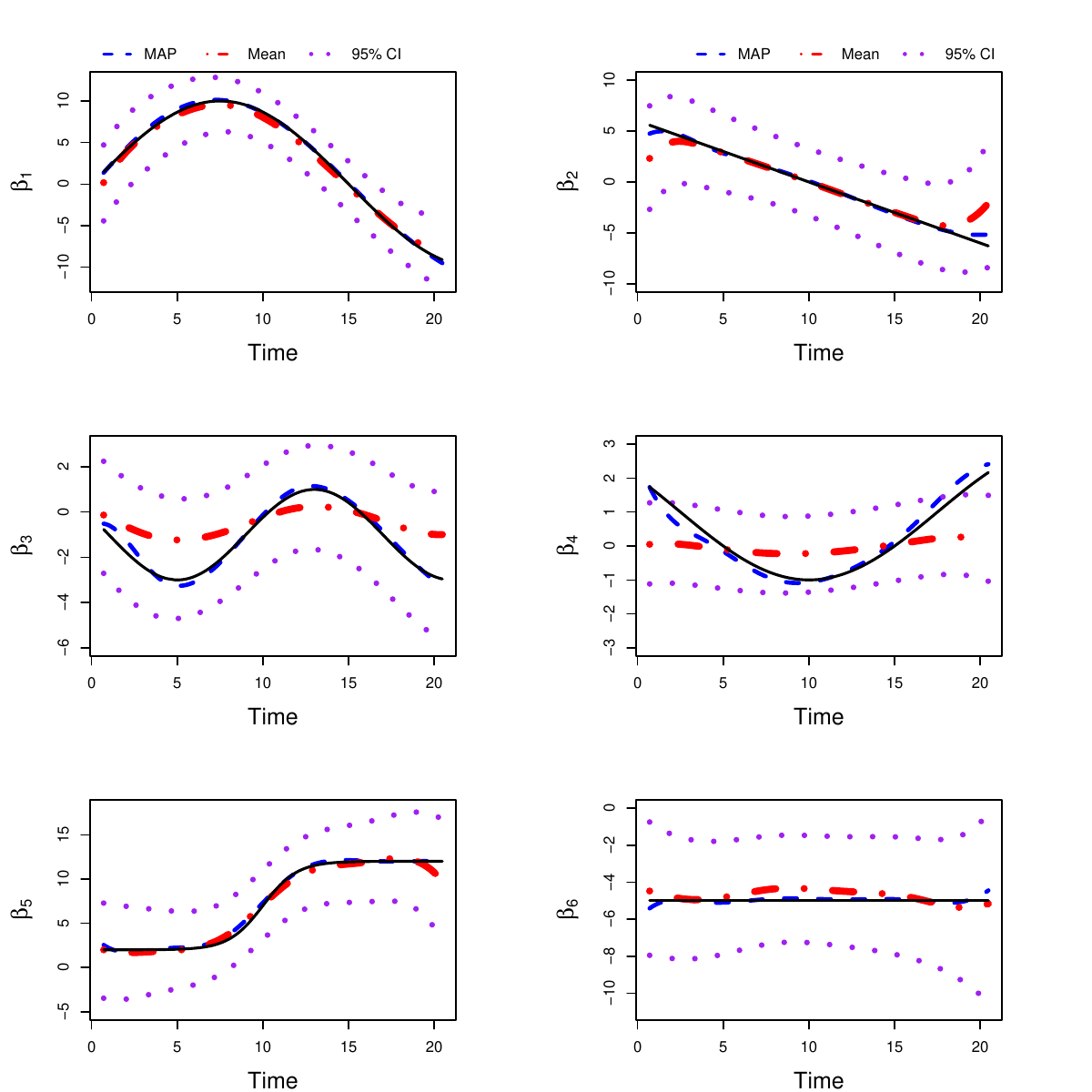} 
	\caption{Plots of the MAP estimates (dashed blue line), posterior mean estimates (dashed red line with dots), and 95\% posterior credible intervals (dotted purple lines) for $\beta_k(t), k = 1, \ldots, 6$, under the NVC-SSL model from one replication of Experiment 2 (i.e. CS within-subject covariance function). The true functions are the solid black lines.}
	\label{fig:NVC_mode_vs_mean}
\end{figure}

We also compared how well NVC-SSL, NVC-gLASSO, NVC-gSCAD, and NVC-gMCP were able to estimate and select the true nonzero varying coefficients $\beta_1, \ldots, \beta_6$. These results are reported in Table \ref{table:ind_beta_point_estimate_results}. In this case, we computed the MSE for the individual functions as $\text{MSE} = N^{-1} \sum_{i=1}^{N} \sum_{j=1}^{m_i} [ \beta_k(t_{ij}) - \beta_{0k}(t_{ij}) ]^2$. Table \ref{table:ind_beta_point_estimate_results} shows that in all our experiments, NVC-SSL had the lowest average MSE, indicating that it estimated the nonzero functions the best. We also kept track of the proportion of the 200 replicates that selected $\beta_1, \ldots, \beta_6$. As shown in Table \ref{table:ind_beta_point_estimate_results}, NVC-gLASSO, NVC-gSCAD, and NVC-gMCP were unable to select the weak signals $\beta_3$ and $\beta_4$ in almost all replications, which explains their lower average sensitivity and MCC in Table \ref{table:point_estimator_results}. In contrast, NVC-SSL selected $\beta_3$ and $\beta_4$ in 96\% to 100\% of the replicates. This verifies that NVC-SSL is especially well-suited for detecting weak signals. Our results in Table \ref{table:ind_beta_point_estimate_results} further reinforce the benefit of having a slab density in the $\mathcal{SSGL}(\lambda_1, \lambda_0, \theta)$ prior \eqref{ssgrouplasso} for NVC-SSL, which helps smaller signals to escape the pull of the spike and thus be detected.

Figure \ref{fig:MAP_estimates} plots the estimated varying coefficients (dashed line) against the true varying coefficients (solid line) for $\beta_k(t), k = 1, \ldots, 6$ from one replication of Experiment 1. Figure \ref{fig:MAP_estimates} shows that the MAP estimator under NVC-SSL is able to capture the true shapes of these functions, including the weaker signals ($\beta_3$ and $\beta_4$). Moreover, NVC-SSL was also able to capture the linear trend in $\beta_2$, the flat regions of the function $\beta_5$, and the constant nonzero (non time-varying) function $\beta_6$. This demonstrates the flexibility of our model and justifies the use of B-splines as the basis functions in \eqref{approximatemodel}.

In Appendix \ref{App:A}, we report additional results for our Experiments 1-5 where we used the MCMC algorithm in Section \ref{exactMCMC} to fit the NVC-SSL model. We conclude that the MAP estimator is superior as a \textit{point estimator} for function selection and estimation in high dimensions. This is demonstrated in Figure \ref{fig:NVC_mode_vs_mean} where the MAP estimator (dashed line) is shown to better capture the shapes of the weak signals $\beta_3$ and $\beta_4$ than the posterior mean (dotted and dashed line). However, posterior \emph{inference} from the 95\% credible intervals (dotted lines in Figure \ref{fig:NVC_mode_vs_mean}) is still quite good, even if the MAP estimator is preferred for point estimation.

\subsection{Performance of MCMC for inference} \label{simulationsII}

In this section, we compare the performance of the exact and approximate MCMC algorithms introduced in Section \ref{MCMC} for inference. We simulated data from $n=100$ subjects. For each $i$th subject, $m_i = 8$ time points were randomly sampled from $\mathcal{U}(0,20)$, leading to a total of $N = 800$ observations. We set $p=1000$, with the true varying coefficients,
\begin{align*}
	& \beta_1(t) = 10 \sin \left( \frac{\pi t}{15} \right),~~\beta_2(t) = 8 \cos \left( \frac{\pi(t-20)}{5} \right),~~\beta_3(t) = 2+\frac{10}{1+e^{10-t}}, \\
	& \beta_4(t) = \cdots = \beta_p(t) = 0,
\end{align*}
i.e. $\beta_1$, $\beta_2$, and $\beta_3$ are nonzero functions and the rest of the varying coefficients are zero. We generated the time-varying covariates $x_k(t)$ the same way as we did in Section \ref{simulationsI}, and we simulated the response variables $y(t)$ from the model \eqref{varyingcoefficientmodel2} under the following within-subject covariance structures (see Experiments 1-5 in Section \ref{simulationsI} for details):
\begin{itemize}
	\item Experiment 6: AR(1) covariance function
	\item Experiment 7: CS covariance function
	\item Experiment 8: SE covariance function
	\item Experiment 9: periodic covariance function
	\item Experiment 10: zero covariance function (i.e. no within-subject correlations)
\end{itemize}
In Experiments 6 through 10, the hyperparameters in the within-subject covariance functions $k_i(t_{ij},t{ij'})$ for each $i$th subject were simulated the same way as those described in Section \ref{simulationsII}.

Each experiment was repeated 200 times. In all replications, we ran both the exact and approximate MCMC algorithms introduced in Section \ref{MCMC} for 2000 iterations, discarding the first 500 iterations as burnin. The remaining 1500 MCMC samples were used to approximate the posteriors and perform uncertainty quantification. Our MCMC algorithms were initialized with the MAP estimator obtained from the ECM algorithm, and all hyperparameters and basis dimensions were the same as those used for the ECM algorithm. With the MAP estimator as our choice of initialization, the effective sample size (prior to burnin) was very close to 2000 for each of the basis coefficients in $\bm{\gamma}$, suggesting that 2000 MCMC iterations was sufficient.  

We used the posterior samples of $\bm{\gamma}$ to estimate the posterior mean for each $k$th varying coefficient as $\widetilde{\beta}_k(t) = \sum_{l=1}^{d} \widetilde{\gamma}_{kl} B_{kl}(t)$, where $\widetilde{\bm{\gamma}}$ denotes the posterior mean of $\bm{\gamma}$. In order to obtain the pointwise 95\% posterior credible intervals for each varying coefficient, we used the 2.5th and 97.5 sample quantiles of the MCMC samples for $\bm{\gamma}$. That is, the 95\% credible interval for each varying coefficient function $\beta_k(t)$ at time $t$ was $[\beta_k^{L}(t), \beta_k^{U}(t)]$, where $\beta_k^{L}(t) = \sum_{l=1}^{d} \gamma_{kl}^{L} B_{kl}(t)$ and $\beta_k^{U} (t) = \sum_{l=1}^{d} \gamma_{kl}^{U} B_{kl}(t)$, and $\gamma_{kl}^{L}$ and $\gamma_{kl}^{U}$ were the 2.5 and 97.5 quantiles for the MCMC samples of $\gamma_{kl}$. 

\begin{table}[t!]
	\centering
	\caption{Simulation results for the exact MCMC algorithm and the approximate MCMC algorithm, averaged across 200 replicates. The rescaled MSE (i.e. MSE $\times$ 100) is reported for the posterior mean. The standard errors for the rescaled MSE and the average width of the pointwise 95\% posterior intervals are reported in parentheses.}
	\label{table:exact_vs_approx_MCMC_results}
	\medskip
	\resizebox{\textwidth}{!}{
		\begin{tabularx}{1.2\linewidth}{l *{11}X}
			\multicolumn{5}{c}{\textbf{Experiment 6: AR(1) covariance function}} \\ \toprule
			& MSE $\times$ 100 & Width & Pointwise ECP & Simultaneous ECP \\ 
			\hline \hline
			Exact & 0.202 (0.049) & 2.128 (0.003) & 0.999 & 0.85 \\
			Approximate & 0.202 (0.049) & 2.423 (0.009) & 1 & 1 \\
			\bottomrule
	\end{tabularx}}
	
	\medskip
	
	\resizebox{\textwidth}{!}{
		\begin{tabularx}{1.2\linewidth}{l *{11}X}
			\multicolumn{5}{c}{\textbf{Experiment 7: CS covariance function}} \\ \toprule
			& MSE $\times$ 100 & Width & Pointwise ECP & Simultaneous ECP \\ 
			\hline \hline
			Exact & 0.174 (0.046) & 2.128 (0.003) & 0.999 & 0.89 \\
			Approximate & 0.174 (0.046) & 2.422 (0.008) & 0.999 & 0.99 \\
			\bottomrule
	\end{tabularx}}
	
	\medskip
	
	\resizebox{\textwidth}{!}{
		\begin{tabularx}{1.2\linewidth}{l *{11}X}
			\multicolumn{5}{c}{\textbf{Experiment 8: SE covariance function}} \\ \toprule
			& MSE $\times$ 100 & Width & Pointwise ECP & Simultaneous ECP \\ 
			\hline \hline
			Exact & 0.199 (0.044) & 2.128 (0.003) & 0.999 & 0.91 \\
			Approximate & 0.199 (0.044) & 2.423 (0.008) & 0.999 & 0.99 \\
			\bottomrule
	\end{tabularx}}
	
	\medskip
	
	\resizebox{\textwidth}{!}{
		\begin{tabularx}{1.2\linewidth}{l *{11}X}
			\multicolumn{5}{c}{\textbf{Experiment 9: Periodic covariance function}} \\ \toprule
			& MSE $\times$ 100 & Width & Pointwise ECP & Simultaneous ECP \\ 
			\hline \hline
			Exact & 0.190 (0.046) & 2.129 (0.004) & 0.999 & 0.87 \\
			Approximate & 0.190 (0.046) & 2.423 (0.008) & 1 & 1 \\
			\bottomrule
	\end{tabularx}}
	
	\medskip
	
	\resizebox{\textwidth}{!}{
		\begin{tabularx}{1.2\linewidth}{l *{11}X}
			\multicolumn{5}{c}{\textbf{Experiment 10: Zero covariance function (i.i.d. errors)}} \\ \toprule
			& MSE $\times$ 100 & Width & Pointwise ECP & Simultaneous ECP \\ 
			\hline \hline
			Exact & 0.136 (0.038) & 2.127 (0.003) & 0.999 & 0.89 \\
			Approximate & 0.136 (0.038) & 2.421 (0.008) & 0.999 & 0.99 \\
			\bottomrule
	\end{tabularx}}
\end{table}

We compared the MSE for the posterior mean functions $\widetilde{\beta}_k(t)$'s obtained from the exact MCMC and the approximate MCMC algorithms. We also compared the average width and the empirical coverage probability (ECP) of the 95\% posterior credible intervals. We looked at both the pointwise ECP (i.e. the proportion of pointwise credible intervals that contained the true value of $\beta_k(t_{ij})$ for each observed time point $t_{ij}$ $1 \leq i \leq n, 1 \leq j \leq m_i$) and the simultaneous ECP. Here, the simultaneous ECP was determined by the proportion of simulations where \emph{all} of the posterior credible intervals covered \emph{all} of the true varying coefficient functions in the entire time domain. It is important to note that in high dimensions, a Bayesian credible set is not necessarily a confidence set \citep{VanDerPasSzaboVanDerVaart2017}. Nevertheless, investigating the ECP is a good way to gauge whether the Bayesian uncertainty intervals are reasonable or not. 

\begin{figure}[t!]
	\centering
	\includegraphics[scale=0.65]{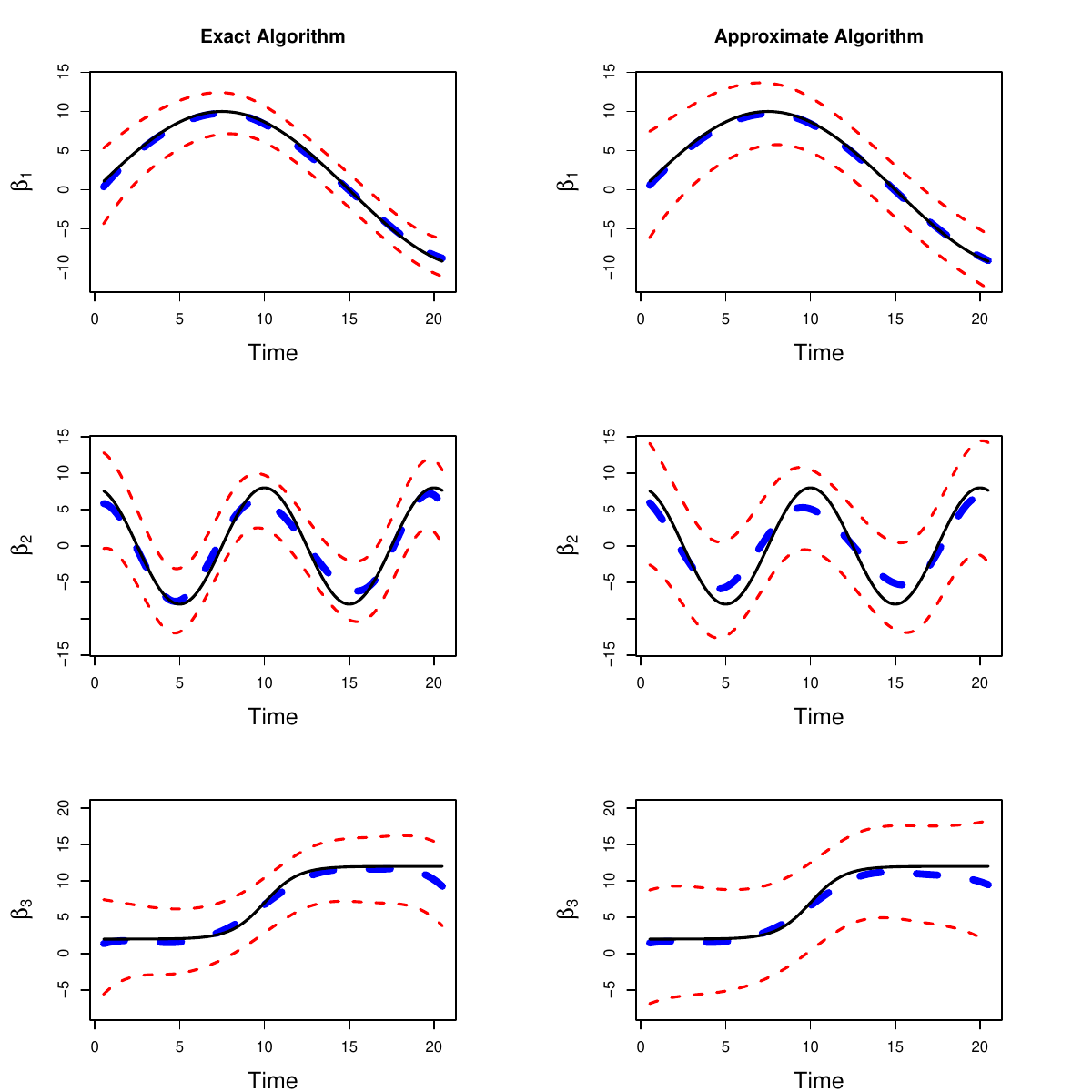} 
	\caption{Plots of the results from one replication of Experiment 9 (i.e. periodic within-subject covariance function) under the exact MCMC algorithm (left three figures) and the approximate MCMC algorithm (right three figures). The true functions $\beta_1(t)$, $\beta_2(t)$, and $\beta_3(t)$ are the solid black lines, the posterior mean estimates are the thick dashed blue lines, and the 95\% posterior credible intervals are the thin red dashed lines.}
	\label{fig:exact_vs_approx_MCMC}
\end{figure}

Our results are reported in Table \ref{table:exact_vs_approx_MCMC_results}. We can see that the average MSE for the posterior means under the exact MCMC and approximate MCMC algorithms were practically identical, which aligns with our theoretical analysis in Section \ref{tradeoffs}. However, the average width of credible intervals were slightly larger for the approximate MCMC algorithm than for the exact algorithm. We theoretically quantified this trade-off in Section \ref{tradeoffs}. 

While the \emph{pointwise} ECP was comparable for both algorithms, Table \ref{table:exact_vs_approx_MCMC_results} shows that the \textit{simultaneous} ECP was considerably higher for the approximate MCMC algorithm. Namely, 99 to 100 percent of simulations in each experiment had credible intervals which contained \emph{all} of the true varying coefficient functions. This can be attributed to the larger size of the uncertainty intervals produced by the approximate MCMC algorithm.

Figure \ref{fig:exact_vs_approx_MCMC} plots the posterior mean varying coefficients for $\beta_1(t)$, $\beta_2(t)$, and $\beta_3(t)$ from one replication of Experiment 9. The ground truth is plotted as a solid line, while the posterior mean and 95\% posterior credible intervals are plotted as dashed lines. The results for the exact algorithm are shown in the left panels and the results for the approximate algorithms are shown in the right panels. Figures \ref{fig:NVC_mode_vs_mean} and \ref{fig:exact_vs_approx_MCMC} show that despite having wider credible intervals, the approximate algorithm still captures the shape of the true varying coefficients. Running the exact algorithm for 2000 iterations also took 2.27 hours for the one replicate in Figure \ref{fig:exact_vs_approx_MCMC}, whereas the approximate algorithm only took only 6.19 minutes on an 11th Gen Intel Core i5-1135G7 processor. In short, the approximate MCMC algorithm gives slightly more conservative pointwise uncertainty intervals but it also provides better \emph{simultaneous} coverage \emph{and} it is much faster and more scalable than the exact algorithm.

\subsection{Timing and efficiency comparisons} \label{simulationsIII}

Here, we report results for the timing and efficiency of the algorithms that we introduced in Sections \ref{optimization} and \ref{MCMC}. To conduct our experiments, we modified Experiment 8 from Section \ref{simulationsII} (i.e. SE within-subject covariance function for $n=100$ subjects and $N=800$ total observations). Namely, we varied $p \in \{ 500, 1000, \ldots, 5000 \}$, but kept all the other simulation settings the same. We again used $d=8$ basis functions for each varying coefficient, leading to a total of $dp \in \{ 4000, 8000, \ldots, 40{,}000 \}$ unknown basis coefficients in $\bm{\gamma}$. We report timing results for the optimal $\lambda_0$ chosen from BIC \eqref{BIC}. To stress that we are in fact dealing with a very high-dimensional problem, we report our results using $dp$ instead of merely $p$.

\begin{figure}[t!]
	\centering
	\includegraphics[scale=0.55]{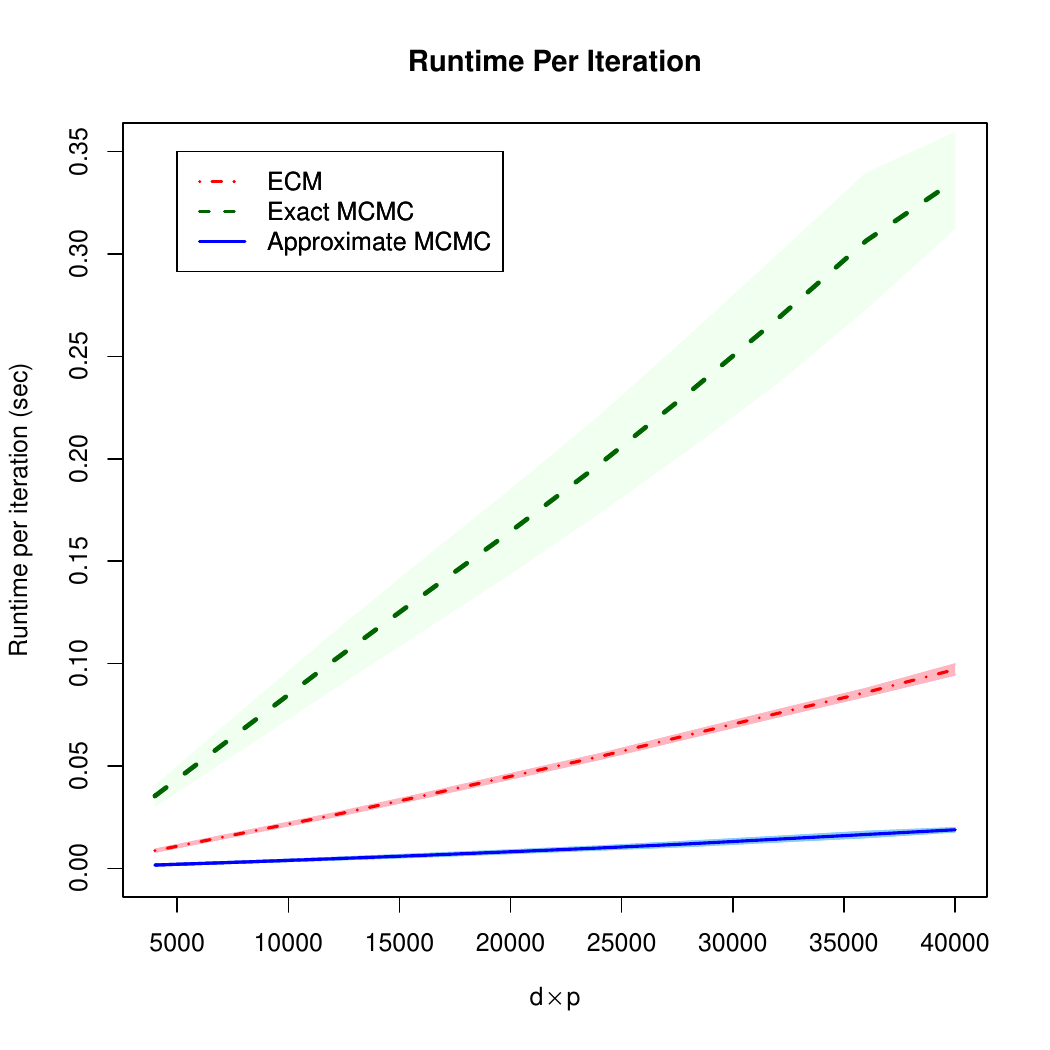} 
	\caption{Plots of the average runtime per iteration (seconds) across 50 replications for the ECM algorithm, exact MCMC algorithm, and approximate MCMC algorithm against the dimension $dp$ of the basis coefficients vector. The lightly shaded areas indicate the regions within one standard deviation of the mean.}
	\label{fig:per_iteration_cost}
\end{figure}

We first compared the average per-iteration runtime for the ECM algorithm of Section \ref{optimization} and the exact MCMC and approximate MCMC algorithms of Section \ref{MCMC} across 50 replicates. Figure \ref{fig:per_iteration_cost} plots the average runtime per iteration for these three methods against $dp$. We see that all methods scale linearly with $p$. In particular, for $dp=40{,}000$, the average runtime for one iteration was 5.83 seconds for the ECM algorithm, 20.16 seconds for the exact MCMC algorithm, and 1.13 seconds for the approximate MCMC algorithm. 

As shown in Figure \ref{fig:per_iteration_cost}, the approximate MCMC algorithm had on average the fastest runtime per iteration, the ECM algorithm has is the second fastest, and the exact MCMC algorithm is the slowest. This matches our earlier complexity analysis, since the computational complexity of the ECM algorithm is $\mathcal{O}(Ndpr)$, where $r$ is the number of iterations it takes to numerically solve for $\bm{\gamma}$ in  \eqref{objgamma}. Meanwhile, the approximate MCMC algorithm in Section \ref{approxMCMC} has worst-case complexity of $\mathcal{O}(Ndp)$. The extra factor of $r>1$ in the ECM algorithm accounts for its slower per-iteration runtime than the approximate MCMC algorithm. With time complexity of $\mathcal{O}(N^2dp)$, the exact MCMC algorithm has the slowest runtime, indicating that in general, $N \gg r$. 

Although the per iteration cost was the fastest for the approximate MCMC algorithm, the ECM algorithm also terminated very quickly. In all of experiments, the ECM algorithm converged within 10 iterations, even when $dp=40{,}000$. On the other hand, we would typically run MCMC algorithms for much more than 10 iterations -- usually for at least a couple hundred iterations. As a result, the \emph{overall} time to complete the MCMC algorithm may still be greater than that for the MAP estimation algorithm. 

\begin{figure}[t!]
	\centering
	\includegraphics[scale=0.62]{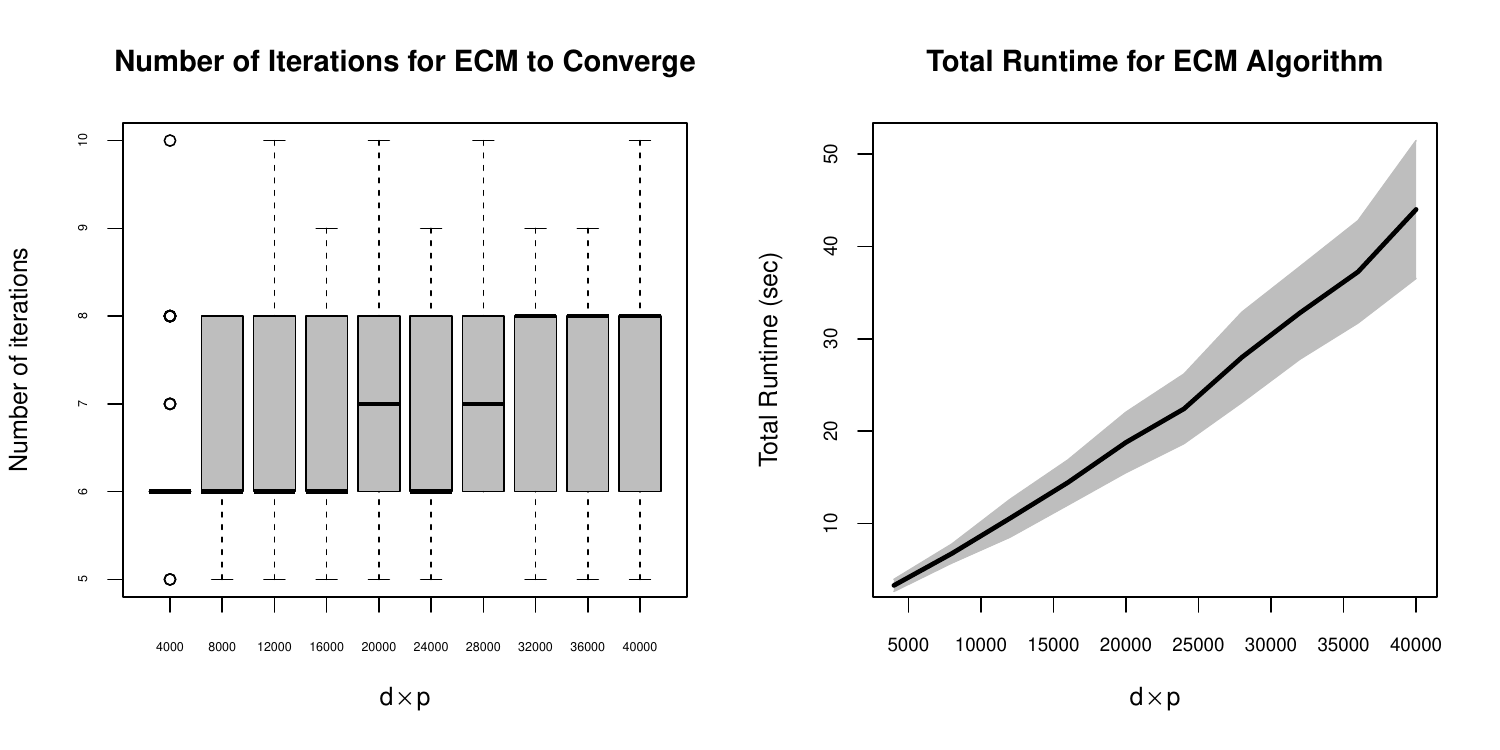} 
	\caption{Left panel: Box plots of the number of iterations it took for the ECM algorithm to converge (50 replications). Right panel: Plot of the average total runtime (seconds) across 50 replications for the ECM algorithm to finish running against the dimension $dp$ of the basis coefficients vector. The lightly shaded area in the right plot indicates the region within one standard deviation of the mean.}
	\label{fig:ECM_plots}
\end{figure}

In the left panel of Figure \ref{fig:ECM_plots}, we plot the box plots for the number of iterations that it took for the ECM algorithm to finish running for each $dp \in \{ 4000, 8000, \ldots, 40{,}000 \}$. The right panel of Figure \ref{fig:ECM_plots} reports the \emph{total} runtime for the ECM algorithm as a function of $dp$. For $dp=40{,}000$, it took on average 7.54 iterations and 44 seconds for the ECM algorithm to converge. These results demonstrate the scalability and computational feasibility of finding the MAP estimator for NVC-SSL.

\begin{figure}[t!]
	\centering
	\includegraphics[scale=0.62]{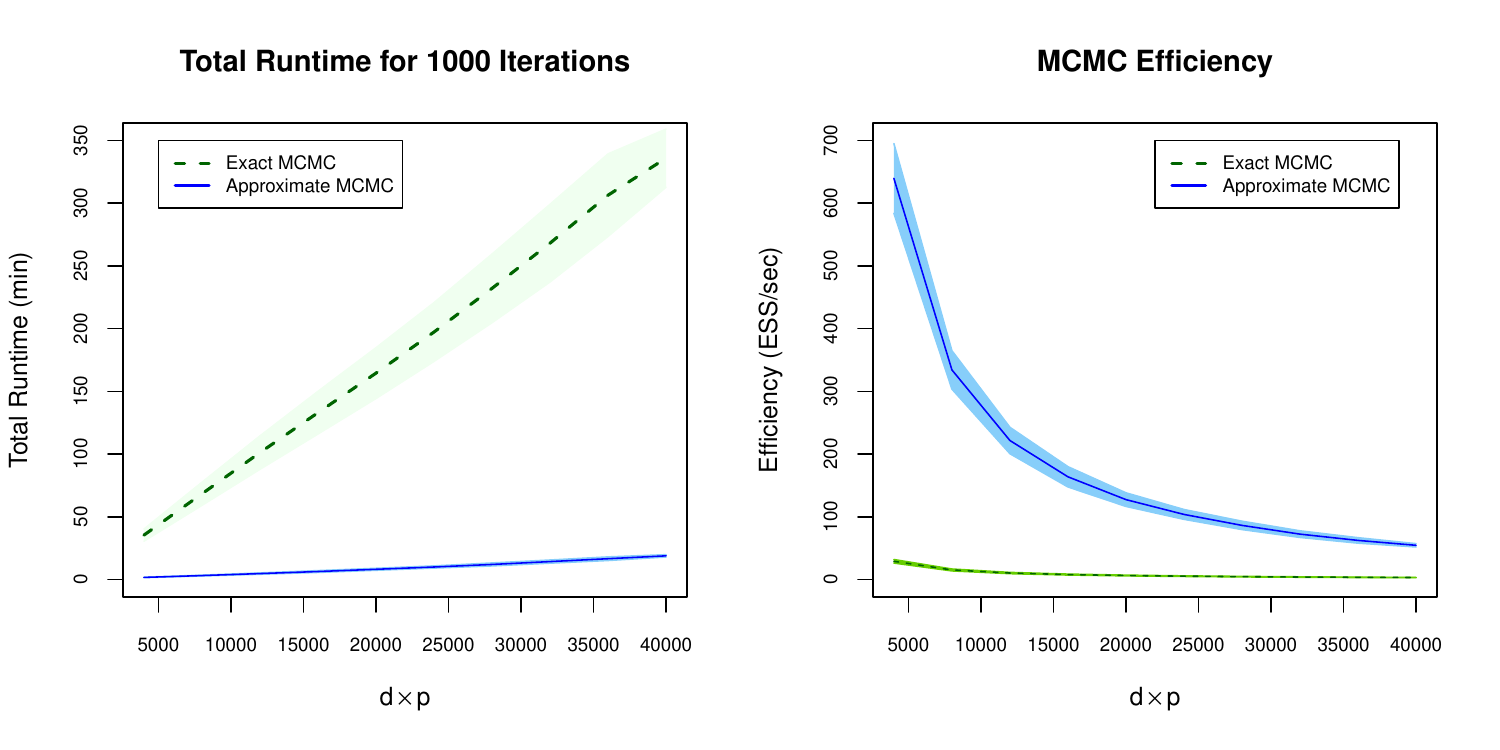} 
	\caption{Left panel: Plots comparing the average total runtime (minutes) across 50 replications for the exact MCMC and approximate MCMC algorithms to run 1000 iterations. Right panel: Plot of the MCMC efficiency (ESS per second) across 50 replications for the exact MCMC and approximate MCMC algorithms based on 1000 iterations. In both plots, the lightly shaded areas are the regions within one standard deviation of the mean.}
	\label{fig:MCMC_plots}
\end{figure}

In the left panel of Figure \ref{fig:MCMC_plots}, we compare the \emph{total} runtime of the exact MCMC and the approximate MCMC algorithms for 1000 iterations. Both algorithms were initialized with the MAP estimator for $\bm{\gamma}$ obtained from the ECM algorithm. For $dp=4000$, the average  total runtime was 1.55 minutes for the approximate MCMC algorithm vs. 34.53 minutes for the exact MCMC algorithm. For $dp=40{,}000$, the average total runtime was 18.82 minutes for the approximate MCMC algorithm vs. 338.91 minutes (or 5.6 hours) for the exact MCMC algorithm. It is clear that the approximate MCMC algorithm provides orders of magnitude speedup, especially when $dp$ is very large.

Perhaps a more transparent way to compare the MCMC algorithms is their efficiency, or their effective sample size (ESS) per second. For correlated MCMC samples, the ESS estimates the number of independent samples that would have given the same precision (or variance) as the MCMC samples. Thus, a higher ESS per second indicates greater MCMC efficiency. We used the \textsf{R} package \texttt{sns} to estimate the ESS for all $dp$ entries in $\bm{\gamma}$ and then took the average ESS for these parameters in $\bm{\gamma}$. In the right panel of Figure \ref{fig:MCMC_plots}, we plot the average ESS against $dp$ for the exact MCMC and the approximate MCMC algorithms. Figure \ref{fig:MCMC_plots} shows that the approximate MCMC algorithm has much higher efficiency. In particular, when $dp=4000$, the average efficiency was 639.65 samples per second vs. only 28.94 samples per second for the exact algorithm. For $dp=40{,}000$, the average efficiency was 54.44 samples per second for the approximate algorithm vs. only 3.05 samples per second for the exact algorithm. Thus, even though we have only approximated the MCMC transition kernel in the approximate algorithm, we have not done so at the expense of efficiency -- in fact, we significantly \emph{increased} the efficiency of our MCMC samples.

\section{Yeast cell cycle data analysis} \label{dataanalysis}

The cell cycle is a tightly regulated set of processes by which cells grow, replicate their DNA, segregate their chromosomes, and divide into daughter cells.  Transcription factors (TFs) are sequence-specific DNA binding proteins which regulate the transcription of genes from DNA to mRNA by binding specific DNA sequences. To better understand how TFs regulate the cell cycle, we applied our proposed NVC-SSL procedure to a dataset of cell-cycle regulated yeast genes and associated TFs. 

The data that we used comes from the $\alpha$-factor synchronized cultures of \citet{Spellman1998} and the CHIP-chip data of \citet{Lee2002}. \citet{Spellman1998} measured genome-wide mRNA levels for 6178 yeast open reading frames (ORFs) over approximately two cell cycle periods, with measurements at 7-minute intervals for 119 minutes (for a total of 18 time points). The data of \citet{Lee2002} contains binding information of 96 TFs which elucidates which TFs bind to promoter sequences of genes across the yeast genome. We aimed to fit the varying coefficient model to these 96 TFs and an intercept function $\beta_0(t)$ representing the baseline change in mRNA over time, i.e.
\begin{equation} \label{genemodel}
	y_i(t_{ij}) = \beta_0(t_{ij}) + \sum_{k=1}^{96} x_{ik}\beta_k(t_{ij}) + \varepsilon_i(t_{ij}),~~i= 1, \ldots, n,~~j = 1, \ldots, 18.
\end{equation}
where $y_i(t_{ij})$ denotes the mRNA level for the $i$th gene at the $j$th time point. Thus, including the intercept function, we have $p=97$ varying coefficients. Like other authors \citep{WangLiHuang2008, XueQu2012}, we also penalized $\beta_0(t)$ in order to ensure the identifiability of all varying coefficients. 

Previous works for fitting \eqref{genemodel} assumed that the error terms $\varepsilon_i(t_{ij})$'s were i.i.d. for all $i$ and $j$ \citep{WangLiHuang2008, WeiHuangLi2011}. However, \citet{deLichtenberg2005} identified 113 yeast genes most likely to be periodically expressed (or to display periodicities over time) in small-scale experiments, including 104 genes used by \citet{Spellman1998}. This suggests that at least some genes display temporal correlation, and the independence assumptions previously used are not appropriate. The NVC-SSL model allows us to flexibly model the genes' temporal correlations by decomposing the error $\varepsilon_i(t_{ij})$ into a functional random effect (where all $m_i$ random effects $\alpha_i(t_{i1}), \ldots, \alpha_i(t_{im_i})$ for the $i$th subject are correlated) and a measurement error term $r_{ij}$, as in \eqref{varyingcoefficientmodel2}. 

Using the datasets in \citet{Spellman1998} and \citet{Lee2002}, we extracted the 104 genes identified as periodically expressed by \citet{deLichtenberg2005}. After excluding genes with missing values in either of the experiments, we were left with $n = 47$ genes, for a total of $N=846$ observations. 

\subsection{Results for variable selection and out-of-sample prediction}

We compared the NVC-SSL, NVC-gLASSO, NVC-gSCAD, and NVC-gMCP models. BIC was used to select the spike hyperparameter $\lambda_0$ in NVC-SSL and the penalty parameter $\lambda$ in NVC-gLASSO, NVC-gSCAD, and NVC-gMCP. To assess their variable selection performance, we fit these models using all $n=47$ genes. We also examined the models' predictive power. To do so, we randomly divided the dataset into 37 training observations and 10 test observations. We fit the NVC models to the training data and then used our fitted models to predict the trajectories of mRNA level $\widehat{y}(t)$ for the 10 test observations and compute the out-of-sample MSPE. We repeated this procedure 200 times, so that we had 200 different test sets on which to evaluate these different methods.

\begin{figure}[t!]
	\centering
	\includegraphics[scale=0.56]{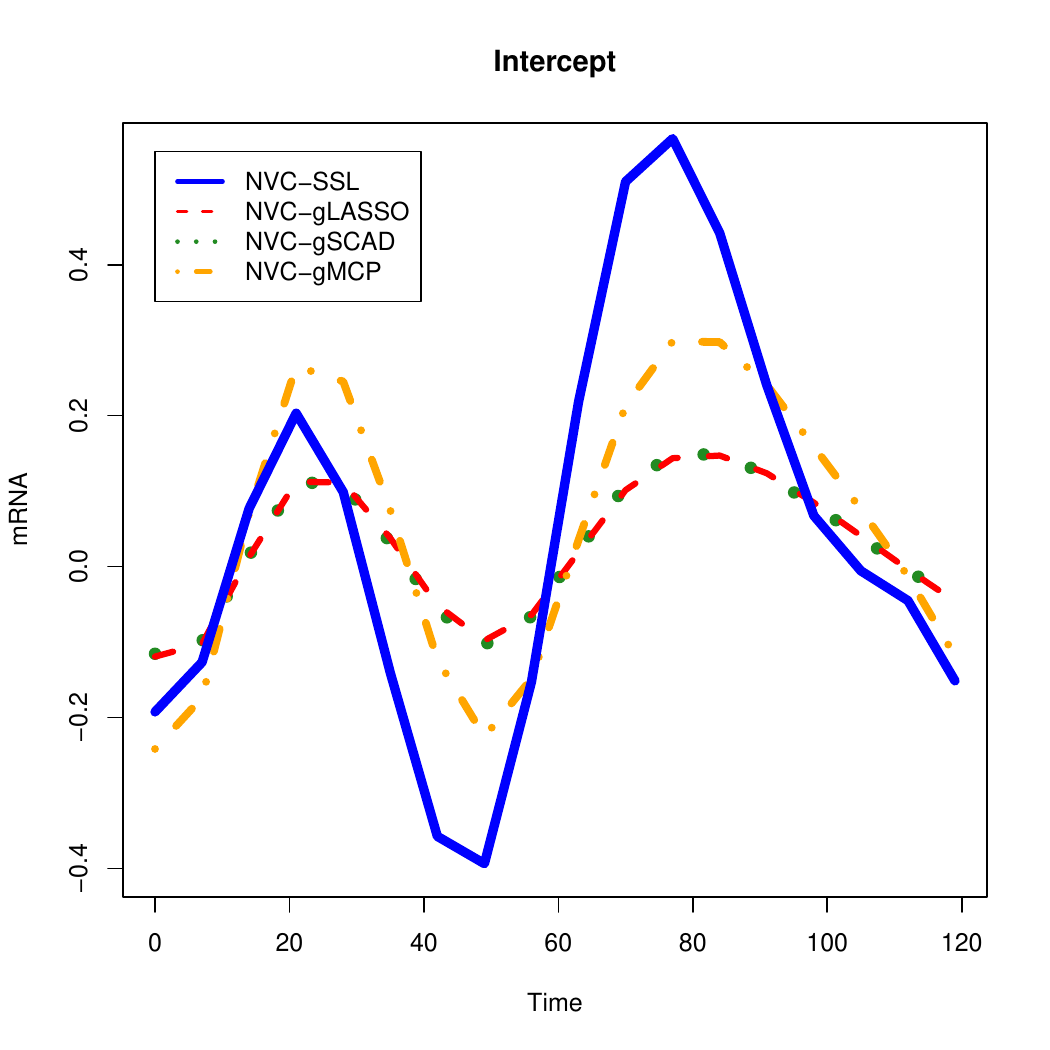} 
	\caption{Plots of the estimated intercept function $\beta_0(t)$ representing the baseline trend in mRNA level over time for NVC-SSL, NVC-gLASSO, NVC-gSCAD, and NVC-gMCP.}
	\label{fig:yeast_beta0}
\end{figure}

All four methods selected the intercept function $\beta_0(t)$. These intercept functions are plotted in Figure \ref{fig:yeast_beta0}. We see that all four methods concluded that there is a baseline periodic trend in mRNA levels over time. However, the NVC-SSL curve for $\beta_0(t)$ is a bit more pronounced and less smooth, with a higher amplitude. 

\begin{figure}[t!]
	\centering
	\includegraphics[scale=0.75]{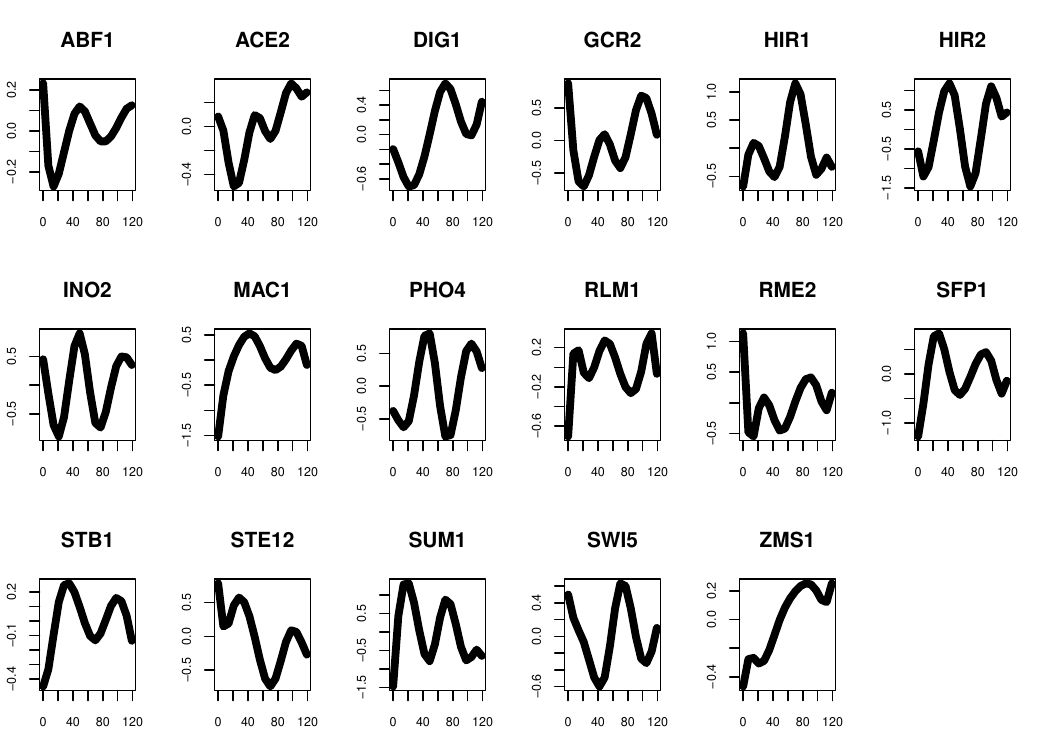} 
	\caption{Plots of the estimated transcriptional effects over time for the 17 TFs selected by NVC-SSL.}
	\label{fig:NVC_SSL_TFs}
\end{figure}

Table \ref{Table:yeast_results} shows our results for the number of TFs selected and the out-of-sample prediction error. The NVC-SSL model selected the most TFs and had the second lowest average MSPE. Figure \ref{fig:NVC_SSL_TFs} gives the names of the 17 TFs selected by NVC-SSL and plots their estimated transcriptional effects over time. NVC-gLASSO, NVC-gSCAD, and NVC-gMCP all selected more parsimonious models, with NVC-gMCP selecting the sparsest model with only three TFs. In addition, NVC-gMCP had the lowest average MSPE. However, the signals in this dataset were rather weak to begin with, so it may not be surprising that the most parsimonious model also had the best predictive accuracy -- a model that always selects the null model on this dataset would likely give a similar predictive performance. We saw from our simulations in Section \ref{simulationsI} that NVC-SSL was better able to detect weak signals, and that may also be the case here.

\begin{table}[t!]  
	\centering
	\caption{Average MSPE on 200 test sets (standard errors in parentheses) and number of transcription factors selected by NVC-SSL, NVC-gLASSO, NVC-gSCAD, and NVC-gMCP.} 
	\begin{tabular}{l c  c }
		\hline
		& MSPE & Number of TFs Selected \\ 
		\hline
		NVC-SSL & 0.512 (0.141) & 17 \\
		NVC-gLASSO & 0.515 (0.129) & 7 \\ 
		NVC-gSCAD & 0.554 (0.290) & 7 \\
		NVC-gMCP & 0.433 (0.167) & 3 \\
		\hline
	\end{tabular} 
	\label{Table:yeast_results}
\end{table}

The NVC-SSL model was able to detect meaningful biological signal in the data.  The cell cycle is an ordered set of events, culminating in cell growth and division into two daughter cells. Stages of the cell cycle are commonly divided into G1-S-G2-M. The G1 stage stands for ``GAP 1.'' The S stage stands for ``Synthesis'' and is the stage when DNA replication occurs. The G2 stage stands for ``GAP 2.'' The M stage stands for ``mitosis,'' when nuclear (chromosomes separate) and cytoplasmic (cytokinesis) division occur.  The NVC-SSL model selected several TFs that have also been shown to be significant at various stages of the cell cycle in the literature. In particular, the NVC-SSL method selected SWI5 and ACE2. \citet{Simon2001} found that the SWI5 and ACE2 proteins activate genes at the end of M and early G1.

The TFs selected by NVC-SSL also included several pairs of syneristic, or ``cooperative,'' TFs that have been reported in the literature \citep{BanerjeeZhang2003, Tsai2005}. These pairs of TFs are thought to cooperate together to regulate transcription in the yeast cell cycle. Among the 17 TFs selected by NVC-SSL, seven of them (ACE2, HIR1, HIR2, STB15, SUM1, SWI5) belonged to cooperative pairs of TFs identified by \citet{BanerjeeZhang2003}, including the complete cooperative pairs HIR1-HIR2 and ACE-SWI5. On the other hand, NVC-gLASSO and NVC-gSCAD only found four genes belonging to cooperative pairs (HIR1, HIR2, SWI5, SWI6) and one complete cooperative pair HIR1-HIR2, while NVC-gMCP found three genes belonging to cooperative pairs (HIR1, STB1, and  SWI5) but no complete cooperative pairs.

\subsection{Variable selection performance with added synthetic noise variables} 

In order to investigate the performance and stability of our variable selection approach in high dimensions, we artificially added 1000 noise variables so that $p = 1097$. For each $i$th gene, these noise variables were randomly generated from a uniform distribution $\text{Uniform}(x_{i,\min}, x_{i,\max})$, where $x_{i,\min}$ and $x_{i,\max}$ denote the minimum and maximum binding information values for the $i$th gene. We then fit the NVC-SSL, NVC-gLASSO, NVC-gSCAD, and NVC-gMCP models with $p=1097$ varying coefficients. With $d=8$ basis functions, we thus had to estimate 8776 unknown basis coefficients $\bm{\gamma}$ in \eqref{approximatemodel}. 

We repeated the above procedure 200 times, adding 1000 artificial noise variables to the original dataset each time. For each of the 200 replications, we recorded the number of true TFs selected and the number of noise variables selected. We also kept track of the TFs that were \emph{always} selected by each method in all 200 experiments. 

\begin{table}[t!]  
	\centering
	\caption{Variable selection results after adding 1000 noise variables to the dataset. The first two columns report the average number of real TFs and the average number of noise variables selected across 200 replications, with the empirical standard error in parentheses. The third column lists the TFs that were selected in all 200 experiments.}  
	\begin{tabular}{l c c c }
		\hline
		& Real TFs & Noise Variables & TFs Always Selected \\ 
		\hline
		NVC-SSL & 6.55 (1.76) & 1.69 (1.15) & HIR1, RME1, SFP1, SWI5  \\	
		NVC-gLASSO & 5.24 (1.75) & 0.09 (0.28) & HIR1 \\ 
		NVC-gSCAD & 6.89 (3.03) & 1.82 (11.68) & HIR1, RME1, SWI5 \\
		NVC-gMCP & 4.40 (3.72) & 8.75 (14.16) & HIR1, SWI5  \\
		\hline
	\end{tabular}
	\label{Table:noise}
\end{table}

Our results are shown in Table \ref{Table:noise}. The NVC-SSL selected on average 6.55 real TFs and 1.685 noise variables. NVC-gSCAD and NVC-gMCP selected more noise variables on average than NVC-SSL. In particular, NVC-gMCP selected an average of 8.75 noise variables and only 4.40 real TFs, indicating that NVC-gMCP performed the worst in terms of being able to exclude noise variables. On this particular dataset, NVC-gLASSO tended to select the most parsimonious model (an average of 5.24 real TFs and 0.09 noise variables), with typically the least number of noise variables selected.

However, Table \ref{Table:noise} also shows that NVC-gLASSO only selected one real TF (HIR1) in all 200 experiments, indicating that variable selection was not as consistent for NVC-gLASSO across the replicates. Although NVC-SSL selected on average 1.685 noise variables, NVC-SSL also exhibited the greatest overall variable selection stability, selecting four real TFs (HIR1, RME1, SFP1, and SWI5) in all 200 replications, compared to three for NVC-gSCAD and two for NVC-gMCP. The four TFs that NVC-SSL selected in all 200 experiments were also selected by NVC-SSL on the original dataset with only $p=97$. Our results demonstrate that variable selection for NVC-SSL is fairly stable in the presence of many known noise variables. 

\section{Discussion} \label{discussion}
In this paper, we have introduced the nonparametric varying coefficient spike-and-slab lasso, a new Bayesian approach for estimation and variable selection in high-dimensional NVC models. The NVC-SSL extends the spike-and-slab lasso methodology \citep{RockovaGeorge2018} to the functional regression setting with dependent responses. NVC-SSL performs simultaneous estimation and variable selection of the functional components. Moreover, the NVC-SSL flexibly models the unknown within-subject covariance structure. This is in sharp contrast to previously frequentist penalized approaches to NVC models which have ignored these temporal correlations entirely or previous Bayesian approaches which have required the prespecification of a parametric covariance structure. Unlike frequentist approaches, the NVC-SSL model also employs a \textit{non}-separable penalty which allows for automatic model complexity control and self-adaptivity to the true sparsity in the data. 

For variable selection and estimation, we introduced an efficient ECM algorithm to rapidly obtain MAP estimates. For uncertainty quantification, we proposed an approximate MCMC algorithm. Both our ECM and approximate MCMC algorithms scale linearly in $p$ and in $N$. We demonstrated through extensive simulation studies and a real data application that our method provides reliable variable selection, function estimation, and uncertainty quantification under a variety of within-subject correlation structures. NVC-SSL is also able to detect weak signals and capture many different function shapes, including functions with flat regions and non-time varying (constant) functions. 

The NVC-SSL enjoys strong theoretical support. However, we have deferred the theoretical treatment of our method to a follow-up paper \citep{Baitheory2023}. \citet{Baitheory2023} gives general sufficient conditions for adaptive posterior contraction in high-dimensional $p \gg n$ Bayesian NVC models (adaptive in the sense that the posterior can adapt to the unknown sparsity level and the unknown smoothness of the varying coefficient functions). The NVC-SSL prior is one particular choice of prior that can be shown to satisfy these sufficient conditions with well-chosen hyperparameters. 

This paper has focused solely on the ``large $p$'' problem, where we implicitly assumed that $N$ was not too large. The ``small $N$, large $p$'' scenario arises in many practical settings such as GWAS studies \citep{LiWangLiWu2015, JohndrowOrensteinBhattacharya2020}. However, it is also worthwhile to explore scalable Bayesian NVC models in the ``large $N$, large $p$'' setting, where \emph{both} $N$ and $p$ could be massive. One possible direction is the prior-preconditioned conjugate gradient (PPCG) method of \citet{nishimura2022shrinkage}. One of the benefits of the approach of \citet{nishimura2022shrinkage} is the fact that it bypasses matrix inversions entirely. In preliminary work, we did attempt to implement a version of the PPCG method for NVC-SSL. However, we found that when $N \ll p$, PPCG was actually \emph{slower} than the \emph{exact} MCMC algorithm that we introduced in Section \ref{MCMC}. This is because \emph{each} MCMC iteration of the PPCG method requires iteratively solving a linear system using conjugate gradient descent (CGD), and the number of iterations it took for the CGD to converge was often greater than $N$. Nevertheless, we believe that PPCG is a useful avenue to pursue if $N$ and $p$ are both large, and the cost of iteratively solving a linear system with CGD is minimal compared to the cost of using direct methods such as Cholesky decomposition.

Another possible direction for scalable uncertainty quantification is to extend the weighted Bayesian bootstrap (WBB) \citep{NewtonPolsonXu2021, NieRockova2022} to the NVC setting. Roughly speaking, WBB methods approximate the posterior by performing MAP estimation on many independently perturbed datasets. Recently, in linear regression with i.i.d. errors, \citet{NieRockova2022} employed WBB to approximately sample from the posterior distribution under spike-and-slab lasso priors. The approach of \citet{NieRockova2022} is shown to scale favorably in both $N$ and $p$. However, WBB requires the observations to be independent, which is \emph{not} the case for the NVC models in this paper. An extension of WBB to the \emph{dependent} data setting considered in this paper is also of interest. 

\section*{Acknowledgments}
The authors are grateful to the Action Editor and the anonymous referee for very constructive feedback which greatly improved this paper. This work was initiated when the first listed author was a postdoc at the Perelman School of Medicine, University of Pennsylvania (UPenn) mentored by the last two authors and the second listed author was an Assistant Professor at UPenn. Dr. Ray Bai was funded in part by generous funding from the National Science Foundation (DMS-2015528). Dr. Mary Boland was funded in part by generous funding from the Perelman School of Medicine. Dr. Yong Chen was supported in part by National Institutes of Health grants 1R01LM012607 and 1R01AI130460.

\bibliographystyle{apa}
\bibliography{VCSSLreferences}

\begin{appendix}

\counterwithin{figure}{section}
\setcounter{figure}{0}    

\counterwithin{table}{section}
\setcounter{table}{0}    

\section{Additional simulation results}  \label{App:A}

In this section, we compare the results obtained from MCMC for Experiments 1-5 in Section \ref{simulationsI} to those obtained from the ECM algorithm. Namely, we assessed the performance of the estimated posterior mean as a point estimate. We also used MPM \citep{BarbieriBerger2004, BarbieriBA2021} to perform variable selection. Unlike the MAP estimator, the posterior mean under the NVC-SSL model is \emph{not} exactly sparse. However, for spike-and-slab models, one can threshold the posterior inclusion probabilities $P(\tau_k = 1 \mid \bm{Y}), k = 1, \ldots, p$, to select variables. These posterior inclusion probabilities can be estimated as
\begin{align*}
	\widehat{P}(\tau_k = 1 \mid \bm{Y}) = \frac{1}{T-B} \sum_{t=B+1}^{T} \tau_k^{(t)},
\end{align*}
where $\tau_k^{(t)}$ is the $t$th MCMC sample drawn for $\tau_k$ in Step 4(a) of Algorithm \ref{algorithm2}, $T$ is the total number of MCMC iterations, and $B$ is the number of burnin samples. In the present context, MPM selects the $k$th varying coefficient $\beta_k(t)$ if $P(\tau_k = 1 \mid \bm{Y}) \geq 0.5$.

We repeated Experiments 1-5 from Section \ref{simulationsI} for 200 replications each, where we used the exact MCMC algorithm in Section \ref{exactMCMC} to estimate the varying coefficients. The MAP estimator was used to initialize the MCMC algorithm, and the hyperparameters were the same as those in the ECM algorithm. We ran the algorithm for 2000 iterations, with a burnin period of 500 samples. The effective sample size prior to burnin was very close to 2000 for each of the basis coefficients in $\bm{\gamma}$, suggesting that the number of MCMC iterations we used was sufficient.

\begin{table}[t!]
	\centering
	\caption{Simulation results for NVC-SSL using the exact MCMC algorithm in Section \ref{exactMCMC} vs. the ECM algorithm in Section \ref{ECM}, averaged across 200 replicates. To better highlight the differences in estimation performance, we rescale the MSE by 100, i.e. we report $\text{MSE} \times 100$ in the first column. The empirical standard error is reported in parentheses following the average. For MCMC, the posterior mean is used to compute the MSE and MSPE, and the MPM is used to select variables and compute Sens, Spec, and MCC. For ECM, the MAP estimator is used to compute all performance metrics. }
	\label{table:MCMC_vs_ECM_overall_results}
	\medskip
	\resizebox{\textwidth}{!}{
		\begin{tabularx}{1.1\linewidth}{l *{11}X}
			\multicolumn{6}{c}{\textbf{Experiment 1: AR(1) covariance function}} \\ \toprule
			& MSE $\times$ 100 & MSPE & Sens & Spec & MCC \\ 
			\hline \hline
			MCMC & 0.670 (0.095) & 6.579 (2.297) & 0.667 (0) & \textbf{1} (0) & 0.815 (0) \\
			ECM & \textbf{0.144} (0.049) & \textbf{4.057} (1.976) & \textbf{0.988} (0.052) & 0.999 (0.002) & \textbf{0.948} (0.069) \\ 
			\bottomrule
	\end{tabularx}}
	
	\medskip
	
	\resizebox{\textwidth}{!}{
		\begin{tabularx}{1.1\linewidth}{l *{11}X}
			\multicolumn{6}{c}{\textbf{Experiment 2: CS covariance function}} \\ \toprule
			& MSE $\times$ 100 & MSPE & Sens & Spec & MCC \\ 
			\hline \hline
			MCMC & 0.679 (0.102) & 6.597 (2.710) & 0.667 (0) & \textbf{1} (0) & 0.815 (0) \\
			ECM & \textbf{0.113} (0.046) & \textbf{4.255} (2.254) & \textbf{0.989} (0.041) & 0.999 (0.002) & \textbf{0.947} (0.067) \\ 
			\bottomrule
	\end{tabularx}}
	
	\medskip
	
	\resizebox{\textwidth}{!}{
		\begin{tabularx}{1.1\linewidth}{l *{11}X}
			\multicolumn{6}{c}{\textbf{Experiment 3: SE covariance function}} \\ \toprule
			& MSE $\times$ 100 & MSPE & Sens & Spec & MCC \\ 
			\hline \hline
			MCMC & 0.663 (0.098) & 6.318 (2.187) & 0.667 (0) & \textbf{1} (0) & 0.815 (0) \\
			ECM & \textbf{0.112} (0.056) & \textbf{4.448} (2.536) & \textbf{0.989} (0.053) & 0.999 (0.002) & \textbf{0.946} (0.071) \\ 
			\bottomrule
	\end{tabularx}}
	
	\medskip
	
	\resizebox{\textwidth}{!}{
		\begin{tabularx}{1.1\linewidth}{l *{11}X}
			\multicolumn{6}{c}{\textbf{Experiment 4: Periodic covariance function}} \\ \toprule
			& MSE $\times$ 100 & MSPE & Sens & Spec & MCC \\ 
			\hline \hline
			MCMC & 0.663 (0.110) & 6.511 (2.324) & 0.667 (0) & \textbf{1} (0) & 0.815 (0) \\
			ECM & \textbf{0.106} (0.042) & \textbf{4.062} (2.134) & \textbf{0.991} (0.045) & 0.999 (0.002) & \textbf{0.965} (0.054) \\ 
			\bottomrule
	\end{tabularx}}
	
	\medskip
	
	\resizebox{\textwidth}{!}{
		\begin{tabularx}{1.1\linewidth}{l *{11}X}
			\multicolumn{6}{c}{\textbf{Experiment 5: Zero covariance function (i.i.d. errors)}} \\ \toprule
			& MSE $\times$ 100 & MSPE & Sens & Spec & MCC \\ 
			\hline \hline
			MCMC & 0.619 (0.098) & 5.198 (2.306) & 0.667 (0) & \textbf{1} (0) & 0.815 (0) \\
			ECM & \textbf{0.062} (0.024) & \textbf{3.027} (2.269) & \textbf{0.999} (0.012) & 0.999 (0.002) & \textbf{0.962} (0.055) \\ 
			\bottomrule
	\end{tabularx}}
\end{table}

\begin{table}[t!]
	\centering 
	\caption{Simulation results for estimation and variable selection of the nonzero varying coefficients $\beta_k(t), k = 1, \ldots, 6$, using the exact MCMC algorithm in Section \ref{exactMCMC} vs. the ECM algorithm in Section \ref{ECM}, averaged across 200 replicates. ``Proportion'' gives the proportion of replicates that selected the varying coefficient. } \label{table:ind_beta_MCMC_vs_ECM_results}
	\medskip
	\resizebox{.94\textwidth}{!}{  
		\begin{tabular}{lccccccccccccc} 
			\multicolumn{14}{c}{\textbf{Experiment 1: AR(1) covariance function}} \\
			\midrule
			\phantom{abc} & \multicolumn{6}{c}{MSE} & \phantom{abc}& \multicolumn{6}{c}{Proportion} \\
			\cmidrule{2-7} \cmidrule{9-14} 
			& $\beta_1$ & $\beta_2$ & $\beta_3$ & $\beta_4$ & $\beta_5$ & $\beta_6$ & &  $\beta_1$ & $\beta_2$ & $\beta_3$ & $\beta_4$ & $\beta_5$ & $\beta_6$ \\ \midrule 
			MCMC & 0.252 & 0.693 & 0.344 & 0.440 & 0.657 & 0.177 & & 1 & 1 & 0 & 0 & 1 & 1 \\
			ECM & \textbf{0.060} & \textbf{0.081} & \textbf{0.093} & \textbf{0.103} & \textbf{0.113} & \textbf{0.042} & & 1 & 1 & \textbf{0.97} & \textbf{0.96} & 1 & 1 \\ 
			\bottomrule
		\end{tabular}
	}
	
	\medskip
	\resizebox{.94\textwidth}{!}{  
		\begin{tabular}{lccccccccccccc} 
			\multicolumn{14}{c}{\textbf{Experiment 2: CS covariance function}} \\
			\midrule
			\phantom{abc} & \multicolumn{6}{c}{MSE} & \phantom{abc}& \multicolumn{6}{c}{Proportion} \\
			\cmidrule{2-7} \cmidrule{9-14} 
			& $\beta_1$ & $\beta_2$ & $\beta_3$ & $\beta_4$ & $\beta_5$ & $\beta_6$ & &  $\beta_1$ & $\beta_2$ & $\beta_3$ & $\beta_4$ & $\beta_5$ & $\beta_6$ \\ \midrule 
			MCMC & 0.254 & 0.728 & 0.349 & 0.440 & 0.656 & 0.181 & & 1 & 1 & 0 & 0 & 1 & 1 \\
			ECM & \textbf{0.066} & \textbf{0.080} & \textbf{0.094} & \textbf{0.094} & \textbf{0.114} & \textbf{0.045} & & 1 & 1 & \textbf{0.965} & \textbf{0.97} & 1 & 1 \\ 
			\bottomrule
		\end{tabular}
	}
	
	\medskip
	\resizebox{.94\textwidth}{!}{  
		\begin{tabular}{lccccccccccccc} 
			\multicolumn{14}{c}{\textbf{Experiment 3: SE covariance function}} \\
			\midrule
			\phantom{abc} & \multicolumn{6}{c}{MSE} & \phantom{abc}& \multicolumn{6}{c}{Proportion} \\
			\cmidrule{2-7} \cmidrule{9-14} 
			& $\beta_1$ & $\beta_2$ & $\beta_3$ & $\beta_4$ & $\beta_5$ & $\beta_6$ & &  $\beta_1$ & $\beta_2$ & $\beta_3$ & $\beta_4$ & $\beta_5$ & $\beta_6$ \\ \midrule 
			MCMC & 0.256 & 0.680 & 0.345 & 0.435 & 0.622 & 0.179 & & 1 & 1 & 0 & 0 & 1 & 1 \\
			ECM & \textbf{0.063} & \textbf{0.076} & \textbf{0.087} & \textbf{0.098} & \textbf{0.106} & \textbf{0.043} & & 1 & 1 & \textbf{0.975} & \textbf{0.96} & 1 & 1 \\ 
			\bottomrule
		\end{tabular}
	}
	
	\medskip
	\resizebox{.94\textwidth}{!}{  
		\begin{tabular}{lccccccccccccc} 
			\multicolumn{14}{c}{\textbf{Experiment 4: Periodic covariance function}} \\
			\midrule
			\phantom{abc} & \multicolumn{6}{c}{MSE} & \phantom{abc}& \multicolumn{6}{c}{Proportion} \\
			\cmidrule{2-7} \cmidrule{9-14} 
			& $\beta_1$ & $\beta_2$ & $\beta_3$ & $\beta_4$ & $\beta_5$ & $\beta_6$ & &  $\beta_1$ & $\beta_2$ & $\beta_3$ & $\beta_4$ & $\beta_5$ & $\beta_6$ \\ \midrule 
			MCMC & 0.246 & 0.681 & 0.343 & 0.438 & 0.623 & 0.183 & & 1 & 1 & 0 & 0 & 1 & 1 \\
			ECM & \textbf{0.065} & \textbf{0.079} & \textbf{0.092} & \textbf{0.098} & \textbf{0.110} & \textbf{0.042} & & 1 & 1 & \textbf{0.97} & \textbf{0.975} & 1 & 1 \\ 
			\bottomrule
		\end{tabular}
	}
	
	\medskip
	\resizebox{.94\textwidth}{!}{  
		\begin{tabular}{lccccccccccccc} 
			\multicolumn{14}{c}{\textbf{Experiment 5: Zero covariance function (i.i.d. errors)}} \\
			\midrule
			\phantom{abc} & \multicolumn{6}{c}{MSE} & \phantom{abc}& \multicolumn{6}{c}{Proportion} \\
			\cmidrule{2-7} \cmidrule{9-14} 
			& $\beta_1$ & $\beta_2$ & $\beta_3$ & $\beta_4$ & $\beta_5$ & $\beta_6$ & &  $\beta_1$ & $\beta_2$ & $\beta_3$ & $\beta_4$ & $\beta_5$ & $\beta_6$ \\ \midrule 
			MCMC & 0.251 & 0.708 & 0.347 & 0.440 & 0.587 & 0.171 & & 1 & 1 & 0 & 0 & 1 & 1 \\
			ECM & \textbf{0.039} & \textbf{0.041} & \textbf{0.045} & \textbf{0.048} & \textbf{0.082} & \textbf{0.026} & & 1 & 1 & \textbf{0.97} & \textbf{0.975} & 1 & 1 \\ 
			\bottomrule
		\end{tabular}
	}
\end{table}

Tables \ref{table:MCMC_vs_ECM_overall_results} and \ref{table:ind_beta_MCMC_vs_ECM_results} report our results from using MCMC to perform estimation and variable selection, contrasted with the results from using the ECM algorithm. Our results show that the MAP estimator obtained from the ECM algorithm gave superior variable selection, both overall (Table \ref{table:MCMC_vs_ECM_overall_results}) \emph{and} with respect to the the six true nonzero varying coefficient functions (Table \ref{table:ind_beta_MCMC_vs_ECM_results}). In particular, the MAP estimator had lower average MSE (both overall and for $\beta_k(t), k = 1, \ldots, p$) across all the different scenarios. Table \ref{table:ind_beta_MCMC_vs_ECM_results} shows that the MPM method consistently failed to select the weak signals $\beta_3$ and $\beta_4$, similar to the competing methods NVC-gLASSO, NVC-gSCAD, and NVC-gMCP (Table \ref{table:ind_beta_point_estimate_results}). This is demonstrated in Figure \ref{fig:NVC_mode_vs_mean}, which shows that the MAP estimator is better able to detect and capture the true shape of smaller magnitude varying coefficient functions than the posterior mean. 

On the other hand, uncertainty quantification from the 95\% posterior credible intervals obtained from MCMC was quite good, with a pointwise ECP of 0.999 in all simulation settings. This is illustrated by the credible bands displayed in Figure \ref{fig:NVC_mode_vs_mean}. We therefore conclude that the NVC-SSL MAP estimator is preferable for the task of \emph{variable selection} -- especially in the presence of weak signals, while the MCMC algorithm is very useful for the task of \emph{uncertainty quantification}.

\section{Proof of Proposition 1}  \label{App:B}

Under the exact MCMC algorithm, the conditional distribution of $\bm{\gamma}$ in Step 6 of Algorithm \ref{algorithm2} has the covariance matrix, 
\begin{align*}
	\bm{\Sigma}_{\bm{\gamma}} = \left( \bm{U}^\top \bm{U} / \sigma^2 + \bm{D}_{\bm{\xi}}^{-1} \right)^{-1} & = \begin{pmatrix} \bm{U}_S^\top \bm{U}_S / \sigma^2 + \bm{D}_S^{-1} & \bm{U}_S^\top \bm{U}_{S^c} / \sigma^2 \\ \bm{U}_{S^c}^\top \bm{U}_S / \sigma^2 & \bm{U}_{S^c}^\top \bm{U}_{S^c} / \sigma^2 + \bm{D}_{S^c}^{-1} \end{pmatrix}^{-1} \\
	& \overset{\Delta}{=} \begin{pmatrix} \bm{A} & \bm{B} \\ \bm{B}^\top & \bm{C} \end{pmatrix}^{-1}. 
\end{align*}
Since $\bm{A} = \bm{U}_S^\top \bm{U}_S / \sigma^2 + \bm{D}_{S}^{-1}$ and $\bm{C} = \bm{U}_{S^c}^\top \bm{U}_{S^c} / \sigma^2 + \bm{D}_{S^c}^{-1}$ are both positive-definite (with smallest eigenvalues greater than or equal to $[\max_{1 \leq k \leq p} \{ \xi_k \}]^{-1} > 0$), we can write
\begin{align} \label{Sigmainv}
	\bm{\Sigma}_{\bm{\gamma}} = \begin{pmatrix} \bm{A}^{-1} + \bm{A}^{-1} \bm{B} (\bm{C} - \bm{B}^\top \bm{A}^{-1} \bm{B})^{-1} \bm{B}^\top \bm{A}^{-1} & -\bm{A}^{-1} \bm{B} ( \bm{C} - \bm{B}^\top \bm{A}^{-1} \bm{B})^{-1} \\ -( \bm{C} - \bm{B}^\top \bm{A}^{-1} \bm{B} )^{-1} \bm{B}^\top \bm{A}^{-1} & (\bm{C}-\bm{B}^\top \bm{A}^{-1} \bm{B})^{-1}  \end{pmatrix},
\end{align}
and the Schur complement $(\bm{C} - \bm{B}^\top \bm{A}^{-1} \bm{B})^{-1}$ is also positive-definite. Thus, noting that $\widetilde{\bm{\Sigma}}_{\bm{\gamma}_S} = \bm{A}^{-1}$ by \eqref{Sigmatilde}, we have that
\begin{align*}
	\widetilde{\bm{\Sigma}}_{\bm{\gamma}_S} - \bm{\Sigma}_{\bm{\gamma}_S} = \bm{A}^{-1} \bm{B} ( \bm{C} - \bm{B}^\top \bm{A}^{-1} \bm{B} )^{-1} \bm{B}^\top \bm{A}^{-1}. 
\end{align*}
But for any $\bm{x} \in \mathbb{R}^{ds}$,
\begin{align*}
	\bm{x}^\top ( \widetilde{\bm{\Sigma}}_{\bm{\gamma}_s} - \bm{\Sigma}_{\bm{\gamma}_s} ) \bm{x} & = \bm{x}^\top \bm{A}^{-1} \bm{B} ( \bm{C} - \bm{B}^\top \bm{A}^{-1} \bm{B} )^{-1} \bm{B}^\top \bm{A}^{-1} \bm{x} \\
	& = \lVert ( \bm{C} - \bm{B}^\top \bm{A}^{-1} \bm{B} )^{-1/2} \bm{B}^\top \bm{A}^{-1} \bm{x} \rVert_2^2 \geq 0. 
\end{align*}
Thus, $\widetilde{\bm{\Sigma}}_{\bm{\gamma}_S} - \bm{\Sigma}_{\bm{\gamma}_S}$ is non-negative definite, i.e. $\widetilde{\bm{\Sigma}}_{\bm{\gamma}_S} \geq \bm{\Sigma}_{\bm{\gamma}_S}$.

Now, using the facts that $\widetilde{\bm{\Sigma}}_{\bm{\gamma}_{S^c}} = \bm{D}_{S^c}^{-1}$, $\bm{\Sigma}_{\bm{\gamma}_{S^c}} = (\bm{C} - \bm{B}^\top \bm{A}^{-1} \bm{B} )^{-1}$, and $\bm{B} = \bm{U}_S^\top \bm{U}_{S^c} / \sigma^2$ by \eqref{Sigmatilde} and \eqref{Sigmainv}, two applications of the Woodbury matrix identity give that
\begin{align*}
	& \widetilde{\bm{\Sigma}}_{\bm{\gamma}_{S^c}} - \bm{\Sigma}_{\bm{\gamma}_{S^c}}  \\ & \qquad = \bm{D}_{S^c} \frac{\bm{U}_{S^c}^\top}{\sigma} \left( \bm{I}_N + \frac{\bm{U}_{S^c} \bm{D}_{S^c} \bm{U}_{S^c}^\top}{\sigma^2} \right)^{-1} \frac{\bm{U}_{S^c}}{\sigma} \bm{D}_{S^c} + \bm{C}^{-1} \bm{B}^\top ( \bm{A}^{-1} + \bm{B} \bm{C}^{-1} \bm{B}^\top )^{-1} \bm{B}\bm{C}^{-1}.
\end{align*}
Now, for any $\bm{y} \in \mathbb{R}^{d(p-s)}$, we have
\begin{align*}
	& \bm{y}^\top ( \widetilde{\bm{\Sigma}}_{\bm{\gamma}_{S^c}} - \bm{\Sigma}_{\bm{\gamma}_{S^c}} ) \bm{y} \\
	& = \bm{y}^\top \bm{D}_{S^c} \frac{\bm{U}_{S^c}^\top}{\sigma} \left( \bm{I}_N + \frac{\bm{U}_{S^c} \bm{D}_{S^c} \bm{U}_{S^c}^\top}{\sigma^2} \right)^{-1} \frac{\bm{U}_{S^c}}{\sigma} \bm{D}_{S^c} \bm{y} + \bm{y}^\top \bm{C}^{-1} \bm{B}^\top ( \bm{A}^{-1} + \bm{B} \bm{C}^{-1} \bm{B}^\top )^{-1} \bm{B}\bm{C}^{-1} \bm{y} \\
	& = \bigg\lVert \left( \bm{I}_N + \frac{\bm{U}_{S^c} \bm{D}_{S^c} \bm{U}_{S^c}^\top}{\sigma^2} \right)^{-1/2} \frac{\bm{U}_{S^c}}{\sigma} \bm{D}_{S^c} \bm{y} \bigg\rVert_2^2 + \lVert (\bm{A}^{-1} + \bm{B} \bm{C}^{-1} \bm{B}^\top)^{-1/2} \bm{B} \bm{C}^{-1} \bm{y} \rVert_2^2~\geq 0 + 0.
\end{align*} 
Thus, $\widetilde{\bm{\Sigma}}_{\bm{\gamma}_{S^c}} - \bm{\Sigma}_{\bm{\gamma}_{S^c}}$ is also non-negative definite, i.e. $\widetilde{\bm{\Sigma}}_{\bm{\gamma}_{S^c}} \geq \bm{\Sigma}_{\bm{\gamma}_{S^c}}$. \hfill $\square$

\end{appendix}

\end{document}